\documentclass[epjH]{svjour}
\pdfoutput=1
\usepackage{graphics}
\usepackage{sidecap}
\usepackage{wrapfig}
\def\BT{Bruno Touschek}
\def\RW{Rolf Wider\o e}
\def\W{Wider\o e}
\def\T{Touschek}
\def\bk{{\bf k}}
\begin{document}
%
\title{Bruno Touschek:  particle physicist and father of the $ e^+e^-$ collider}

\author{Luisa Bonolis\inst{1}\fnmsep\thanks{\email{luisa.bonolis@roma1.infn.it}} 
\and Giulia Pancheri\inst{2}\fnmsep\thanks{\email{giulia.pancheri@lnf.infn.it}}}
%
\institute{AIF, History of Physics Group, Via Cavalese 13, 00135 Rome (Italy) \and INFN Frascati National Laboratories, Via E. Fermi 40, 00044 Frascati (Italy)\\  
and\\
  Center for Theoretical Physics, Laboratory for Nuclear Science,  \\
  Massachusetts Institute of Technology,  Cambridge, Massachusetts 02139\\
}

%
%
%
%

\abstract{
This article gives a brief outline of the life and works of the Austrian physicist Bruno 
Touschek, who conceived, 
proposed and,  fifty years ago,  brought to completion the construction of AdA, the first electron-positron storage  ring. The events which led to the approval of the AdA project and the Franco-Italian collaboration which confirmed the feasibility of electron-positron storage rings will be recalled.
We shall   illustrate Bruno Touschek's formation both as a theoretical physicist and as an expert in particle accelerators during the period between the time he had to leave the Vienna Staat Gymnasium in 1938, because of his Jewish origin from the maternal side,  until he arrived in Italy in the early 1950s and, in 1960,  proposed  to  build AdA, in Frascati.  The events which led to  \T 's collaboration with  Rolf Wider\o e in the construction of the first European betatron will be described. The article will make use of  a number of   unpublished as well as previously
unknown documents, which include an early correspondence
with Arnold Sommerfeld and  \BT 's letters  to his family in Vienna from Italy,  Germany and Great Britain. 
The impact of \T 's work on  students and collaborators from  University of Rome will be illustrated through  his work on  QED infrared radiative
corrections to high energy $e^+e^-$ experiments and the book {\it Meccanica Statistica}. }

\maketitle

\tableofcontents
\date{\today}
\listoffigures
\section{Introduction}
\label{sec:introduction}
 Bruno Touschek was born in Vienna on February 3rd, 1921 and died in  Innsbruck, on May 25th, 1978. He was the  theoretical physicist who had the vision 
to propose the construction, and bring to completion, the first  electron-positron storage ring, in Italy in 1960. In barely one month, between February  and March,
 Touschek explored the feasibility of experimenting the physics of $e^+ e^-$ annihilation processes and prepared the actual design for AdA \cite{ada}, whose name comes from the italian acronym Anello di  Accumulazione, namely storage  ring. He then went on to propose a bigger and higher energy  machine named ADONE \cite{adone}, where multiparticle production was first observed
\cite{multiparticle} and the discovery \cite{tingjpsi,richterjpsi} of the $J/\Psi$  was confirmed  \cite{frascatijpsi}.
 The history of AdA and ADONE, and how they came  to be, both in  the mind and in the actions of Bruno Touschek, is a  story which passes
 through all of  Europe, in a geographical and  historical sense as well. Through this 
brief outline of  Bruno Touschek's life, we  shall see how European scientists 
overcame the  past and built a new world of knowledge and discoveries. 

The major source on \T 's life and scientific work is the biography written by   Edoardo Amaldi, published as  a CERN Report in 1981 \cite{amaldi81}, and,  in its Italian version, in Quaderni del Giornale di Fisica \cite{amaldi82}, one year later.  Amaldi's work is unchallenged in its breadth and completeness. However, thirty years have passed since this biography appeared, and, although \BT 's figure and accomplishments have been the subject of many articles and works, most of these are in Italian, and \T 's name is only vaguely remembered by physicists outside Italy. Some perspectives have also changed. The great colliders which would definitely  establish the Standard Model of elementary particles, as we know it at the beginning of LHC era, had not started operating when Amaldi wrote his biography, and  fundamental  discoveries, like those of the W and Z bosons in the 1980s,  had not yet taken place.  Another important source on \T's life  is Rolf Wider\o e's autobiography, edited by Pedro Waloshek and published only in  1994 \cite{wideroe}. \W 's autobiography  contains many long passages about his relationship and his work with Touschek during the war years. In addition, as we describe in detail in the subsection of this introduction dedicated to the sources, new material has appeared which sheds light on some periods of Touschek's life.

\subsection{General outline}
Touschek's life can be roughly divided into four main periods, which span through  the 
Second World War, and  were spent in different European countries, namely in Austria, where he was born, in Germany  both during  and soon after the war, in Scotland, Glasgow,  where   he obtained his doctorate, and then in Italy  from 1953. After moving to Italy,  other important travels in his scientific and personal life include  a period in   France, at Orsay   in 1962 and 1963, a few months at CERN  in Geneva, during his last year, and  the final return to   Austria, where he died in May 1978.

  In order to understand the relevance of Touschek's scientific contributions  and put him in a historical perspective, we start  Sec.~\ref{sec:ada} by recalling  the  major discovery of the $J/\Psi$ in 1974, and then  focus 
 on the birth of electron-positron collisions,  describing  \T 's early years in Rome and how the proposals to build AdA  and  ADONE came to be.  
 We shall   highlight   the importance of Touschek's  work on electron-positron storage rings, positioning his work in Frascati within  the international background efforts. Using unpublished documents and letters from the Archives in Rome and in France, we  will  include a description of the second stage of AdA's work, which took place   at Orsay and  confirmed  the feasibility of electron-positron colliders.   This section includes also a  brief reminder of the work on electron-positron colliders taking place in Soviet Union around the same time. As for  the contemporary American efforts, 
which led to the construction of the Stanford Positron-Electron Accelerating Ring (SPEAR),  which discovered the $J/\Psi$, we refer the reader to the  complete and detailed description  found in \cite{Paris2001}. For more details on AdA's birth and the early work in Frascati  we shall refer to   \cite{amman,bernardini2004,bonolis2005a}.

In Sect.~\ref{sec:beforeitaly} we   go back to   the period in Bruno Touschek's life between the time he had to leave the Vienna Staat Gymnasium, because of his Jewish origins from the maternal side, until  he arrived in Italy in 1952. During this time he met  \RW \  in Germany in 1943,  and started working with him on the theory of the betatron. 
We shall  present an  in-depth discussion of the early years
, 1938--1947, spent between  Austria and Germany. These  years include the war period, and the years in G\"ottigen, where  he obtained  his diploma in physics and established a relationship with Werner Heisenberg. We will then illustrate the years 1947--1952, when he moved to Glasgow, Scotland, where he received  his Ph.D. and where he remained as a lecturer until the end of 1952, when he moved to Rome University. These different periods of his life reflect the unusual  circumstances under which he studied and became  both a  theoretical physicist and   an expert in accelerating machines. 
A number of  previously unknown documents, which give new insights and provide a key to understand much better the distinctive nature of his later scientific activity, will be  discussed in  Subsect. ~\ref{subsec:sources}, dedicated to the available sources.

 In Sect. ~\ref{sec:resummation} we  illustrate Touschek's work in Rome after the proposal to build  AdA and ADONE, and  describe his work on radiative corrections to high energy electron-positron experiments, summarizing the development of soft photon summation in QED and its relevance to present day physics. We shall also mention another instance of   \BT 's influence on the development of  research in theoretical physics at Sapienza University of Rome,\footnote{Sapienza University of Rome was  simply known as  University of Rome until 1981, when a second State University  was established in Rome, in  Tor Vergata, and the old name of  ``La Sapienza'' was reintroduced in common usage.}
through   the book {\it Meccanica Statistica}, with some comments kindly provided to us  by \T 's student and co-author G. Rossi.

  In Sect.~\ref{sec:letters}, in order to complement the narrative of Touschek's war years, 
we   publish  the complete translation of the two post-war letters describing Touschek's imprisonment in Germany and the shooting incident which occurred on the way to the Kiel concentration camp in Spring 1945.
 
\begin{figure*}
\resizebox{0.5\textwidth}{!}{\includegraphics{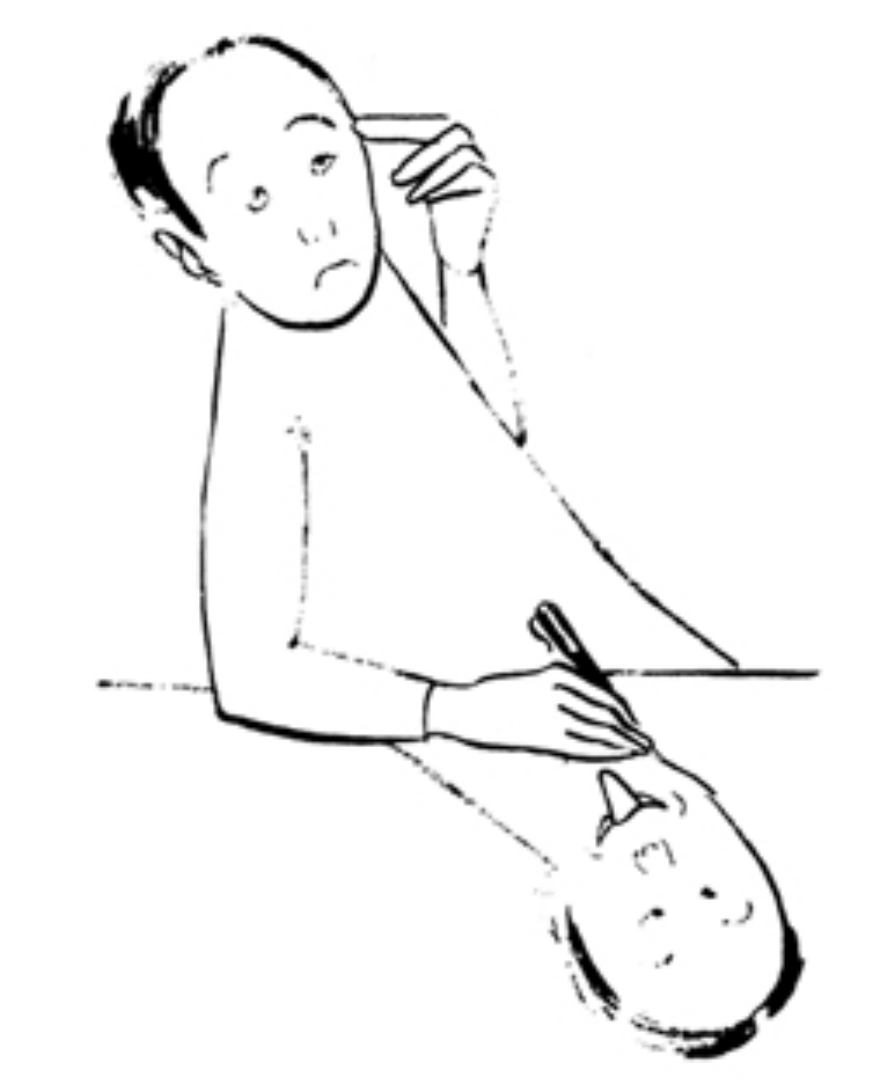}}
\resizebox{0.5\textwidth}{!}{\includegraphics{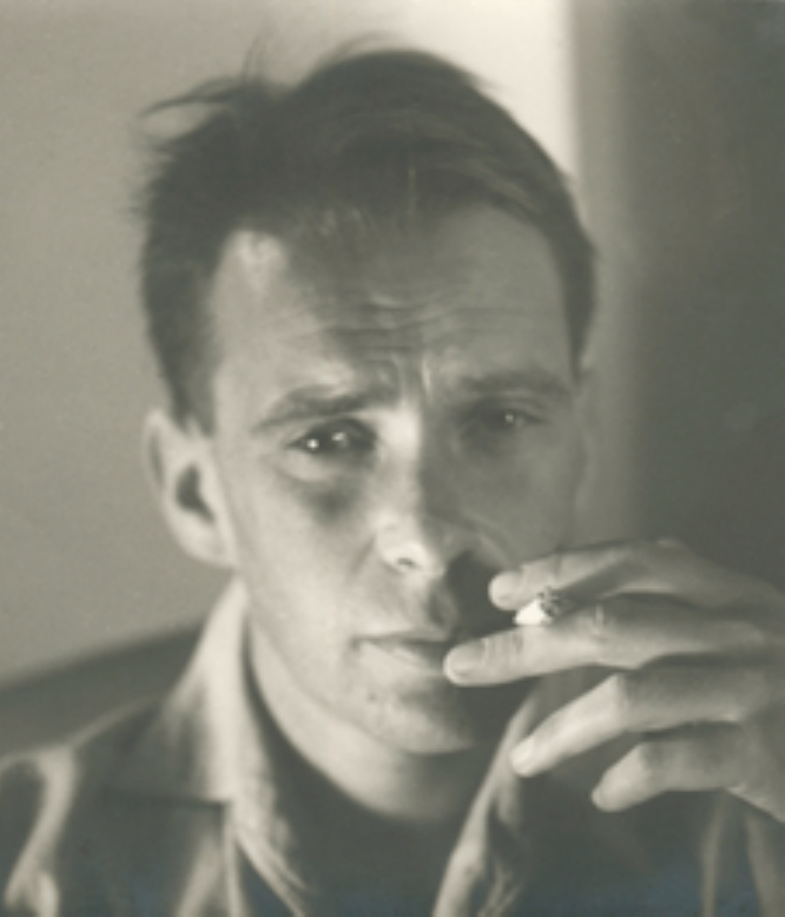}}
\caption{A drawing of T.D. Lee by \BT \ on the left and, on the right,  Bruno Touschek in Rome in 1955.}
\label{fig:tdlee-sigaretta}
\end{figure*}

 \BT \ had inherited from his mother an  artistic bend, which, coupled to   his unique sense of humor, often produced remarkable comments on contemporary life. We show in Fig.~\ref{fig:tdlee-sigaretta} one such drawing as well as a 1955 photograph of Bruno Touschek in Rome. Some  of his best known drawings can be found in \cite{amaldi81,amaldi82}.  In Sect.~\ref{sec:drawings} we reproduce a short selection of unpublished drawings, recently made available to us by  Touschek's family.

 Our present work is not, nor   could  be, an alternative to  \cite{amaldi81,amaldi82}. Our aim is to highlight some aspects of \T 's life which complement Amaldi's work,  through   Touschek's personal papers and the  newly found material contributed by Touschek's family and  with a new perspective on Touschek's contribution to particle physics through the work of his students and collaborators.

\subsection{Available sources }
\label{subsec:sources}

To prepare this article we have made use of both secondary and primary sources.
 The major published source about \BT\  is Edoardo Amaldi's  work \cite{amaldi81,amaldi82}, 
 based on information from \BT \  himself and   from his  friends and collaborators.  When Bruno's health started failing in the spring of 1978, Amaldi gathered  Bruno's  recollections  of his life prior to the arrival in Italy  in a set of notes, which he started  on  February 28,  1978.\footnote{ 
``[Today] at 13 I have gone to see  Bruno Touschek at La Tour Hospital, in Meyrin [\dots] Bruno has started to tell me a part of his life  and I have  taken some notes, which  I am relating here  in an attempt  to reorganize right away what he told me in his extraordinary Italian, incisive and  concrete [\dots]''. Typescript in Amaldi Archive, Physics Department,   Sapienza University of Rome, Box 524, Folder 6.}  
 These  
 notes, drafted under 
\T 's  supervision,  are maintained in the Amaldi Archive in  the Physics Department of Sapienza University  of Rome and  constitute the first nucleus of  the basic work on \BT's life. 
Other publicly  available sources with extensive material can be found in  a collection of memories by scientists who had known \BT \ in Rome or in Geneva  \cite{greco2005}.  New material with direct video interviews was collected during the preparation of the movie {\it Bruno Touschek and the art of physics} \cite{movie}.

  Information on the events which led to \T  's work on the betatron can also be found in  Wider\o e's autobiography \cite{wideroe}, which was not available to Amaldi.  Some of the material concerning \BT 's work on the first  European betatron and the war years, was known to Amaldi through  the letters sent  by \W  \ to Amaldi in 1979, in response to  his   queries about \T  's life, and can be found in  \cite{amaldi81}. \W 's authobiography however throws light on other relevant aspects and we have made use of it in preparing this article.  Likewise, other  important sources of personal recollections about \T \ are  contained in the correspondence between Amaldi and a series of people who had known \BT\ during his life in Germany and in Glasgow. Together with other documents, these letters are preserved in Amaldi Archive, and of course were not used in their  entirety in \cite{amaldi81,amaldi82}.
We have made  use of this source, but have also accessed other sources  which had not been available to Amaldi.  
In  preparing   the biography, Amaldi was not able to use Touschek's 
scientific  papers which were collected and catalogued only in subsequent  years. A complete catalogue of these papers, maintained  in the Physics Department of Sapienza University in Rome, has been published only in 1989 \cite{battimelli89}. 

A major  novel   primary source on Touschek's life is  material obtained  courtesy of Touschek's family and still preserved by his  wife, Mrs.  Elspeth Yonge Touschek, who has kindly made it available to the authors of this article. This material, which had never been examined before, includes  so far unknown  personal documents from his life prior to the war, a great number of drawings and, notably, more than 100 letters sent by Touschek to his father and stepmother during the period 1939--1971.\footnote{The letters are usually addressed to the parents, occasionally only to the father.}
  These letters, generally very long and detailed, have been very important in  establishing  the precise chronology of certain  crucial periods in \T 's life, as well as  for  understanding his early formation as a physicist. 
  
   Excerpts from  \T 's letters to his father     throw new light  on  his  initial involvement  with  the project of the German betatron and his collaboration with Rolf  \W . 
  Using these letters 
  and   other family  documents, 
we have  been able   to clarify  some contradictions between     Touschek's  biography by Amaldi \cite{amaldi81,amaldi82} and  other published sources.  In such instances, as highlighted below, 
    \T 's letters, as the  oldest primary documents, have been used to establish the sequence of events.

A source of confusion has  been  
  the year in which \T  , as a private external student, obtained  his high school diploma ({\it matura}),  which we can now place in February 1939. Other contradictions include the date when he  left Austria and moved to Germany  under Arnold Sommerfeld's protection,   the date of \T 's  imprisonment at the end of the war, and the shooting accident on the way to the Kiel concentration camp. Amaldi is dating Touschek's imprisonment at the beginning of 1945, whereas \W \ in a letter to Amaldi dates  it in November  1944 and, in his autobiography,  November--December 1944.\footnote{Touschek himself, in a  {\it Curriculum Vitae}  prepared after 1970, states to have  spent in prison  the first four months of 1945.} This confusion is clarified by Touschek's letters,  one   dated March 13, 1945, which is the last received by his parents before being arrested for espionage, and   two post-war letters,  dated June 22nd  and November 17th, 1945.
These letters give a detailed, occasionally  day by day description of the period  between March 13th and April 30, when he was set free, and establish  the  sequence of events during the dramatic  months  preceding the end of World War II.

  A series of letters exchanged during the 1950s with Wolfgang Pauli during  the 1950s, is preserved in Pauli's Archive at CERN. Together with other scientific correspondence Touschek had at the time, notably with Max Born and Werner Heisenberg,   they highlight  Touschek's relationship with the fathers of modern physics in Europe. Courtesy of the Deutsches Museum in Munich,  we have also retrieved a small group of 
letters 
exchanged 
with Arnold Sommerfeld between the end of 1941 and the beginning of 1942. These letters, which are mainly on scientific issues,  throw  new light on his previously known relationship with Sommerfeld, and on the latter's role in helping Touschek to move from Vienna to Hamburg  where he could  continue his studies and have his first research experiences, and where nobody knew  of his Jewish origin. 

 In addition to the above sources, various articles about the history of science in Europe after the war \cite{demariaetal} and of accelerator physics in Italy \cite{amman,bernardini2004,bonolis2005a,valente,amman1997,bernardini1989,bernardini1991,bernardini1997,cabibbo1997,bonolis2005b,bonolis2007} shall also be recalled in other sections of this article.

At the end of this Introduction, we note that    \T 's publications have also been an important source of information on his scientific  achievements and we have referred to some of them in connection with  specific episodes. We refer the interested reader to Amaldi's biography \cite{amaldi81,amaldi82} for the  complete list.
 \section{Touschek and the AdA proposal}
\label{sec:ada}

\subsection{When it all came together}

On November 11th, 1974 the simultaneous announcement of the discovery of the
 $J/\Psi$ \cite{tingnobel} by two research groups \cite{tingjpsi,richterjpsi}
 led respectively by  Sam Ting and Burt Richter, opened a new era in particle
 physics. This far reaching discovery was
 made at a traditional proton machine in Brookhaven National Laboratory and at
 a relatively new type of accelerator,  the electron-positron collider SPEAR 
in Stanford Linear Accelerator Center (SLAC). The discovery was  confirmed  three days later by the Italian 
physicists in Frascati \cite{frascatijpsi}, who, following a telephone call 
from the United States,\footnote{From \cite{tingnobel} one reads: ``On November
 11 we telephoned G. Bellettini, the director of Frascati National Laboratories
 [in Italy], informing him of our results. At Frascati, they started a search 
 on 13 November and called us back on 15 November to tell us excitedly that
 they had also seen the J signal [\dots]''.} immediately started searching for the
 new particle at the Frascati storage ring ADONE.

 Although Frascati\footnote{The Frascati National Laboratories, founded in 1957 by INFN, the Italian Institute for Nuclear Physics,  will be hereafter generically referred to as "Frascati".} had been a pioneer in electron-positron collisions, the Frascati electron-positron collider ADONE had been designed to operate at a lower energy than SPEAR, and, indeed at an energy below the $J/\Psi$. It took a second telephone call, this time from the West coast \cite{mariolia}, to give the exact indication of how far one needed to push the machine energy to see the incredibly high counting rate which signaled the presence of a very narrow resonance at $\sqrt{s}=3.1\ GeV $ and thus confirm the discovery. Ten days later, a second, somewhat more massive particle (the $\Psi$'), clearly related to the first one, was discovered at SLAC, and many other discoveries followed in rapid succession. What has come to be known as the  ``November Revolution'', showed that matter-antimatter collisions in a laboratory setting could compete with the traditional proton machines and were a formidable tool  
 for discovering  new particles.  From then on, experiments performed at electron-positron machines  consolidated thinking about basic forces and about quarks as building blocks of matter, and  changed approaches to performing experiments in high-energy physics.

ADONE had been built in Frascati following an earlier, smaller, prototype
 named AdA.  Both names carry with them \T 's sense of humour. When the name AdA was chosen, \T \ wrote ``My aunt Ada (which is short for Adele) had just died, so that   one could  now justly  say with conviction `Ada is dead long live AdA','' whereas  the name ADONE  (Italian for Adonis) for the higher energy machine which followed AdA in Frascati,  suggested    higher energy and much bigger dimensions, as well as an aspiration to beauty.\footnote{B. Touschek, ``A brief outline of the story of AdA'', excerpts from a talk delivered by Touschek at the Accademia dei Lincei on May 24, 1974 (typescript, B. Touschek Archive, Physics Department, Sapienza University of Rome, Box 11, Folder 92.5).}

The road leading to matter-antimatter collisions in the laboratory, had been laid out  thirty
 years earlier, when two European scientists, the Norwegian Rolf Wider\o e and
 the Austrian born Bruno Touschek, had met in war-ravaged Germany, 
collaborating on the building of a 15 MeV betatron. As recalled in his
 autobiography \cite{wideroe}, it was  \RW, who first thought of having two beams of particles collide head-on in order to maximize the energy available, and even patented his idea at the time.\footnote{See original copy of the patent, submitted on September 8, 1943, which at
the moment was kept secret and appeared only after the war.
``Gleichzeitiger Umlauf von negativen und positiven Teilchen
(Kernreaktionen)'' preserved in Ernst Sommerfeld's personal papers at
Deutsches Museum Archive, NL148,001.  Arnold Sommerfeld's son Ernst, was
an engineer and a lawer specialized in patenting.} He discussed the matter with Touschek during one of their meetings, but it was Touschek, who, in early 1960,  applied the idea of the kinematic advantage to beams of particles of opposite charges and 
 actually proposed and built, in  Frascati National Laboratories, the first electron-positron storage
 ring, named AdA, the Italian acronym for Anello di Accumulazione, literally 
Storage Ring.

The story of this achievement is an illustration of how physics ideas
 start and develop.

\subsection{Italy and the construction of AdA}
At the beginning of the 1950s, modern physics in Italy was resurging  after the disasters of the war, building its post-war blossoming 
on the  
 tradition established 
during the 1930s by personalities like Enrico Fermi, Bruno Rossi, Franco Rasetti, and their  pupils and collaborators, among them Edoardo Amaldi, the youngest of the group.  Fermi left Italy in 1938,   while Amaldi  remained and,  after the war,  together with Gilberto Bernardini,\footnote{Gilberto Bernardini, 1906--1995,
was the first president of the INFN from 1953 to 1959. He directed the CERN Proton Synchrotron experimental research group in the period 1957--1960, and 
was one of the founders of the European Physical Society and its first president until 1970.}
 took upon himself the task of  reconstruction 
of 
Italian physics.   In 1949, when Amaldi became  director of the  Guglielmo Marconi Physics Institute in Rome, he was already  a leading figure for the reemergence of physics in Italy and in  Europe. In the 1950s  he  played  a major role in the birth of CERN and the European Space Agency, as well as in promoting scientific research and science policy at an international level \cite{RubbiaAmaldi}.  The fullfilment of  Fermi's 
 dream of having an accelerator and a laboratory for nuclear research \cite{battimelli1997} became  Amaldi  and Bernardini's  plan for reconstruction, and  a 
 project for the construction of a 1100 MeV  electron synchrotron was put in action. The project was led by the 33 year old  Giorgio Salvini, who would later become the first director of the National Laboratories built in Frascati near Rome during the second half of the 1950s.\footnote{G. Salvini, born in  1920, from 1965 until 1971 was president of INFN, founded in 1953.  In 1959 he gave a series of physics lectures on public television,  which stimulated university  enrollment in physics  and are still remembered by that generation. He has been President of the Accademia dei Lincei and Italian Minister for Research.}   Touschek's arrival in Rome in these years became one of the building blocks of the fullfilments of these dreams.
 
  Touschek   had often come  to  Rome
 because of the presence of his maternal aunt Adele, named Ada.  Visiting from Glasgow in 1951, he  hoped to 
  spend a sabbatical year in Italy.\footnote{During the summer of 1951 he had been in Rome,  
and enquired about the possibility of what he felt might  be a stimulating sojourn in Rome. See \BT\  to his parents from Glasgow, november 8, 1951 (Bruno Touschek's personal papers preserved by Elspeth Yonge Touschek).}
 On September 15, 1952, he was officially offered a position by Edoardo Amaldi, within the newly founded Istituto Nazionale di Fisica Nucleare (INFN).\footnote{E. Amaldi to \BT\, September 15, 1952, Amaldi Archive, Sapienza University of Rome, Box 143, Folder 4, Subfolder 2.}

\begin{figure}
 \centering
\resizebox{0.8\textwidth}{!}{\includegraphics{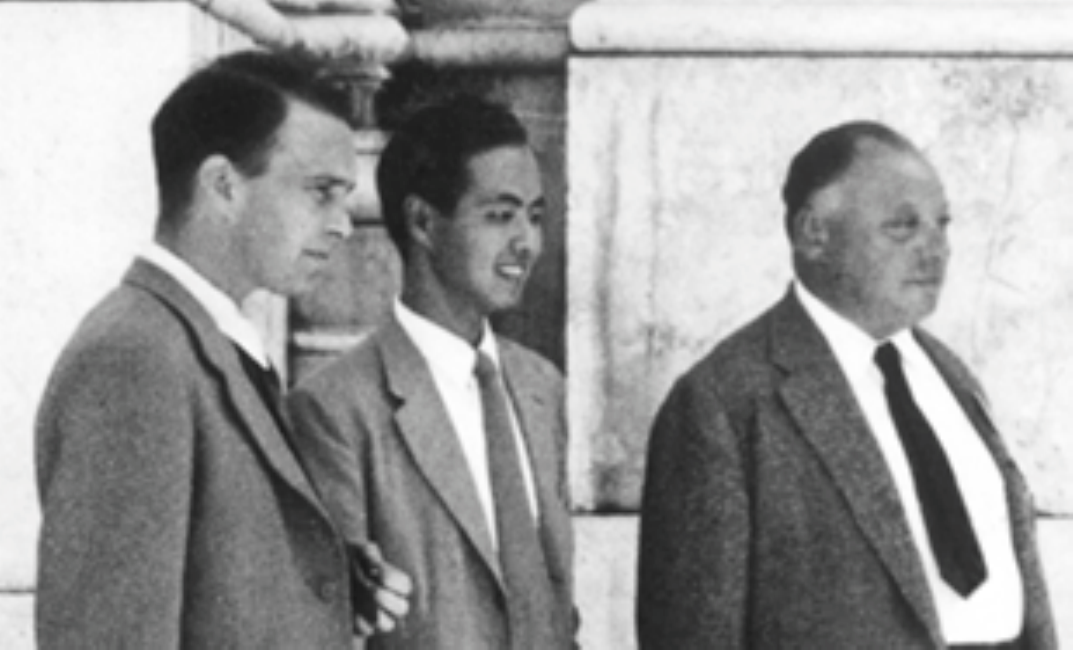}}
\resizebox{0.8\textwidth}{!}{\includegraphics{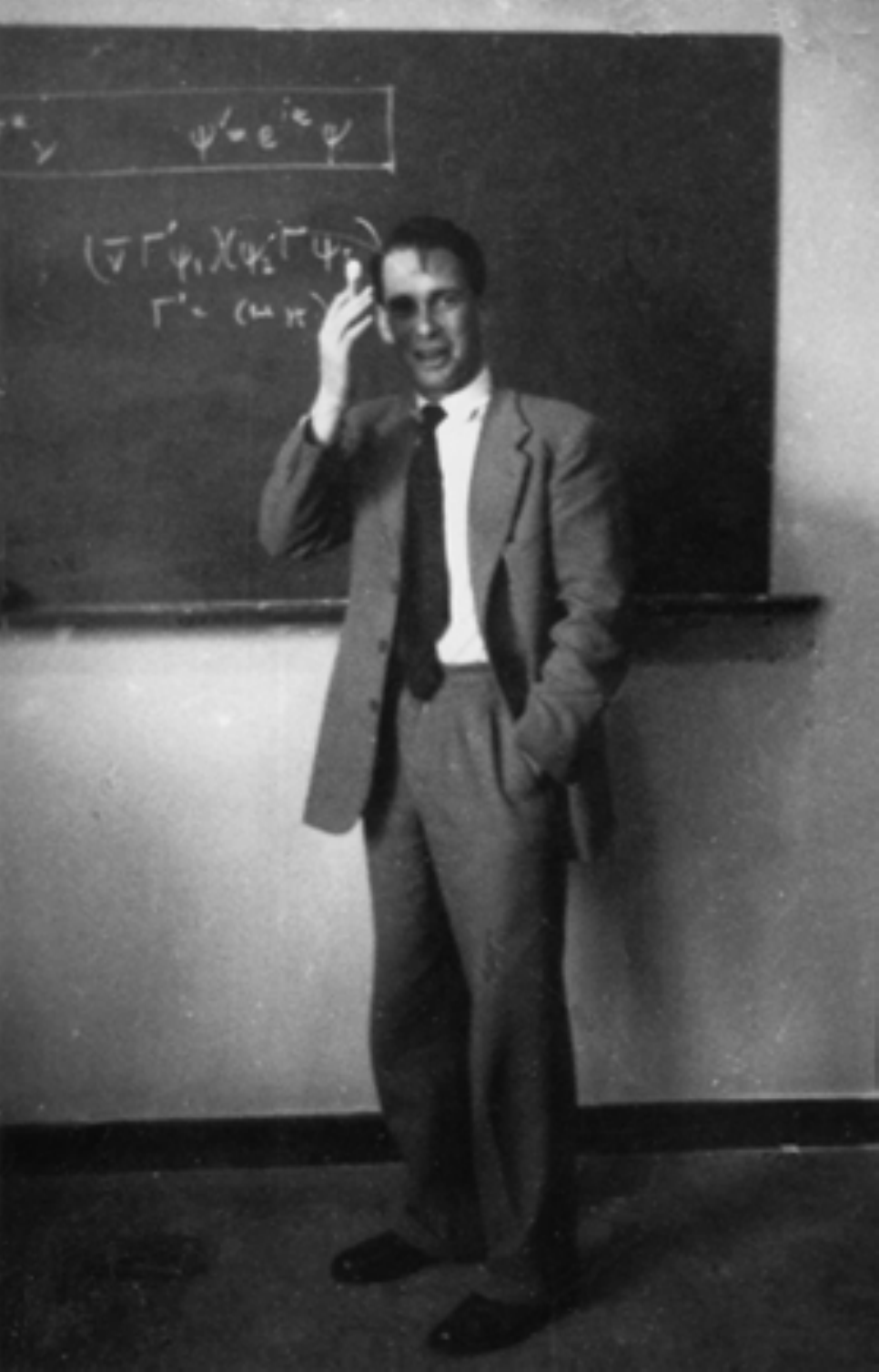}}
\caption{At top \BT \ with T.D. Lee and Wolfgang Pauli during  the 1957 Padua-Venezia International Conference on Weak Interactions.
On the bottom, Bruno Touschek in front of the blackboard in the late 1950s.}
\label{fig:touschekleepauli}
\end{figure}

He immediately  became  very active in  the life of the Rome University Physics  Institute, joining discussions and seminars  and bringing with him the  impressions from a life outside the restricted confines of Italy, the personal acquaintance with the great German and Austrian physicists, Sommerfeld, Pauli, Heisenberg. 
He became good friends with the Amaldi  family
and  fully  immersed  himself in particle physics and in  the debates about parity 
non conservation and invariance under various symmetries, in particular writing  papers on chiral symmetry transformations  \cite{btpaperParity1,btpaperParity2}, and time reversal \cite{btpaperTime}.  In Fig.~\ref{fig:touschekleepauli} we show him during a Conference on Weak Interactions together with T.D. Lee and  Wolfgang Pauli.\footnote{ \T \ was in correspondence with  Pauli until 1958, when  Pauli died, and with T.D. Lee until the mid 1970s. A 1972 letter to T.D. Lee is reproduced in \cite{greco2005}.} 

At the same time, his work on accelerators during the war and in Glasgow during the planning and construction of the 300 MeV Glasgow   synchrotron, made him curious and interested in the Frascati enterprise.  Thus, when  discussions started about building  machines which could test quantum electrodynamics or find new physics,  he was mentally prepared
 toward {\it realizing the impossible} and {\it  thinking the unthinkable}, 
as Carlo Rubbia, who was also in Rome at the time, says in \cite{movie} and in  \cite{crBTML}. 

By the fall of 1959  the first Italian electron syncrotron had reached a full regime of experimentation. With the Frascati machine, Italy had become competitive at an international level, especially with U.S. high energy physics. However, people were already speculating on ultra-high-energy accelerators beyond the Alternating Gradient Synchrotron at Brookhaven National Laboratory  and the  Proton Synchrotron (PS) at CERN, which were beginning  to operate for physics. 
Touschek  had followed the  design and construction of the Italian machine, immediately writing 
a paper with Matthew Sands on the alignment errors in a strong-focusing synchrotron \cite{TouschekSands}.
As he himself recalled years later, when the Italian synchrotron started operating ``new preoccupations arose [\dots] it was felt that if Frascati wanted to keep abreast, something big and new had to be planned.''\footnote{ B. Touschek, ``A brief outline of the story of AdA'', excerpts from a talk delivered by Touschek at the Accademia dei Lincei on May 24, 1974 (typescript, B. Touschek Archive, Physics Department, Sapienza University of Rome, Box 11,  Folder 92.5, p. 5).}
A series of seminars  was  held in Frascati National Laboratories in order to discuss proposals for experiments with the  electron synchrotron, with the CERN Proton Synchrotron  and aiming at developing new lines of research for entering in a new phase.\footnote{E. Amaldi to B. Touschek, November 21, 1959 and G. Salvini to B. Touschek, November 21, 1959 (B. Touschek Archive, Physics Department, Sapienza University of Rome, Box 1, Folder 3).}
 On February 17 1960, Touschek was invited to  a meeting dedicated to the creation   of a theoretical physics group. As he later wrote,\footnote{B. Touschek, AdA and ADONE are storage rings (incomplete typescript, B. Touschek Archive, Physics Department, Sapienza University of Rome, Series III, Section IV, Folder 11, 3.92.4, p. 7.)}
Touschek did not like this idea,  ``It smelled of what in Germany was known as the `Haustheoretiker,' a domesticated animal, which sells itself and what little brain he has to an experimental institution to which it has to be  `useful' [\dots]. I was, however attracted by the possibility of learning how a big enterprise like Frascati worked [\dots]''.
 Instead, he   
 put forward an 
idea, which had been floating around in conferences \cite{salviniregenstreif} or discussions   but had never  been taken  into serious consideration.\footnote{For example, on the occasion of the first international conference on high-energy accelerators held in Geneva in 1956, during the discussion following the Session ``New ideas for accelerating machines'' Giorgio Salvini commented as follows: ``When we have 2 beams, one of positive particles and one of negative particles (travelling in opposite directions), can we expect extra focusing by the magnetic field of one beam acting on the other, or will the  particles simply collapse?" \cite{salviniregenstreif}.}
 Touschek proposed  to forget the  electron synchrotron, which, with its 1100 MeV energy, was  competitive with  the  most powerful of its kind in the world (the other two being at Cornell and at Caltech). To his colleagues, eager and waiting to start their planned experiments, he suggested   transforming  the newly built machine  in a single ring for observing collisions between electrons and positrons.

 The road which would lead to collide together positrons and electrons had started a few month earlier, during a seminar held in Rome by Pief Panofsky, the Director of Stanford High Energy Physics Laboratory (HEPL).
At the time Gerry K. O'Neill, W. Carl Barber, Burton Richter, and Bernie Gittelman  were discussing the construction of an electron-electron collider, following a proposal \cite{O'Neill:1958gk,Barber:1959vg} by Jerry O'Neill of Princeton.  The colliding-beam approach had been first proposed in 1956 by the collaboration Midwestern Universities Research Association and was  based  on the recently discovered fixed-field alternating gradient focusing. The construction of such a complex machine, which was expected to be capable of stacking intense beams of protons with interesting reaction rates, was never approved, but Donald Kerst,  the leader of the Midwestern Universities Research Association group, who had built the first betatron, proposed to use \textit{two} such machines with the beams colliding \cite{kerst}.
O'Neill was interested in proton-proton collisions at high center-of-mass energy \cite{oneil}, but thought that the goals of high intensity and colliding beams could be combined using an ordinary synchrotron to accelerate the particles, and  then accumulating them in two rings which met tangentially, so that the two stored beams could be brought into collision. Stanford High Energy Physics Laboratory  had an ideal source of electrons in the Mark III linear accelerator,  and  O'Neill's approach, with two electron storage rings with one common straight section, was implemented at Stanford starting from 1958. This approach  was presented by O'Neill at CERN in June 1959  \cite{oneil59a,oneil59a1,oneil59b}, in order to perform high-precision experiments and  check the predictions of Quantum Electro-Dynamics.

In the fall of 1959, Pief Panofsky came to Italy and gave  seminars in  Frascati\footnote{ The list of  seminars held in  Frascati during  the period  June 1959--1960, gives  October 26th, 1959, as date of   Panofski's seminar  ``On the two miles linear accelerator'' (G.P. personal collection, courtesy of V. Valente).} and Rome.  During the  seminar in Rome, as  remembered by Nicola Cabibbo  \cite{cabibboinmovie} and Raoul Gatto \cite{gatto2004}, Panofsky presented and discussed the Stanford-Princeton electron-electron collider being built  on the Stanford campus. Touschek's speculations emerged during discussions following the seminar. He was fascinated by the quantum properties of the vacuum 
 which could be probed  through the basic process of vacuum polarisation, and he immediately stressed the relevance of electron-positron reactions proceeding through a state of well-defined ``minimal'' quantum numbers. Touschek's outstanding idea was that, because of symmetry, opposite charges can be stored in one single ring, and made to collide head-on, provided that their masses are equal:  ``Bruno Touschek came up with the remark that an $e^{+}e^{-}$ machine could be realized in a single ring, {\it because of the CTP theorem}'' \cite{cabibbo1997,cabibboinmovie}. The discussions about a single ring are also remembered by Raoul Gatto \cite{gatto2004}, who was at a time a young assistant professor in Rome and who was  to head the Frascati theory group for a while. He remembered that  ``There was a lot of discussion on technical issues on machine building. The question of a single ring instead of two was certainly discussed. I remember how Bruno kept insisting on CPT invariance, which would grant the same orbit for electrons and positrons inside the ring''.\footnote{R. Gatto to L. Bonolis, December 2, 2003.}

Electron-electron collisions would allow to test the photon propagator in the space-like region;  however, Raoul Gatto \cite{gatto2004} recalled that ``Answering to a question, Panofsky mentioned that, to test the electron (rather than the photon) propagator, electron-positron collisions would have been suitable through observation of 2-photon annihilation, but that such a development could present additional technical difficulties and that for the moment it  had been postponed." 
An electron-positron collider would  have actually allowed to explore the time-like photon region as well as the electron propagator in the space-like region through the two-photon annihilation. However, in Touschek's views, the really exhilarating perspective would be the annihilation into hadrons. But  one needed positrons to collide with electrons; thus, as Touschek remarked:  ``The challenge of course consists in having the first machine in which particles which do not naturally live in the world that surrounds us can be kept and conserved.''\footnote{B. Touschek, ``A brief outline of the story of AdA'', excerpts from a talk delivered by Touschek at the Accademia dei Lincei on May 24, 1974 (B. Touschek Archive, Physics Department, Sapienza University of Rome, Box 11, Folder 92.5).}
After the lively discussions following Panofsky's seminar, Touschek continued to reflect on the possibility of performing  an experiment at the cutting edge of research in physics, and especially on the related difficulties. During his opening address at the  already mentioned Frascati meeting of February 17, 1960, he immediately remarked he had thoroughly thought of  what should be the ``future goal'', which might attract theoretical physicists at Frascati Laboratories: ``an experiment aiming at studying electron-positron collisions.'' According to Touschek, such an experiment could be realized modifying the newly  built electron synchrotron.\footnote{Frascati, February 17, 1960, Report of the Meeting (Courtesy of F. Amman).} However, 
``This proposal was not very tactful in front of a meeting of people who had built the machine and were proud of it and others who had spent years in preparing their experiments and were eager to bring them to a conclusion,''  as he  himself  recalled years later.\footnote{ B. Touschek, ``A brief outline of the story of AdA'', excerpts from a talk delivered by
Touschek at the Accademia dei Lincei on May 24, 1974 (B. Touschek Archive, Physics
Department, Rome University Sapienza, Box 11, Folder 92.5), p. 5.} 
  At the end, Touschek stressed the importance of working at improving the intensity (``probably we need to inject with a linear accelerator, rather than with a Van de Graaff''), and listed three  main items as preliminary to the realization of such a project:  Intensity of the beam, Extraction of the beam, Acceleration of the positrons.
 
\begin{figure}
\begin{center}
\resizebox{1.0\textwidth}{!}{
\includegraphics{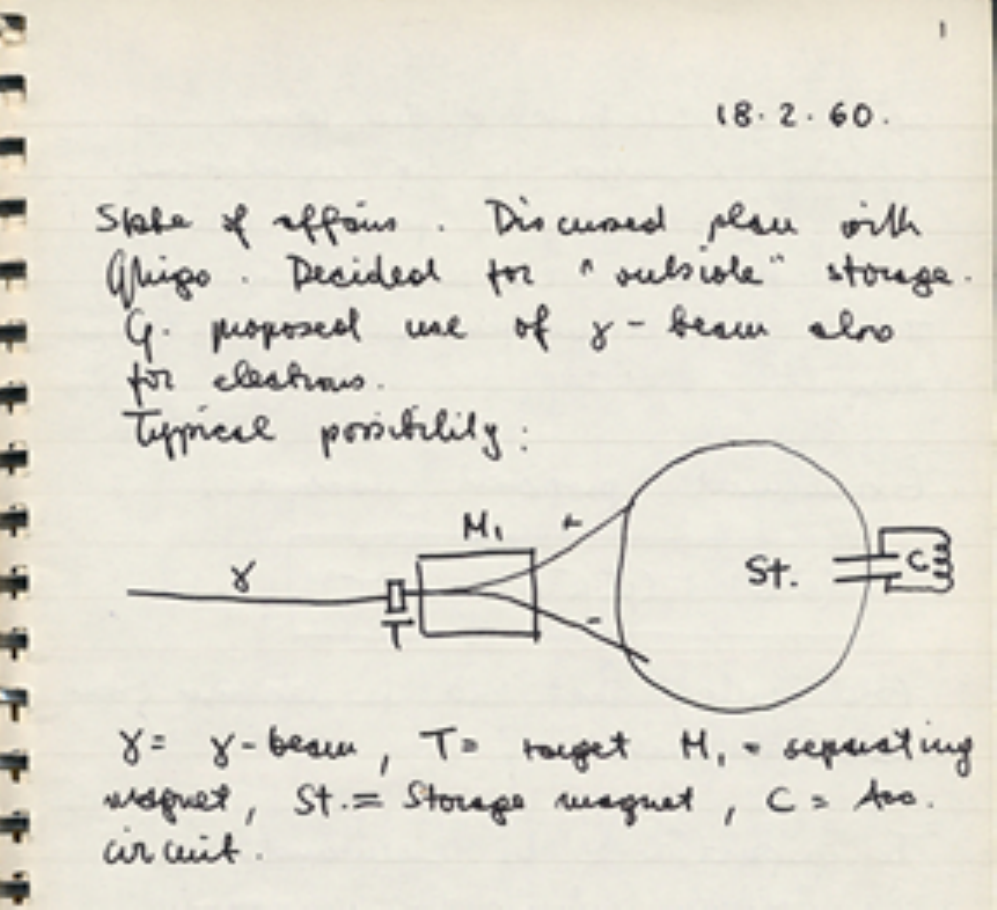}}\caption{First page of Bruno Touschek's first notebook, started on the day immediately following the Frascati meeting, where he had proposed  to construct an electron-positron storage ring  (Bruno Touschek Archive, Physics
Department, Sapienza University of Rome).}
\label{fig:SRnotebook}
\end{center}
\end{figure}

During the full discussion following \T 's talk, Giorgio Ghigo, Machine Director of the electron synchrotron, observed that the synchrotron was not a machine which could be easily converted, and that it was ``probably easier to construct an {\it ad hoc} 250 MeV machine, in order to carry out  the experiment suggested by Touschek.'' At the end of the meeting, all participants agreed that Touschek's  suggestion  deserved a deeper study, and Giorgio Salvini, the director of Frascati Laboratories, stressed that they must   go ahead with it.\footnote{Frascati, February 17, 1960, Report of the Meeting. For an account of the early days of AdA and ADONE see \cite{amman}.}
Touschek immediately took up the challenge: the following day, on February 18, 1960, he started a new notebook which he entitled ``SR'', Storage Ring, where he explored the physics of the proposed $e^{+}e^{-}$ storage ring.\footnote{B. Touschek, ``SR Notebook'', B. Touschek Archive, Physics Department, Sapienza University of Rome, Box 11, Folder 88. For a description of the SR-notebook content, see \cite{bonolis2005a}, pp. 28--35.} 
We reproduce in Fig.~\ref{fig:SRnotebook} the first page of his first notebook.

In an incomplete manuscript  in which he sketches some of the reasons which led him to propose building an electron-positron storage ring, Touschek stated 
that:  

\begin{quote}
\small

The outstanding motive was my conviction that the plan was workable. As a theoretical physicist I had played with the symmetry properties of particle physics, which had become the centre of attention in 1957 through the discovery of the `non conservation of parity' by Lee and  Yang [\dots].  

Another reason to prefer this type of effort to the more orthodox proposals of building a bigger and better machine (either for electrons or protons) is the following [\dots]  the atomic nucleus is held together by strong interactions. They make very messy physics, theoretically, because there is no method of calculating them, perturbation methods breaking down, just because they are strong. [\dots]. Weak interactions --- on the other hand --- seemed hardly feasible, just because they are weak and the events they produce are therefore very rare --- at least in the energy range, which we could hope to be available at Frascati by, say,  1964 [\dots]. We know that at least one stable strongly interacting particle -- the proton -- exists, that it interacts with the electromagnetic field [\dots]. The mere existence of the proton will therefore dictate a modification of pure electrodynamics and the proton itself can call into the fray all its strongly interacting friends. True electrodynamics can therefore not be indifferent to the existence of that part of the physical world which interacts strongly and  noisly [\dots].

The third motive was more of a challenge than a reason: positrons unlike electrons are not constituents of ordinary matter. They have to be produced artificially [\dots].

The fourth argument was demagogic rather than physical[\dots]. Equal charges require two rings, opposite charges can be stored in one ring, provided that their masses are equal. Italy being a poor country cannot afford an experiment which requires two rings. If we cannot even afford one ring we have the synchrotron which can be converted into one.''\footnote{B. Touschek, ``A brief outline of the story of AdA'', excerpts from a talk delivered by Touschek at the Accademia dei Lincei on May 24, 1974 (B. Touschek Archive, Physics Department, Sapienza University of Rome, Box 11,  Folder 3.92.5).}

\end{quote}
\normalsize

\begin{figure}
\begin{center}
\resizebox{1.0\textwidth}{!}{
\includegraphics{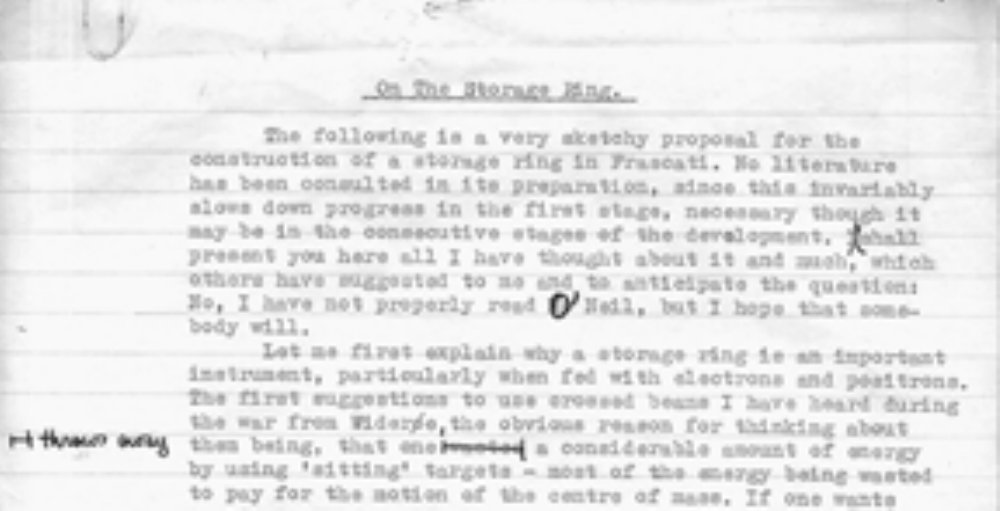}}
\caption{The opening paragraph  of the proposal for the construction of AdA by B. Touschek  (B. Touschek Archive, Physics Department, Sapienza University of Rome).}
\label{fig:Adaproposal}
\end{center}
\end{figure}

By March 7 Touschek had thoroughly sketched the main lines of his idea, and  he exposed a full project for a 250 MeV beam energy machine for positrons and electrons, with a 100 cm diameter, during an epoch-making seminar held in Frascati.\footnote{In the previously mentioned list of seminars held in Frascati in the years 1959--1960, \T 's seminar  has the title  "Anelli di cumulazione per l'urto elettrone positrone".}
 His colleagues  were particularly impressed by ``the extreme beauty of the `time-like one-photon channel' dominating, to first order of QED, the production of final states,''\cite{bernardini97}. Striking everbody's imagination was  the creative character of $e^{+}e^{-}$ collisions  providing  a state of pure energy through a channel with  well-defined quantum numbers, with no bias towards one form of matter or another and no  complications from the ``messy physics'' of strong interactions in the final state.   During the time between his first suggestion of February 17 and the March 7 seminar, he had  been  thinking  about the physics which could be extracted from such machine, 
and  the possibility to probe the ``quantum vacuum and the frequencies at which it resonates'' \cite{bernardini2004}.
 In a typescript document entitled ``On the Storage Ring'', which appears to have been  prepared in view of the seminar, Touschek presented ``a very sketchy proposal for the construction of a storage ring in Frascati.''\footnote{ B. Touschek, ``On the Storage Ring'', typescript, B. Touschek Archive, Physics Department, Sapienza University of Rome, Box 11, Folder 3.86.1.}
  In Fig.~\ref{fig:Adaproposal}  we reproduce the beginning 
 of this proposal.
He recalled that it was from \W \ that he had heard  the first suggestion to use crossed beams, 
and pointed out the advantages of using  beams consisting of electrons and positrons which disappear in the final state: ``This means that much more information can be gained by much fewer events.''  At this stage Touschek defined a little better his project (``{\it I prefer to think of it as an experiment rather than as a machine} [\dots]''  [our emphasis])
and proposed to study the reactions 

\begin{eqnarray}
e^{+}e^{-}& \rightarrow &2\gamma \\
e^{+}e^{-} & \rightarrow & \mu^{+}\mu^{-}\\
e^{+}e^{-} & \rightarrow & \pi^{+}\pi^{-} (2\pi^{0})
\end{eqnarray}

He proposed to use the first process  as a ``monitoring process'', i.e. one with  a well defined cross-section which would give rise to a calculable  number of events, which he estimated should be  at least one event per second and which could allow to then measure the cross-section for other processes. In these notes, he points out that ``the first of the processes  listed is [...] predominantly back-to-forward in the c.m. system and in these preferred directions no radiative corrections are expected''. This comment throws light on what will become later his main preoccupation for   ADONE experiments, i.e., higher order radiative corrections.

\begin{figure}
\begin{center}
\resizebox{1.0\textwidth}{!}{
\includegraphics{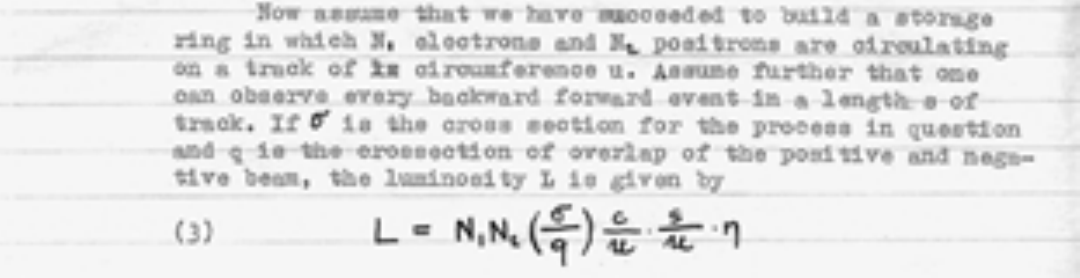}}
\caption{The expression for the interaction rate, which \T \ called the luminosity, from  the manuscript of the proposal for the construction of AdA by B. Touschek. Symbols as described in the text.}
\label{fig:luminosity}
\end{center}
\end{figure}

  After calculating  the  cross-section  for annihilation into photons as $\sigma=6.3\times 10^{-30} \ cm^2$, he  estimated the number of events observed by such ``luminosity" monitor, namely a monitor for events whose signal was given  by two photons, writing the formula shown in Fig.~\ref{fig:luminosity}, 
where $N_{1,2}$ is the number of particles  circulating along a circular track of radius $u$, $c$ is the speed of light, $s$ is the length of the track where events can be observed and recorded, and  $q$ the 
cross-section of overlap between the electron and positron beam.  $\eta$, explained Touschek, ``is an enhancement factor, which is due to the bunching of the beams and which can rise to 4 by a proper choice of the point of observation. { $L$} is measured in events/sec.  If one can get $q$ down to 1 cm$^{2}$, with  $u=300$  $cm$ and $s=10$ $cm$ one gets, with $\eta=4$ and $N_{1}=N_{2}=N$, $N \ge 10^{11}$, and this is the number of particles which one would want to circulate in the ring.'' 

 The  number of events per second,  which he had referred to as a ``luminosity'',  had been his  very first preoccupation. The  term ``luminosity'' appears to have been used  for the first time in Touschek's work, and  was probably inspired by his proposal  of a two photon state as  the monitor for event rate.  The interaction region, for instance, is  referred to as ``luminous region''  in a discussion of the luminosity monitor at SPEAR \cite{luminosityoneil}. Presently, the word  ``luminosity'' is used to indicate the proportionality factor between the number of events and the cross-section.
   Previous estimates of the interaction rate in storage ring machines had been put forward  by O'Neill
   in the context of  the storage ring for protons, which O'Neill had proposed a few years earlier \cite{oneil}, and subsequently in unpublished reports on building the electron-electron Princeton-Stanford storage ring \cite{O'Neill:1958gk,Barber:1959vg}. Touschek had certainly been  inspired by O'Neill's work, although, as read from  Fig,~\ref{fig:Adaproposal}, he may not have been fully aware of the details.

In discussing the indicated processes, Touschek referred to a ``recent paper of Cabibbo and Gatto''.\footnote{B. Touschek, ``On the Storage Ring'', typescript, Touschek Archive, University Sapienza, Rome, Box 11, Folder 3.86.1.} 
Indeed, after Panofsky's seminar, discussions about the concrete possibility of $e^{+}e^{-}$ physics were circulating in Rome and Frascati. Very soon,  at the beginning of 1960, papers triggered  by these discussions   appeared. Among the young theoretical physicists in Rome University, Nicola Cabibbo and Francesco Calogero had been the first students to graduate with Bruno Touschek. 
The first paper to be submitted for publication in  {\it The Physical Review Letters}  was  
 written by Laurie M. Brown and  Calogero, and received on  February 5. It calculated 
the effect of the pion form factor on the photon propagator 
\cite{Brown1960}. The second one, received February 17 and published 
in the same number of {\it The Physical Review Letters}, was written by Nicola Cabibbo 
and Raoul Gatto \cite{Cabibbo1960a}.\footnote{Gatto recalled how they sent the paper to Physical Review Letters holding out ``a very faint hope that the work would be accepted."R. Gatto to L. Bonolis, November 24, 2003.}
The Gatto and Cabibbo paper was  later  completed with detailed calculations of the processes which could be explored at such machines,  and became known in Frascati as {\it la bibbia} [the Bible] \cite{Cabibbo1960b}.
 
In the meantime, preliminary studies  after Touschek's seminar did not show any insurmountable barriers. A week later, on March 14,  the decision  to go ahead with the project was taken and eight  million lire were initially allocated for the proposed experimental device, later raised to twenty  million (about thirty two thousand  dollars at the time). It was decided that Touschek would be leader of the  experiment, Giorgio Ghigo cooperating for the technical problems and Carlo Bernardini\footnote{Carlo Bernardini, born in 1930, has been a prime collaborator of Bruno Touschek in the AdA enterprise. He is Professor Emeritus at Rome University Sapienza, has been Senator of the Italian Parlament, and chief Editor of ``Sapere''.} for  theoretical aspects.\footnote{The meeting was attended by F. Amman, C. Bernardini, N. Cabibbo, R. Gatto, G. Ghigo, G. Salvini and B. Touschek.} Their competence has been established 
through  the successful enterprise of building the Frascati electron synchrotron in record time. On March 16 Touschek and Ghigo prepared a first sketch of the program ahead, with a rough estimate  of the working times, and an outline of the  the main characteristics of the magnet, the vacuum chamber, and the radio frequency  cavities. \footnote{The preliminary draft of this proposal was jotted down by Touschek (B. Touschek, ``Proposta d'esperienza'',  two manuscript pages, Bruno Touschek Archive, Physics Department, Sapienza University of Rome, Box 11, Folder 3.87.) and later fully elaborated by G. Ghigo into the final document dated March 22, 1960 (B. Touschek Archive, Physics Department, Sapienza University of Rome, Box 11, Folder 3.80).}  The order for materials was placed on April 20.
On April 3 Touschek had inaugurated what would become his ``AdA logbook'': the very first issue he discussed  was  ``The vacuum''.\footnote{B. Touschek, AdA Notebook, B. Touschek Archive, Physics Department, Sapienza University of Rome, Box 11, Folder 3.89. The original document is still preserved by Elspeth Yonge Touschek.}

\begin{figure*}[h!]
\resizebox{0.5\textwidth}{!}{\includegraphics{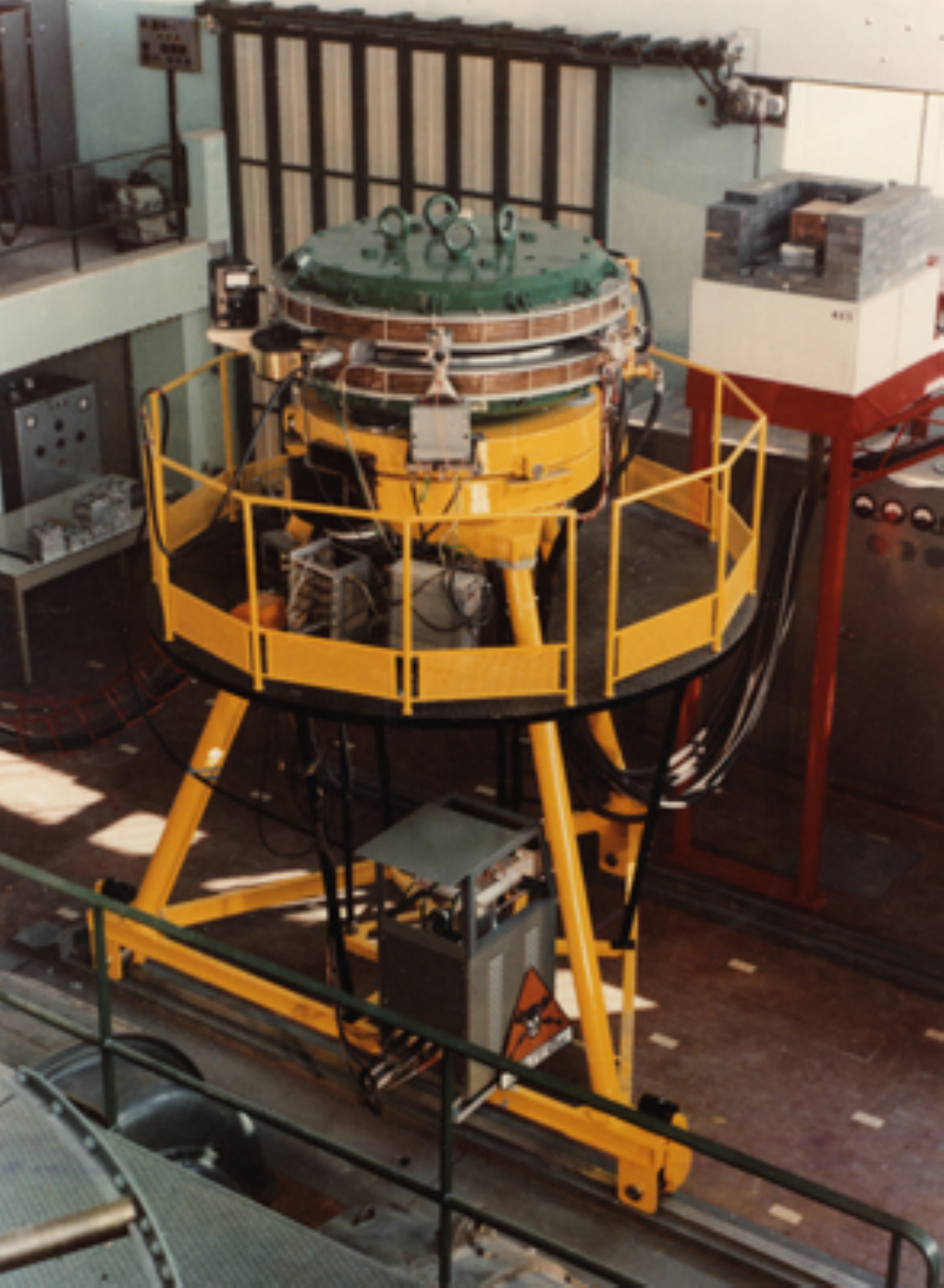}}
\resizebox{0.5\textwidth}{!}{\includegraphics{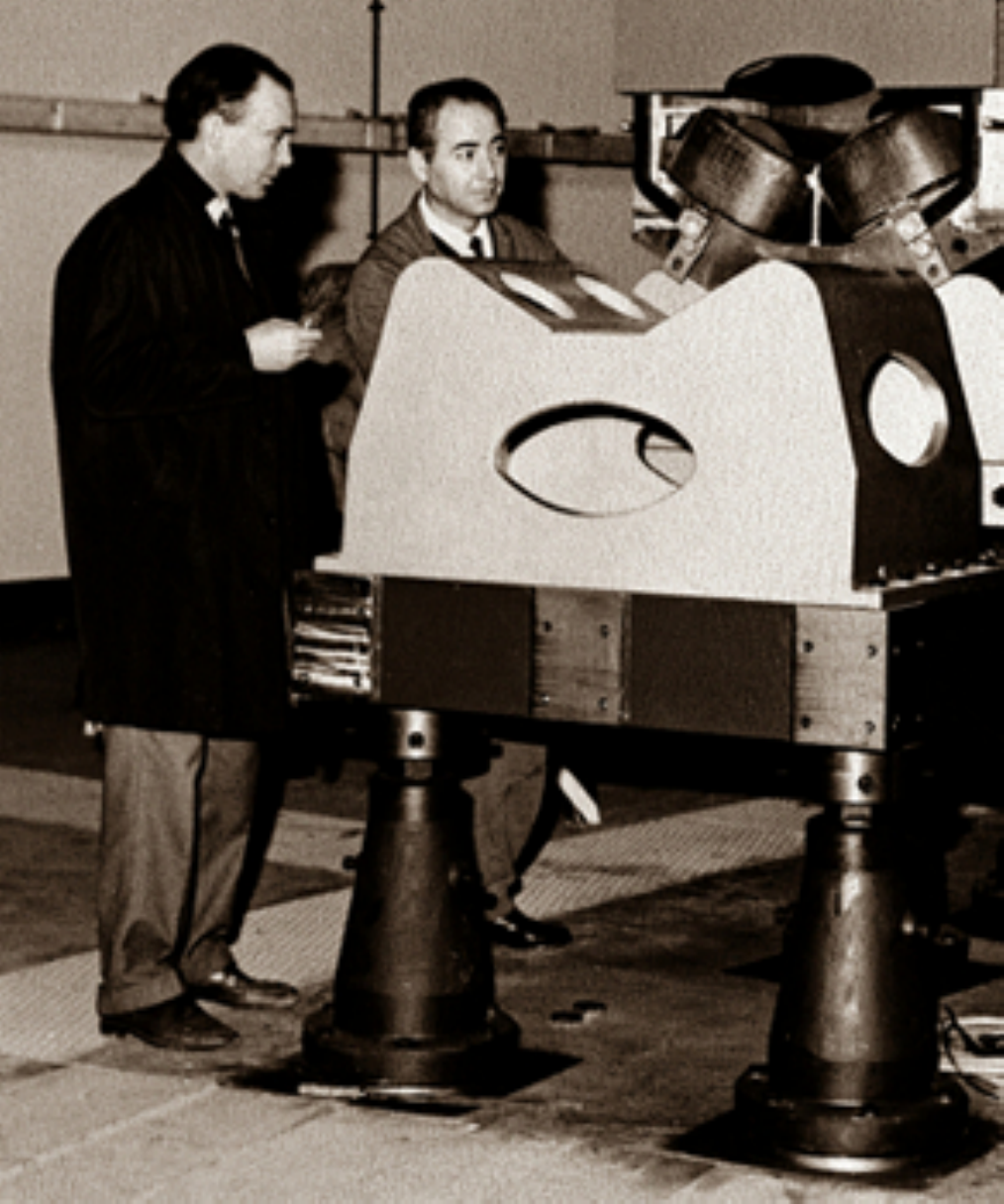}}
\caption{The first electron-positron storage ring AdA,
in 1961 (left). The synchrotron was used as an injector, until AdA was moved to Orsay in July 1962. Bruno Touschek  in  Sala Magneti (Magnet machine shop) during the construction of ADONE (right).}
\label{fig:ada}
\end{figure*}

The sequence of events to follow is a measure of  the enthusiasm and passion of all the scientists working  on the project in Frascati at the time.  On November 7 1960, ``The Frascati storage ring'', an article describing  AdA, the first matter-antimatter collider, was received by {\it Il Nuovo Cimento} \cite{ada}: the small machine had a 65 cm radius and a beam energy of 250 MeV.  On November 9, two days after submission of the paper, Touschek prepared a manuscript entitled ``ADONE -- a Draft proposal for a colliding beam experiment.''\footnote{ ``ADONE -- a draft proposal for a colliding beam experiment'', typescript, B. Touschek Archive, Physics Department, Sapienza University of Rome, Box 12, Folder 3.95.3.}
 The proposal for the construction of a storage ring with beam energy of  1.5 GeV was presented at the annual meeting of the INFN in Frascati and  on January 27, 1961, F. Amman, C. Bernardini, R. Gatto, G. Ghigo and B. Touschek presented an  Internal Laboratories Report  with the title ``Storage ring for electrons and positrons (ADONE)'' \cite{adone}.\footnote{This report is reproduced as Appendix A in \cite{greco2005}.} In February 1961 a study group was formally set up with the task of preparing a first estimate of the feasibility and costs of such a project. 
Soon after, on February 27, 1961, {\it only one year} after Touschek's proposal,  the first electrons had been accumulated in AdA using the Frascati electron synchrotron as an injector.  In Fig.~\ref{fig:ada}
we show a period photograph of AdA on the installation platform which allowed it to reach the level of the electron beam from the  synchrotron. At right a photograph of Bruno Toushek during the construction of ADONE, in 1964.

\subsection{AdA at Orsay}
AdA had been  built by a small team of physicists and engineers led by Touschek, namely Carlo Bernardini, Gianfranco Corazza, Giorgio Ghigo and Giancarlo Sacerdoti (who took care of the magnet), and  Mario Puglisi, Antonio Massarotti and Dino Fabiani  helping  to design the radio frequency cavity. Corazza, who had already taken care of the electron synchrotron vacuum chamber, was able to obtain  an unprecedented vacuum in a volume as big (for the time) as AdA's doughnut. Ghigo was  in charge of the overall assembly and operation of  AdA.

Actually the machine could not really work very well,  because, in order to reach a luminosity sufficient to establish that electron-positron collisions had taken place, it  needed a better injector.  It thus happened that AdA was  transferred   to France,  at LAL, the Laboratoire de l'Accelerateur Lin\'eaire at  Orsay, near Paris, where a high intensity linear accelerator (LINAC) was available.  This transfer allowed to prove  the feasibility of electron-positron collisions and launched  the era of electron-positron physics.

  It had all started with a  
 visit to Frascati by  Pierre Marin  \cite{marin2009}.\footnote{ Pierre Marin, 1927-2002, played  a principal role in the French  effort to   accelerator building in post-war Europe, including research  with ACO, built  at the Laboratoire de l'Accelerateur Lineaire, LAL in Orsay, near Paris.} 
 Pierre Marin, after   a period spent at CERN,  was trying to find his own research direction and was  told  
 that in Frascati ``[il] se passait des choses qui intriguaient les esprits.'' In summer  1961,\footnote{ In \cite{marin2009} Marin dates his visit  in  August (1961), elsewhere in  early September, while letters obtained courtesy of Jacques Ha\"issinski and  exchanged between Frascati  and LAL, indicate 
 July 1961.} together with George Charpak, Pierre Marin visited Frascati, 
 and    was soon caught by   the  enthusiasm of the Italian team about their machine, ``un vrai bijoux''.  Marin's visit was quite successful and a collaboration was envisaged.\footnote{ On september 19, 1961  Marin  prepared a report of his visit, where the possibility of bringing AdA to Orsay is already considered. In this report, agreed upon with Ruggero Querzoli, then in charge   of  Frascati experiments, Marin describes  the future programs of AdA and ADONE. Listing   one of AdA's objectives, Marin writes ``S'il s'av\'erait qu'il ne se puisse \^{e}tre r\'ealis\'e \`a Frascati, A.D.A. serait transport\'e \'a Orsay aupr\`es de l'Acc\'el\'erateur Lin\'eaire.'' Copy of this report,  which is preserved at Orsay,  Archives of the Linear Accelerator Laboratory, was  kindly provided by Jacques Ha\"issinski.}   Letters were exchanged  to allow two or three French scientists to come to Frascati. During a second visit, this time in early 1962,  Marin   became  appraised of the   disappointment about the lack of sufficient luminosity in AdA, which was  due to the poor performance  of the synchrotron  as an injector.  Pierre Marin recalls \cite{marin2009} that, as he started describing how good the linear accelerator in Orsay was, Touschek and Bernardini  put forward the possibility  of   bringing  AdA to Orsay and obtain a higher luminosity thanks to the high intensity LINAC. They  proposed it  to Marin and more letters were exchanged between  the two laboratories.\footnote{ Letters exchanged between Rome, Frascati and Orsay  envisaged  
    a collaboration
    between Orsay and Frascati  which  would include  future experiments  at   ADONE. On December 22, 1961, Andr\'e Blanc-Lapierre, the new LAL Director, wrote to Italo Federico Quercia, Director of Frascati Laboratories in order to organize a  visit of two or three scientists  from Orsay, where  they were doing preliminary studies for a 1.3 GeV storage rings for electrons and positrons.
On January 16 Amaldi (then Director of INFN) wrote to Blanc-Lapierre about the visit of the French scientists and about the importance of   coordinating a future collaboration in the best possible way. As mentioned in a letter written by Blanc-Lapierre to Quercia on January 23, 1962, the  date for the visit was fixed for February 5. The visiting group was formed by Fran\c{c}ois Fer, Pierre Marin and Boris Milman. And indeed, on February 12, Fer thanked Touschek for  ``information and advices about storage rings''  received during their stay in Frascati.   On February 21 Touschek wrote  to his collaborators as well as to Fernando Amman (who was leading the ADONE project) and to Amaldi, about a meeting dedicated to the Franco-Italian collaboration to be held the following Saturday. A new letter sent on April 4 by  Blanc-Lapierre to Quercia, mentioned having met Carlo Bernardini and Fernando Amman in Geneva. During their conversation Bernardini had envisioned that  AdA might be moved to Orsay next June (B. Touschek Archive, Physics Department, Sapienza University of Rome, Box 1, Folder 4, and letters kindly provided by Jacques Ha\"issinski and preserved in LAL Archives).}
   According to the correspondence, by early April a decision about the date had  
  been taken, and  the transfer of AdA to Orsay was prepared. Andr\'e Blanc-Lapierre, the LAL Director, invited  Touschek  to give a seminar.\footnote{  Andr\'e Blanc-Lapierre to B. Touschek, April 11, 1962 (B. Touschek Archive, Physics Department, Sapienza University of Rome, Box 1, Folder 4).} At the beginning of July, AdA was packed on a big truck, which would have to cross the Alps with a fully evacuated  beam pipe, and batteries, lasting about three days,  to power the vacuum pumps:  at that time one  needed months to reach the required pressures of order of $5\times 10^{-10}$ $torr$ and one could not waste  precious time obtaining the vacuum anew.\footnote{ There is some confusion in the literature  about the date of AdA's arrival in Orsay. In \cite{bernardini2004}, a typo dates  it    in 1963,   Pierre Marin talks of ``d\'ebut 1962'', whereas  Ha\"issinski  \cite{haissinski1965} gives the date of  summer  1962. As a matter of fact, on June 28, 1962 Touschek was writing to F. Perrin that ``a second convoy [with AdA and the  evacuated  doughnut] will presumably  leave Rome  on the 4th of July  and should arrive in Paris on the 7th.''  (B. Touschek Archive, Sapienza University of Rome, Box 1, Folder 4).}  
  As the truck was ready to go, Touschek jumped in to try  the driving, and immediately hit a lamp post, luckly without damage neither to himself nor to the truck and its precious cargo \cite{bernardini2004}. 
 
 AdA owes its crossing   the border between France and Italy   to a number of high level diplomatic  interventions. 
    Touschek recalled the custom officer asking: ``What's inside?", pointing at the doughnut. To which it  was truthfully  answered : ``Nearly absolute vacuum."\footnote{\BT, ``Convegno Adone'', manuscript dated May 5, 1974, prepared on the occasion of the meeting organized at the Accademia dei Lincei on May 24. Bruno Touschek Archive, Physics Department, Sapienza University of Rome, Box 8,  Folder 61.}  But   humor alone would  not suffice to convince the border authorities, and a number of letters  and telephone calls between France and Italy had to take place before AdA, with its high vacuum still intact, could enter France. Jacques Ha\"issinki, who would do his doctoral thesis \cite{haissinski1965} on the AdA results,  writes \cite{haissinski1998} : ``It took the intervention of Francis Perrin, then the Haut Commissaire \a l'energie Atomique, to get over this hurdle.'' Carlo Bernardini \cite{bernardini2007}  remembers also the intervention from the Italian side, with Amaldi, in Rome, calling  the Italian  Ministry of Foreign Affairs, and, through this, the French authorities.

In early  July 1962, AdA was  in Orsay  and, by August,  installation was completed.  The  installation  had its dramatic moments \cite{haissinski1998}. While being hauled  by  heavy crane in its place in the so called medium energy hall, AdA was almost smashed against a wall. It was Pierre Marin, alerted by  yellings from the Italian group, who ran and pushed the buttons which averted the crash.  Another time, one of  the Cherenkov detectors, while  being moved close to the AdA ring, tipped over and broke Pierre Marin's foot. None of this quenched the enthusiasm and drive of the  Franco--Italian group, which was  joined in the fall 1962 by 
Jacques Ha\"issinski. Once   in Orsay,    AdA   started  making collisions thanks to the linear accelerator  and the dedication  of the group. 
But soon,  an unexpected effect  appeared  to limit the machine luminosity.  What  is now called {\it Touschek effect},  
manifested itself through a progressive decrease in the beam life
time while  the number of stored particles increased. Still one of the effects which limit the beam life time in accelerators,\footnote{L. Evans, The Large Hadron Collider, talk given during the {\it Bruno Touschek Memorial Lectures}, Frascati, November 30th,  2010.}  it was explained by Touschek   as an intrabeam scattering effect, which changes the lepton momenta and throws the scattered particles off the designated orbit
 \cite{touschekeffect}.  The effect was unexpected, also  because of a wrong evaluation of the vertical beam size, assumed to be 1000
 times  larger in a multiple rather than single damped scattering regime;  the
wrong assumption was corrected by Carlo Bernardini,  in an unpublished note  \cite{bernardini62}, and fitted the Touschek calculation.

\subsection{AdA's legacy}
 AdA showed the feasibility of electron-positron
collisions \cite{bernardini64}, and opened the way to  the machines which would discover new fundamental particles and bring the experimental confirmation of the Standard Model.
 Soon after the proposal for AdA  appeared, proposals to build more powerful
colliders, in the USA  and Europe, were put
forward, with higher energy and higher luminosity, among them ACO in France \cite{marin2009},  and SPEAR  \cite{Paris2001,pellegrini,richter} in the USA.\footnote{ 
 SPEAR was completed in 1972. See  also \cite{pellegrini} for a source book on the development of colliders, and \cite{richter} for an account on the rise of colliding beams made by Burton Richter, one of the protagonists of the field. It is however to be mentioned that Richter, even if giving due credit to the importance of ADONE for the development of new physics with colliders, still made the following sharp statement about AdA on p. 269 of his essay: ``In my opinion, AdA was a scientific curiosity that contributed little of any significance to the development of colliding beams (there is one exception; a beam-loss mechanism now called the Touschek effect was discovered) [\dots]''. This sentence sounds too dismissive of the influence of AdA's work, in light of other assessments, as, for instance,  in  \cite{marin2009}.
For discussion on this topic see \cite{bonolis2005a}, especially p. 50.} 
In Italy the  construction of ADONE, proposed as early as January 1961 \cite{adone},  
was under way  by the end of the year: orders,  financed  by the Comitato Nazionale per l'Energia  Nucleare (CNEN)\footnote{ The Frascati Laboratories, operational  in 1957, were staffed and  partly financed by  INFN, with CNEN  owning the grounds and financing the major projects.} were issued for various parts,  and  construction was started in 1963 \cite{valente2007}.
 Similar efforts took place also in the Soviet Union, where   Gersch Budker\footnote{ G.I. Budker was called Andreij Mikhailovic by his friends, as for instance in \cite{bernardini2007}.} and his collaborators had been active since the midfifties  in electron-electron colliders,  and later in electron-positron collisions. Because of the internal political situation of that period,  early records of Russian  efforts in electron-positron studies   are very scarce. Their work   became partly known to scientists from outside the Soviet Union   in 1963,  in  occasion of  the Dubna Conference on  High Energy Accelerators, where Budker gave a talk in Russian on the activity on particle colliders   which had taken place in the Soviet Union since 1956.   According to  \cite{marin2009}, none of the foreign scientists present to his talk understood Russian, however, and, although there was a simultaneous translation,\footnote{Courtesy of A. Skrinsky, INP Director since 1977.} not much of his presentation was understood. From the  Conference Proceedings \cite{dubna63},  published in Russian in 1964, one can see that Budker  and his collaborators indicated  and described three types of activity: electron-electron, electron-positron and proton-proton collisions. After the conference, a few foreign visitors were invited to see the newly established Institute of Nuclear Physics  (now Budker Institute of Nuclear Physics) of the Siberian Division of the USSR
Academy of Sciences in  Novosibirsk, three thousand  kilometers to the East of Moscow. Pierre Marin \cite{marin2009} remembers the astonishment in seeing VEPPII, the electron-positron accelerator with an energy of 700 MeV, in such an advanced  stage of preparation  to indicate that work on it must have been taking place for quite some time. For electron-positron collisions, nothing of this had appeared  before the Conference.
After the Conference Proceedings arrived in 1964, a French translation of Budker's  talk was commissioned in Orsay,  as well as an English translation exists \cite{USAEC1965}. The French translation    remained practically ignored until chance saved  it from destruction  in the year 2000.\footnote{A copy of the 
French  translation of Budker's talk was sent  by Pierre Marin to Carlo Bernardini in a letter dated December 26, 2000. In this letter Marin also notices that the literature quoted in these Proceedings only refers to electron-electron studies.} In these Proceedings, Budker dates the start of electron-positron work in late 1958, but this appears unlikely, as also discussed in \cite{amaldi81}.\footnote{According to Amaldi  (\cite{amaldi81}, footnote 114, p. 78), ``it appears reasonable to conclude that the activity on $e^+e^-$ storage rings was started [in Soviet Union] after 1961 and no document is known which proves the contrary.''}  Presently,  Russian contemporary sources would rather place  the start not earlier than 1959.
Since then, a number of contributions have appeared which put forward  some dates in the Russian development of  electron-positron collisions \cite{Budker1994,baier2006}. In  \cite{baier2006},   the Russian theorist Vladimir Baier  vividly recalled  how the appearance in 1960 of the {\it Nuovo Cimento} paper by the Frascati group \cite{ada} strengthened their resolve to build an electron-positron collider. In this recollection 
he writes to have   started working  on electron-positron collisions at the end of October  1959.\footnote{Baier describes a discussion with Budker on October 28th, 1959, during which  he first proposed physics with electron-positron collisions, and writes that he started working on it on the following day.  }
 Even if  the theoretical work started as early as in the case of the Frascati group, the Russian progress in building a working accelerator  was   delayed\footnote{Data taking at VEPPII started  in 1966 \cite{baier2006}.} by  moving from the Kurchatov Institute in Moscow to the Novosibirsk site and by  the secrecy surrounding its construction. In addition,  serious technical problems  with the vacuum slowed down quite a bit the work of the Siberian team.\footnote{Such problems were well known, according to Carlo Bernardini. Private communication to the authors.}
 
 The legacy of Touschek's work   appears very clearly from \cite{baier2006}, where  the impact of the successful experimentation in Orsay, which had been followed by the ACO and ADONE projects, and the discovery of the {\it Touschek effect},  are described as  clear signs that electron-positron collisions had  become ``very respectable''  and a race to obtain useful physics results could now start.

The above is a short and limited account of Touschek's work on AdA and how the road which led to the discovery of the $J/\Psi$ and to the  establishment of   the Standard model had been opened. We shall now describe  Bruno Touschek's life prior to  his arrival in Italy and  the dramatic experiences  and  scientific   exchanges  which brought him to the Physics Institute of Rome University and to the construction of AdA.

\section{Touschek's life prior to arrival in Italy}
\label{sec:beforeitaly}
\subsection{Early years: 1938 -- 1941}

Bruno Touschek was born in Vienna, on February 3rd 1921 from Franz  Xaver Touschek,\footnote{The family name
Touschek comes from Moravia, which, with capital Brno, belonged to Czech country
and Slovakia and  now belongs to Czech state.
The name is also current  in the southern part called
Sudety.} an officer in the Austrian army, and Camilla Weltmann, who belonged to a Viennese Jewish family
 prominent in artistic and intellectual circles like the Vienna Secession and the Wiener Werkst\"atte. Camilla Weltmann died when Bruno was ten years old, but he kept frequent and 
intense relations with his maternal family, in particular with his aunt Adele,
 called Ada, who lived in Italy, married to an Italian businessman,  and had a villa on the Alban hills southeast of Rome. Early letters to his parents (the father had remarried) testify to
 his love for Italy and  Rome in particular. In 1938, however, everything changed forever for  the Vienna Jewish community.  The Nuremberg Laws of September 15, 1935, had established that a person  was considered Jewish if at least three grandparents were of Jewish religion. 
When on 12  March 1938  German troops entered Austria, German laws became also Austrian laws. With annexation  to Germany,  political and racial criteria were quickly applied, and    Jewish students were soon not allowed anymore to attend classes in the Gymnasiums.

Only two of \T 's grandparents were of Jewish religion, still, as \T \ was ready to start the last year of high school (gymnasium) in the fall of 1938, he was told that he could not attend his classes anymore
  nor take the final {\it (matura}) examination, the obligatory passage to the university, together with his friends and schoolmates. He had to  abandon the Piaristen Gymnasium. 
  However, he was still able  to obtain his diploma, by registering for the exam as  a private student,
 and, in February 1939, he 
 graduated  from the Vienna Schottengymnasium. The exam had been anticipated, because the Austrian authorities wanted to have as  many young officers as possible ready for the war \cite{amaldi81,amaldi82}.\footnote{We note here that in \cite{amaldi81,amaldi82} \BT \ 
  is said to have received his {\it matura} in February 1938, but this makes
 the chronology very confused: it is unlikely that he   could  have been expelled from the Lyceum 
until the {\it Anschluss} took place, and this establishes that he must have
 obtained his {\it matura } in  1939 as   confirmed by  a document and also stated in  \BT 's Curriculum Vitae (CV) still preserved by Touschek's wife Elspeth Yonge Touschek.}
 
After passing the exam, as the custom was, Bruno went to Rome, in Italy, visiting his aunt Adele, called Ada,  for 
 the  vacation traditionally following the end of the high school years.
From  letters exchanged with his father, we learn that he was waiting for a visa which would allow him to go to Great Britain and study Chemistry in Manchester.\footnote{Touschek wrote very often to his parents (his father had remarried after Bruno's mother's death) and gave many details of his everyday life. The whole correspondence, which spans between 1939 and 1971, is still preserved by Touschek's wife Elspeth Yonge Touschek.}
   At the same time, 
     he also considered 
   studying engineering in Rome, and   started attending some classes, 
following 
``with enthusiasm'' the 
course of Mathematical Analysis taught by  the mathematician Francesco Severi. 
But  after  his   return to Vienna for a family vacation  in  Summer 1939,
everything changed.
 
 In September 1939 World War II started with Hitler's invasion of Poland. Bruno's  plans of studying abroad were completely upset, and in October he enrolled
 to study physics and 
mathematics in 
  the 
University of Vienna,  where he  attended   ten courses and passed the exams,  (mostly in 
Mathematics and Theoretical Physics).\footnote{In his CV, he says  to have attended and taken exams for ten  courses, among them Physics Laboratory I and II, Differential Equations and Exercises, Mathematical Seminar, Rational Mechanics, 
 Theory of Functions and Exercises, Introduction to Theoretical Physics, Statistical Theory of Heat, Thermodynamics  (``Curriculum Vitae del Prof. Bruno Touschek'', B. Touschek personal papers preserved by Elspeth Yonge Touschek).}

In May 1940, as  he was beginning to emerge as a brilliant student,
university life  also  closed for him  ``for racial-political reasons''.\footnote{ ``Breve Curriculum Vitae del prof. Bruno Touschek'', Amaldi Archive, Physics Department, Sapienza University of Rome, Box 524, Folder 4.}
He continued his studies at home reading books 
 loaned to him by Paul Urban, then  a young assistant professor at University of Vienna.\footnote{
  Paul Urban, who had obtained his Ph.D. in physics and mathematics at the University of  Vienna in 1935, was Hans Thirring's assistant at the Institut f\"ur Theoretische Physik of the University of Vienna. Urban later became Professor of Theoretical Physics at University of Graz and, in 1962,  founded the  Schladming Winter School of Theoretical Physics.}
In a letter written to Amaldi after Touschek's death, Urban recalls how he would invite  \BT \ and  other students to his house, where they could study and, occasionally, also have meals prepared  by Urban's mother.\footnote{Paul Urban to Edoardo Amaldi, June 3, 1980, Amaldi Archive, Sapienza University of Rome,  Box 524, Folder 4, Subfolder 4.  Reconstruction of this period is mainly based on documents and correspondence preserved in Box 524 of the just mentioned source, and in Touschek's correspondence with his parents and with Sommerfeld.}
 It is during this period that,
  with Urban's help, he studied the first volume of Arnold Sommerfeld's treatise {\it Atombau und Spektrallinien} \cite{sommerfeld}.\footnote{ Since its appearance  in 1919 and up to the beginning of 1940s, the first volume of Sommerfeld's  treatise underwent several editions (1919, 1921, 1922, 1924, 1931), and we do not know which one Touschek used.}  Having found some small errors he wrote a letter to Sommerfeld  encouraged by Edmund Hlawka, a
professor of Mathematics at the University of Vienna, who later became
worldwide known for his works in number theory. According to Urban,
they always discussed with Hlawka difficult issues.\footnote{ P. Urban to E. Amaldi, June 3 and September 16, 1980, Amaldi Archive,
Sapienza University of Rome, Box 524, Folder 4, Subfolder 4. Information on this
episode is contained in the first biographical notes taken by Amaldi on February
28, 1978 and based on  Touschek's personal recollections (E. Amaldi, Typescript with
handwritten notes, Amaldi Archive, Physics Department, Sapienza University of Rome,
Box 524, Folder 6). However, the first letters written by Touschek to
Sommerfeld are apparently missing, so that we are not able to reconstruct the
dates of the first contacts between Touschek and Sommerfeld.} 
 Sommerfeld  suggested   him  to read the second volume as well \cite{sommerfeld2}.  Later Touschek recalled how he had learned quantum theory from the second volume of Sommerfeld's treatise and how he had also ``tried Dirac's famous book'', both of which ``lean heavily on wave mechanics.''\footnote{B. Touschek, ``Remarks on the influence of Heisenberg on physicists'', undated manuscript (Bruno Touschek's personal papers  preserved by Elspeth Yonge Touschek).}

But life for Urban's
prot\'eg\'es soon became difficult, because Urban's principal, Theodor Sexl, was continuously requested by the University to take measures concerning   students not complying with the racial laws. At last,
 to help Touschek to   continue his studies  {\it incognito} away from Vienna, Urban decided to  put him in contact with
influential physicists of the time from whom he hoped to have some help
in finding a position for Touschek in Germany. The occasion arose from a
seminar on the tunnel effect, on which Urban had worked at that time, and
which he was going to present in Munich in the presence of Sommerfeld
and other prominent physicists.\footnote{This issue is mentioned in a letter written by A. Sommerfeld to P. Urban,
on October 17, 1941, a copy of which was sent by Urban to Amaldi with a letter
where he told the whole story (P. Urban to E. Amaldi, June 6, 1980, Amaldi
Archive, Sapienza University of Rome, Box 524, Folder 4, Subfolder 4). }

 Apart from his theoretical works on the quantum theory of the atomic structure, Sommerfeld had given a fundamental contribution to the new physics in training more than a generation 
of Germany's best theoretical physicists. His students included Wolfgang Pauli, 
Werner Heisenberg, Hans Bethe, Peter Debye, Walter Heitler, Isidor Rabi, Rudolf Peierls,
 Gregor Wentzel; many of them were to become Nobel Prize winners. In 1928, nearly one-third of all full professors of theoretical physics in the German-speaking world were SommerfeldÕs pupils.
   Sommerfeld  
  had always openly supported Einstein and his physics when the latter had been attacked.
   After his retirement in 1935, he had held his position during the long selection process for his successor, which  
  ended in December 1939, when at last Wilhelm M\"uller was  appointed to Sommerfeld's chair of theoretical physics \cite{beyerchen}.\footnote{ Sommerfeld had repeatedly suggested Heisenberg as his favorite successor, and  the final  choice was much criticized, also because M\"uller was not a theoretical
physicist, and had not even published in a physics journal, but  was a supporter
of  Deutsche Physik.}

  At the time of Urban and Touschek's visit, Sommerfeld had  found hospitality   in Klaus Clusius'  Institute of Chemical Physics, where he ran a small seminar followed by  friends and admirers.  Urban held his seminar on November 24, 1941, and brought Touschek
with him in order to ask Sommerfeld's support and help. A few days later, on December 2, 1941, Urban wrote  to Sommerfeld,  sending greetings for his birthday and thanking for the hospitality.\footnote{ P. Urban to A. Sommerfeld, December 2, 1941. Deutsches Museum Archive,
NL 89,017.} 
  The latter answered immediately
and mentioned the possibility of a position for Touschek in Hamburg,
suggesting that ``he could attend the good courses given by Prof. Lenz
and Harteck.''{\footnote{ See A. Sommerfeld to P. Urban, December 2, 1941. Copy of the letter was
sent by Urban to Edoardo Amaldi, Amaldi Archive, Physics Department, Sapienza University of Rome, Box 524 Folder 4, Subfolder 4.} 
  On December 20 Touschek thanked Sommerfeld for  ``his
friendly handling of the Hamburg matter''; at the same time he discussed
some aspects of the scattering theory, clearly referring to Sommerfeld's
{\it Atombau}, and quoting the existing literature.\footnote{ B. Touschek to A. Sommerfeld, December 20, 1941, Deutsches Museum
Archive, HS 1977-28/A,343. This is the first of the letters written by Touschek
preserved in Sommerfeld's correspondence, but there are
clear hints in the text showing that he must have written at least once or even twice before the
letter of December 20. Unfortunately, as far as we know, these letters got lost. No letter written by Sommerfeld is preserved among Touschek's papers.} 
We know what Sommerfeld answered to Touschek's letter of December 20 thanks to the stenographic text that he must have dictated to his secretary and which can be found at the end of the letter. The opening lines confirm that Touschek had already written before that date: ``After the commendable addenda  of your  letter of Dec. 20,  everything seems clear [\dots]'' (``Nach den dankenswerten Erg\"anzungen Ihres Briefes vom 20ten Dec.  scheint alles klar'').\footnote{Sommerfeld continued remarking that ``[\dots] after having worked so deep into this direct approximation method, you should give a positive turn [\dots]'' (`` [\dots] so sollten Sie, nachdem   Sie sich so tief in diese direkte Ann\"ahrungsmethode eingearbeitet haben, die positive Wendung geben''). Here Sommerfeld referred to Touschek's observation that ``the development of $\alpha_{f}$ was not granted in the first approximation, and asked him to discover why Sexl and Urban's formula showed a better agreement with observations if compared with Mott's shortest version [\dots]''. The transcription of Sommerfeld's answer is attached to the letter preserved at the Deutsches Museum Archive, NL 089,015. Up to now this is the only evidence we have for Sommerfeld's answers to Touschek's letters.}
 
 Touschek wrote again at
the end of the year reporting on his calculations and telling that he had
found only ``a few errors (except IV ¤) [\dots]''.\footnote{ B. Touschek to A. Sommerfeld, December 31, 1941. Deutsches Museum
Archive, HS 1977-28/A,343.}
 All these discussions were often related to the new edition of  {\it Atombau} which Sommerfeld was preparing and which appeared in 1944. Sommerfeld thanked Touschek and Urban at the end of the preface \cite{sommerfeld2}. However, Urban was rather annoyed because Touschek, without mentioning the matter
in advance, had taken the initiative to write Sommerfeld about his own ideas
on issues previously discussed with Urban himself. In a letter written by Urban
to Sommerfeld on January 4, 1942,\footnote{ P. Urban to A. Sommerfeld, January 4,
1942. Deutsches Museum Archive, HS 1977-28/A,343.} 
the former complained
about Touschek's behavior and distanced himself from any future opinion his
pupil might express about scientific questions of common interest. In the opening
lines of his answer of January 9, Sommerfeld stressed how sorry he was
that he had someway been responsible for the misunderstanding.  He added that, 
he himself had written to Touschek with the aim of proposing him the job in
Hamburg (``manufacturer seeks assistants''). He remarked that, if  in
this occasion Touschek had also written about the scattering problem, was not
to be considered ``as an arbitrariness or even a lack of sincerity''  towards Urban,
and in fact, he added, he had sent greetings for him and for Sexl. He encouraged
Urban to get over his distrust and made clear that he would be very pleased if
Urban would let Touschek continue to work with him on those tricky issues. He
closed the letter quoting a sentence from the Bible about the necessity of loving
each other.\footnote{A. Sommerfeld to P. Urban, January 9, 1942. Deutsches Museum
Archive, NL 089,015.}

 Sommerfeld's interest and benevolence towards Touschek should not surprise. A   great scientist and a great teacher,  Sommerfeld could only  view with favor a brilliant student like \BT, eager to continue his studies and so interested in theoretical physics. Indeed, Sommerfeld's wider interest in helping the young Austrian, is made clear by a letter he wrote in that same period, early autumn of 1941,  to the ``State (Reich) Minister'' (``Staatsminister'') Friedrich Schmidt-Ott about the state of physics research in Germany,  and about the necessity of concentrating funds on a few relevant research projects. 
Sommerfeld's  main worry was the  state of theoretical physics: ``The difficulty is only to find adequately trained young theoreticians in Germany, after the awful bleeding that the previously flourishing theoretical physics has experienced during the last years. The state of things is really desolate. Planck's chair is still unoccupied. On my chair a man was called, merely for Party's interests,  who  never had any involvement with modern physics and who does not show any interest in it.''\footnote{A. Sommerfeld to F. Schmidt-Ott, September 24, 1941.  Deutsches Museum Archive, NL 089,015.}

This unpublished correspondence provides an  insight on Sommerfeld's role in helping Touschek to find a work in Hamburg, which would allow him  to earn his living.  At the same time, he was  entrusting him to physicists who would not betray him and would help him to continue his studies. On January 12 Touschek announced to Sommerfeld that he had been very happy in receiving ``a friendly letter from Dr.  Jobst'' and that he had begun to read more experimental books on electronic  devices (Br\"uche, Recknagel).\footnote{ B. Touschek to A. Sommerfeld, January 12, 1942, Deutsches Museum
Archive, NL 089,015. As far as we know this is the last letter Touschek wrote
to Sommerfeld before the end of the war.  G\"unther Jobst worked for the Allgemeine Deutsche Philips Industrie which had been founded in Berlin with the name Philips Deutschland in 1926, and had very soon acquired in Hamburg the C.H.F. M\"uller company, known as R\"ontgenm\"uller.}

\subsection{Clandestinity in Germany. 1942-1944}
\label{sec:2}
During this period,  Bruno Touschek put the basis of his knowledge of the physics of accelerators, which  brought   him later to construct   AdA. He learned about accelerators through his encounter with  the Norwegian engineer \RW , who  built  the first  European betatron \cite{wideroe}.
 The story of \T's encounter with \W \ and  their collaboration in the construction of the betatron belongs to the history of accelerators and will be briefly described in this section.

 In  early 1942, Touschek  started his new life  in Germany. 
  On February 26, 1942, Touschek was writing to his parents from Munich: ``Sommerfeld has recommended me to Prof. Harteck, Lenz and M\"oller in Hamburg, who can  help me to work without impediments   in the local university".\footnote{Paul Harteck, a well known Austrian born chemical physicist, was director of
the Department of Physical Chemistry. Wilhelm Lenz had been Sommerfeld's student as well as his assistant in Munich. From 1921 he was at University of Hamburg as Ordinarius Professor of
Theoretical Physics and Director of the Institute of Theoretical Physics. In
Hamburg Lenz had the collaboration of Wolfgang Pauli, who was his assistant,
and Otto Stern, so that the Physics Institute became an international center for
nuclear physics.}

By March 4 Touschek, writing 
from the  Hotel Reichshof in Hamburg, was looking for an apartment.   He also mentioned that Sommerfeld had asked him to make some calculations and to control corrections to the Italian edition of his {\it Atombau}.\footnote{The Italian edition never appeared, most probably because of problems connected with the difficult situation due to the war.}
 In Hamburg, Touschek found  the possibility of continuing his studies. From various sources and so far unpublished letters to his parents,\footnote{In addition to the  letters written to his parents from 1941 until early spring 1945, other  direct sources of information 
  for this period of Touschek's life are \W 's biography  \cite{wideroe}, the recollections written by Rolf Wider\o e to Edoardo Amaldi in 1980 and some notes on Touschek's personal  remembrances taken by Amaldi when the latter was already very ill in Switzerland. Other informations are   drawn from different versions of \T 's  Curriculum Vitae preserved in Touschek's and Amaldi's papers at the Physics Department of Rome University Sapienza, as well as in personal papers kept by his wife Elspeth Yonge Touschek.} we know that
  during 1942 Touschek attended classes at the University of Hamburg   as an auditor not regularly registered, and studied theoretical physics under the guide of Wilhelm Lenz who held a course on relativity.\footnote{ Some physicists remembered how specific directives had forbidden mentioning  Einstein's name in lectures and even in published articles. However universities did not cease to teach relativity theory, even if  the description ``the electrodynamics of moving bodies'' was generally used (see \cite{beyerchen}, pp. 169--170).}

  Other courses were taught by J. Hans D. Jensen, who had been Lenz's student.\footnote{J. Hans D. Jensen had received his doctorate at the University of Hamburg under Wilhelm Lenz, and
since 1937 he was Privatdozent (unpaid lecturer) at the University of Hamburg,
where he collaborated with Paul Harteck.} 
Another prominent physicist, Otto Stern, when Touschek arrived in Hamburg, was not  teaching there anymore.
  Jensen and Lenz, being related to Sommerfeld, could be trusted  
  in supporting Touschek in his semi-clandestine life in Germany, where his Jewish origin was not as easily detectable as it had been in Vienna.  
  
To earn his living in Hamburg, Touschek worked at Studiengesellschaft f\"ur Elektronger\"ate, a society associated to Philips, with the aim of developing an electronic valve, a precursor of the ``klystron''.  Sommerfeld himself had got him this job, through G\"unther Jobst, who 
had been Sommerfeld's student after World War I.\footnote{B. Touschek to A. Sommerfeld, Wien, January 12, 1942, Sommerfeld Archive Deutsches Museum HS 1977-28/A,343; ``Curriculum Vitae del prof. Bruno Touschek'' and ``Curriculum Vitae'' (Bruno Touschek personal papers); ``Breve Curriculum Vitae del prof. Bruno Touschek'', Amaldi Archive, Sapienza University, Rome, Box 524, Folder 4, Subfolder 4.  See also correspondence of the period.}  

Toward  the end of 1942 Touschek moved to Berlin, 
where  started to work at a small electronic devices firm, (L\"owe) Opta Radio, where he had   the task of developing oscillographic valves for the radar,\footnote{ See text of a patent with the title ``Einrichtung zur Bestimmung der Geschwindigkeit eines Geschosses''; discoverer: Touschek, Berlin, and presented by Opta Radio Aktiengesellschaft (B. Touschek Archive, Physics Department, Sapienza University of Rome, Box 3, Folder 5).} 
 a pivotal event in    \T 's life, which led him to the encounter with \W \ and accelerator physics, as we shall now narrate.

\BT \  was introduced to the Opta Radio by a 
young woman
  he had met on the train to Berlin and  who worked there  with Karl A. Egerer,  who was also editor-in-chief of the scientific magazine 
  {\it Archive f\"ur Elektrotechnik}. 
  When in Berlin, 
   he tried to continue his studies, and followed some lessons at the  University. Together with G\"ottingen and Munich, Berlin was one of the three prominent centers where modern physics had been developed since the beginning of the 20th century.
It was considered the citadel of German physics, where Max Planck had founded quantum theory in 1900. Von Laue, who had been his pupil, held there an extraordinary (ausserordentlicher) professorship in theoretical physics and had always taken public stand against the 
 government dismissal politics. Touschek attended his lectures  on the theory of superconductivity, on which the latter had made prominent contributions.

 It was  in Berlin that \BT 's life  crossed path with \RW 's. At first \T \   came to know \W \ because of 
  the work with Dr. Egerer  in the editorial department at 
{\it Arkiv f\"ur Electrotechnik},  to which \W \ had submitted an article on the betatron. 
 Later they met in Hamburg, 
 and started collaborating on the construction of the first European betatron. From their encounter,  the road to electron-positron colliders started. 
 In his autobiography \W \ , 
 races the birth of storage rings to the summer  1943 when he was in Norway observing clouds collide in the sky, and relates later  discussions with \T \ about the possibility of  head-on-collisions  between oppositely charged particles.  Such discussions about head-on-collision where   also remembered by \T \  in his first notes about the building of AdA, as one can read in  Fig.~\ref{fig:Adaproposal}. \W 's photograph is shown in fig.~\ref{fig:wideroe}.

\begin{wrapfigure}{r}{0.5\textwidth}
\begin{center}
\resizebox{0.5\textwidth}{!}{
\includegraphics{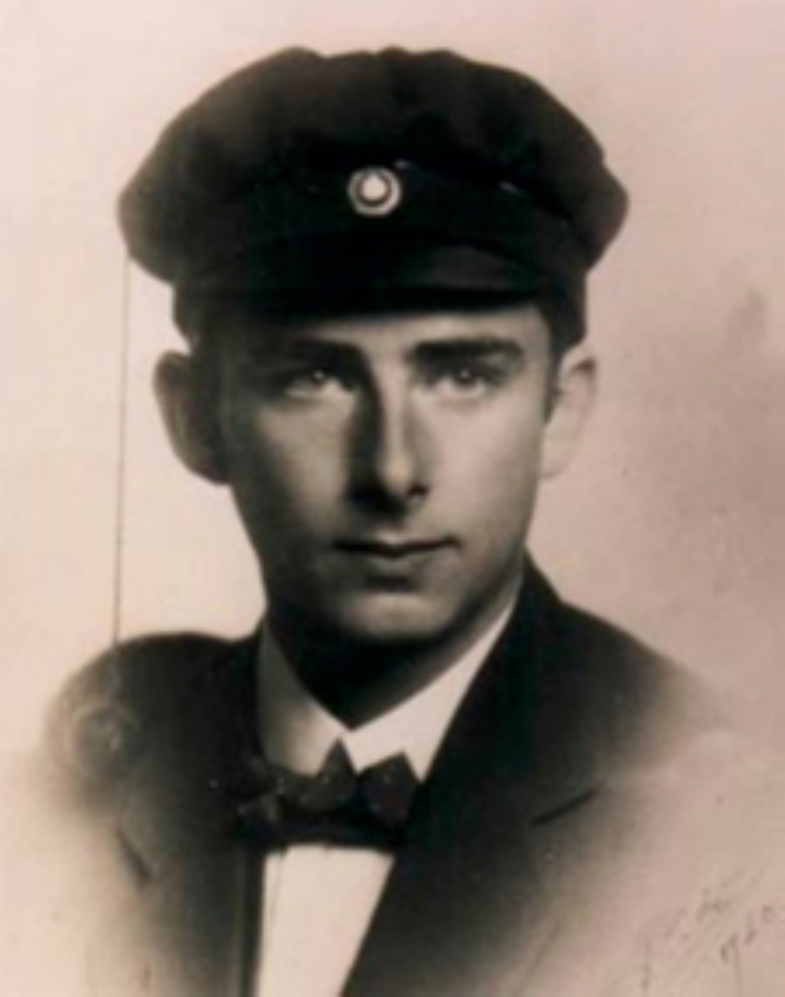}}
\end{center}
\caption{Rolf \W \ at 18,  from \cite{wideroe}. }
\label{fig:wideroe}
 \end{wrapfigure}
In 1942,  \RW\ was living in Oslo, working for  Norsk Elektrisk og Brown
Boveri (NEBB), and his article,   submitted on September 15, 1942, and subsequently published in 1943 \cite{wideroe42}, 
 discussed  Kerst's recent results on the construction of a 2.3 MeV betatron and presented new ideas for the future as well as a design for a 100 MeV betatron. 
 
  \W \ was the author of a seminal paper
 where he had presented  the design of the first functioning linear accelerator and  described the Strahlentransformator, a new device which later became known as betatron \cite{wideroe1928}. The article was the result of his thesis work entitled ``\"Uber ein neues Prinzip zur Herstellung hoher Spannungen".\footnote{In his article \W \ described the principle of 
   alternating current-based multiple acceleration  which he had used for constructing the worldÕs first linear accelerator. At the same time,
    he presented  the first comprehensive description of the principles for the betatron, including the fundamental 
    2:1 equation ${d{\bar B}_{j}\over dt}=2{dB_{f}\over dt}$ for the space-averaged field strength in the core and the guiding field at the orbit, that permits simultaneous acceleration and maintenance of the orbit at a constant radius and which later became known as ``the \W \ relation.''
   Some years later Ernest O. Lawrence found a copy of \W's thesis on the  linear accelerator in a library and drew 
   from it the idea to build the world's first cyclotron in 1939 introducing a fixed magnetic field to obtain a circular path 
   for the accelerated particles.}
Before the war,  work on development of the betatron has been actively pursued in the USA, where  D.W. Kerst and R. Serber  had then   operated the first successful betatron at the University of Illinois and the results had been published  in {\it The Physical Review} \cite{kerst1940}. Almost by chance \cite{wideroe}, because 
all American journals were banned in Germany occupied Norway,   \W \ had come to know  of Kerst's article  
and  found that, in this article, Kerst  had referred to  the work of his thesis.

  Stimulated by Kerst's article,  \W \ had then decided  to  write his own proposal for the construction of new betatrons  with higher energies and intensities and in the Autumn of 1942 submitted   to the
 {\it Archiv  f\"ur Elektrotechnik}  an  article where he discussed  a   project for a  100 MeV betatron 

Early in 1943, \T \ came to know about \W 's proposal and wrote to him.\footnote{In a letter to his parents, dated February 15th, 1943,  \T \ mentions reading  what
appears to   be Wider\o e's article, which in fact   was published  at the beginning of that year.} 
Thus    a life-long friendship and collaboration started  between these two men, brought together by different personal adverse circumstances  to work on particle accelerators.

    The   article must have also   reached  the attention  of the   German Air Force which had been  interested in financing the betatron to explore this technology and 
    obtain highly focused high  intensity X-rays bundles   (that they hoped coud be used   to kill the pilots of enemy aircraft), something often referred to as ``death-rays".\footnote{It is possible that Egerer was involved in this. In the 15th February 1943 letter, after  mentioning reading ``a stupid article" (\W 's?), \T \ mentions that   Egerer makes crazy plans (``w\"uste Pl\"ane"), like a possible assignment from the Wehrmacht
to build [...]  a cyclotron.}
 Thus it happened that, one day, as \W \ recounts in his autobiography \cite{wideroe}, in March or April 1943,  he was approached  in Oslo by some German Air Force officers, and asked  to go with them to Berlin for a matter of importance to his brother Viggo \W. 
  At that time, Wider\o e's brother Viggo, a pioneer of Norwegian aviation, had been arrested in Norway for activities against the Nazi regime and sentenced to 10 yearsÕ hard labour in concentration camps in Germany. The whole family  was  obviously   
concerned about  his well being. In order to 
secure help for his brother,   \W ,  
  initially unaware of the military uses  envisaged by the German authorities, 
 accepted to  cooperate in  
the development of betatron technology.  The project, which was financed by the Reichsluftfahrtministerium (RLM), the  Ministry of Aviation of the Reich, was given high priority and adequate resources especially with regard to  funding and staff \cite{brustad98}.

After reading \W 's article, \BT \ wrote to  him \cite{amaldi81} about some mistakes he thought were contained in  the relativistic treatment of the stability of the orbits and a correspondence ensued, which  has not been retrieved.  We know however that, on June 17th 1943,  he was writing to his parents about a meeting at RLM with Wider\o e (whom he calls ``Mein Norweger'' and does not mention by name probably for reasons of secrecy)  concerning  some points of physics. It appears that Touschek  must have convinced Wider\o e about his ideas, because he wrote  that something would be changed in the project and that he hoped to write two or three papers on the subject of their discussion, part of which would probably be published as RLM Internal Reports owing to secrecy problems.\footnote{In the June   letter he   also mentions his studies on group theory, on which he hoped  to become an expert by the end of the war.}  
From  the above,  it would appear that a considerable amount of theoretical work was carried out by \BT \  already by June 1943.   On  the betatron,  \T 's contribution to the theoretical aspects of the project was highly valued and  recognized   by 
  \W , as  described in details in a letter to Edoardo Amaldi and later in his autobiography.\footnote{``He was of great help to us in understanding and explaining the complications of electron kinetics. Especially the problems associated with the injection of the electrons from the outside to the stable orbit where they are being accelerated. Touschek showed that this process could be described by a Painlev\'e differential  equation [\dots]''. R. Wider\o e to E. Amaldi, November 10, 1979, Amaldi Archive, Physics Department, Sapienza University of Rome, Box 524, Folder 4, Subfolder 2.} 
 Similar considerations concerning \T 's contributions to the design work for the betatron  can be found in \cite{kaiser47}.\footnote{``In collaboration with the design work of Wider\o e, a considerable amount of theoretical work was carried out by Touschek which was known to have been of invaluable aid in the development of the 15-Mv accelerator.'' The list of works carried out by Touschek in 1944--1945 is mentioned in a document preserved in Wider\o e's archive at Eidgen\"ossische Technische Hochschule Z\"urich  (Handschriften und Autographen der ETH-Bibliothek, 175
Rolf Wider\o e, ``Akten, Korrespondenz und andere Dokumente zu Werk und Leben'', available at http://e-collection.ethbib.ethz.ch/). This work was never published nor mentioned by Touschek in any Curriculum Vitae.}

After an  initial period, during which he  visited Germany a few times,
 \W \ had progressed far enough with his betatron studies and, 
 in late August,   moved   to Germany  for the actual construction of the betatron.  He started working in Hamburg in  August 1943   in collaboration with the physicist Rudolf Kollath and then with \T , whom he invited to join for the theoretical work.  In Hamburg \W \  had his first contacts with Richard Seifert, who was the owner and director of a medium sized factory already manufacturing devices for X-rays since 1897, only two years  after   R\"ontgen's discovery. 
   \W \ and his group  realized that the  place  best suited to build the betatron was at the  C. H. F. M\"uller factory, which produced big X-ray-tubes and radio-valves, and had thus a great  experience in glass-blowing and vacuum techniques, 
     the basic tools of accelerator designs at the time. Incredibly, the so called R\"ontgenm\"ullerÕs  building had survived the bombings, so that it was possible to start constructing the betatron already in  autumn 1943. 
Working in Hamburg was not easy during the war. Although   Hamburg was regarded as a ``relatively safe'' place after the intensive  bombing attacks of the early summer, being the major port of the North as well as an industrial center, Hamburg was also the site of the oldest dynamite factory and continued to be the target of several strategic allied bombing missions.\footnote{Since the mid 1942, the United States  Army Air Forces had arrived in the United Kingdom, and a combined offensive plan had been  put in operation for bombing Germany. 
The combined strategic bombing offensive began on 4 March 1943. During the summer, the battle of Hamburg,  codenamed Operation Gomorrah, took place. Commencing on the night of July 24, 1943, what was  later considered the heaviest assault in the history of aerial warfare, continued until August 3.  Hundreds tons of bombs were dropped which included incendiary bombs. 
During the course of five air  attacks, which almost completely destroyed Hamburg's centre and some outer areas, over two-thirds of Hamburg's population fled the city, but 40,000--50,000 people were, and over  a million of civilian were left homeless.} New air attacks often forced the betatron group to flee to the basement, and wait until danger had passed. There was always a big question as to whether the betatron-tube was still sealed and  the vacuum still intact.  Throughout this period, Touschek kept
 sending to his parents detailed accounts of his daily life and the devastating effects of bombing on  German cities. 
At the end of a long letter written to his parents on November 27, 1943, completely dedicated to the description of his life in Berlin which was bombed night after night, he made a drawing of destroyed  buildings, which we reproduce in  Fig.~\ref{fig:bombardamento}.

 On March 20, 1944, Touschek announced to his parents that Wider\o e had required his presence in Hamburg ``in order to accelerate the work.'' Thus,  starting from April-May 1944, Bruno started moving between Berlin and Hamburg. 
\begin{figure}
\begin{center}
\resizebox{0.8\textwidth}{!}{
\includegraphics{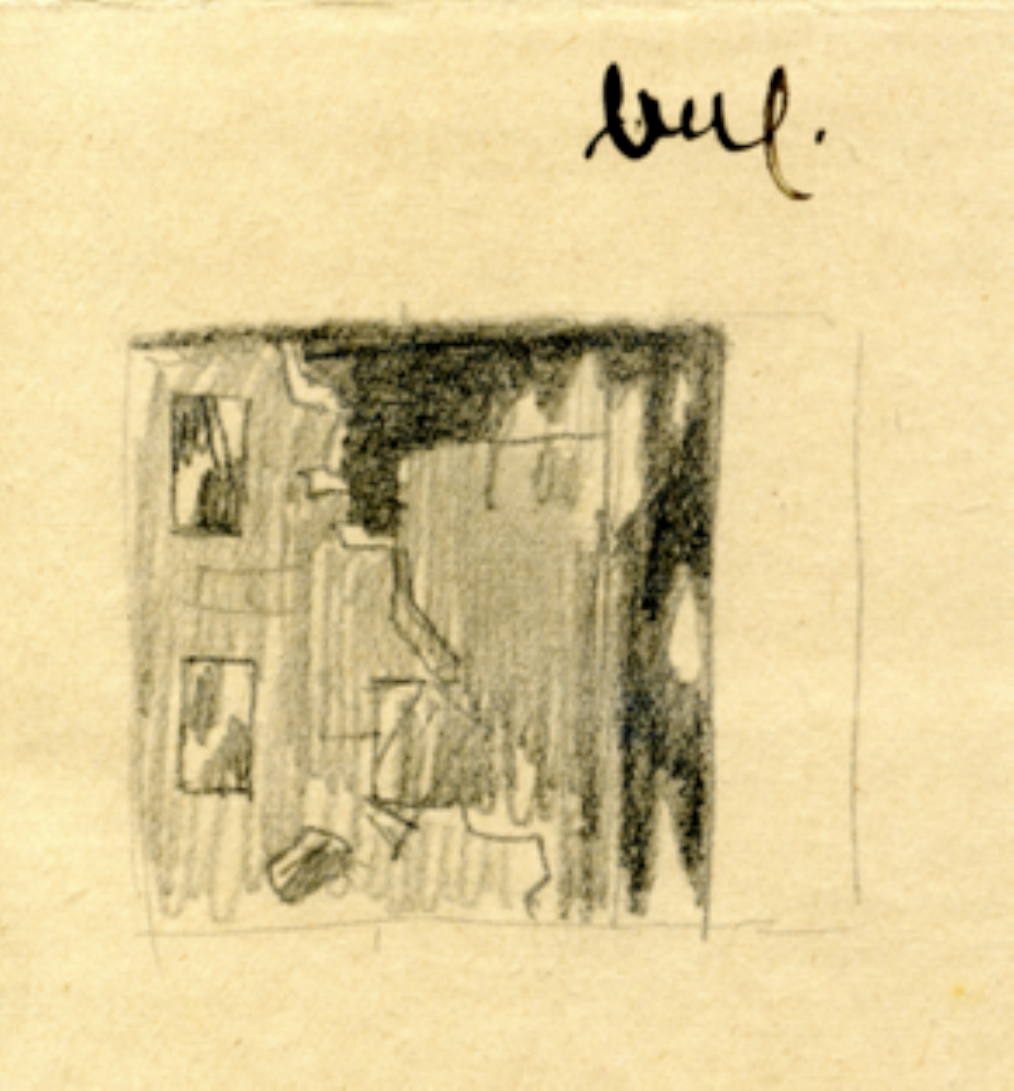}}
\caption{Touschek's drawing of a bombed building  from a letter to his father, November 27, 1943. }
\label{fig:bombardamento}
\end{center}      
\end{figure}
During the summer of 1944, the betatron was put in operation by \W \ and Kollath. The first test runs showed that the bremsstrahlung produced had an energy of 12-14 MeV. On July 8, 1944, Touschek wrote to his parents: ``I like Hamburg, I am not studying so much  [\dots]. I am always invited to very interesting conferences and seminars. However, this is all what I am doing  for my future. \W \ and I are planning to write a book on the Rheotron,\footnote{Rheotron was one of the names used to indicate the betatron.} in the meantime he has collected a lot of  material [\dots]''. On July 20 he wrote again: ``The war must be going to end soon {[\dots]''.} On July 30,  after a new bombing raid during the previous night,  he reassured them:  ``For your peace of mind, I can  say that nothing happened during the night between Friday and Saturday: all I wanted to do was to be able to { sleep [\dots].} Earlier in the week I have presented my works on nuclear theory (Yukawa theory) to Prof. Lenz for him to check them. Lenz is keen that we meet every week to discuss and the first meeting is fixed to be next Saturday, with a very selected { audience [\dots].} I  worked all { night [\dots]''. }

In these letter there is mention of repeated bomb alarms. On August 14 he was hoping to visit his parents in Vienna, and explicitely mentions his starting a vacation on August 28:
``[\dots] if you agree, I shall come to Vienna, and it will be for 3 weeks. Send your next letter to Berlin, I will arrive on  24th.''  In a letter written on September 11 he mentions his return journey, and includes the drawing reproduced in Fig.~\ref{fig:1944}.
\begin{figure}
\begin{center}
\resizebox{1.0\textwidth}{!}{
\includegraphics{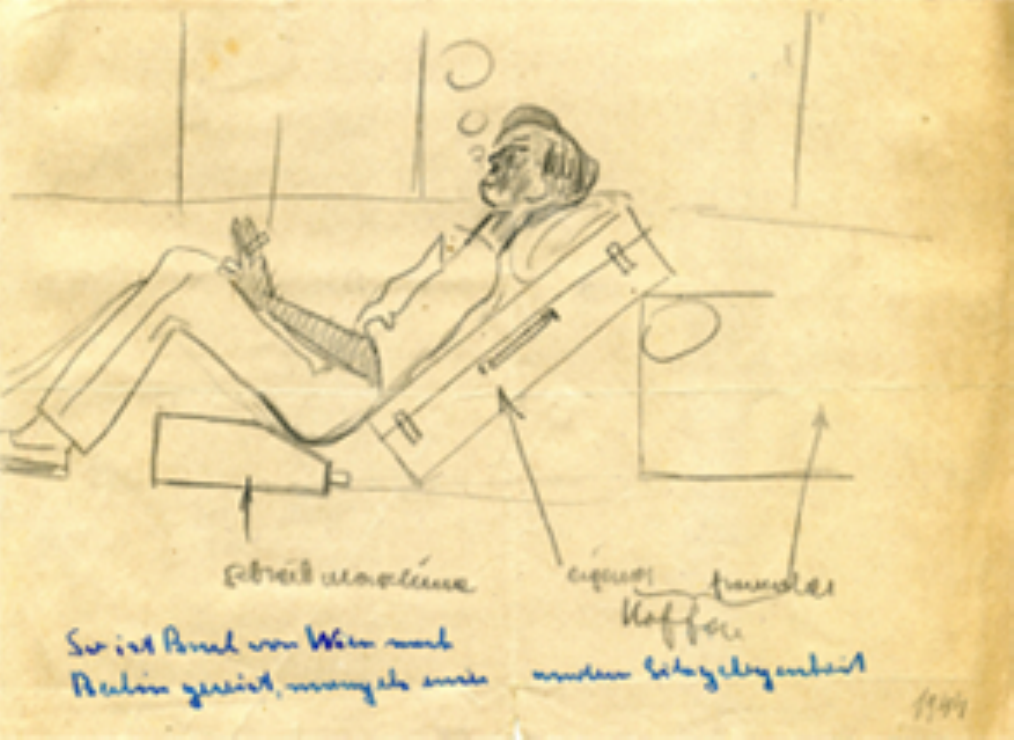}}
\caption{Unpublished drawing by  \BT, extracted from a letter to his father, dated  September 11, 1944. The caption reads: 
 ``Thus is Burl  [family nickname for Bruno]  travelling from  Wien to Berlin  for lack of seating-accomodation'', and the arrows in English translation read as ``typewriter'' and ``my own
[and] somebody else's suitcase''. }
\label{fig:1944}
\end{center}
\end{figure}

From \W 's autobiography \cite{wideroe} we learn that by the autumn of 1944 a meeting took place at the Kaiser-Wilhelm Institute in Berlin, during which the activity with the betatron was presented. According to \W, Heisenberg or Walter Gerlach must have organized it.  It had become clear that the machine could be an interesting research tool both for nuclear physics and  for medical applications. 
In the meantime, following Max Steenbeck's proposal, Konrad Gund built a 6 MeV betatron in the Siemens-Reiniger factory.  Initially, this betatron had problems with the vacuum and did not work, as recalled by \W; it was taken to G\"ottingen after the end of the war and later employed for medical uses.

 Spanning the period from September 11, 1944 until March 13, 1945, Bruno's letters to his parents describe  his  life  and work on the betatron and with Wider\o e's group.
 The March 13 letter is the last to reach his family  before the end of the war and describes  Bruno's  frenetic activity during the previous  days and the  anxious movements between Hamburg and Kellinghusen, where the betatron had been moved according to the  orders. In fact, as the British Army was approaching Hamburg, the German Aviation Ministry had ordered the betatron to be moved away, to Kellinghusen, near Wrist,  some 40 km north of Hamburg, and  the group formed by Wider\o e and his collaborators Rudolf Kollath, Gerhard Schuman and Touschek, had  been busy with transportation of the last materials, away from Hamburg and to Kellinghusen.  
 In the March 13th letter,  
 we read  that Touschek had been going back and forth during  the whole day, so that his last words in the letter written from   Kellinghusen, and probably terminated on the following day, were: ``In the meantime Wednesday has arrived, and I am really exhausted.''  
 No more letters reached \T  's family until the end of the war.  Following  this last letter,  there are two telegrams sent from Kellinghusen to \T 's father in Vienna,  the first from Bruno himself, sending his new address in Kellinghusen, and a second,  signed by an unidentified sender, Erhorn, who writes    that the son is well, but cannot answer their queries.  
 On October 22nd, 1945, a letter dated  June 22, 1945 reached Vienna, followed in November  by a second one, dated November 17, 1945. These post-war letters constitute a very important document, both about \T 's life, but also about conditions in Germany while the war was coming to a close.  They  provide a clear sequence of dates about  Touschek's movement during that dramatic period.   This period includes  \T 's imprisonment, the subsequent march towards the Kiel  concentration camp, and the   shooting episode, which we shall  describe below, and which probably saved Touschek's life. They also establish clearly that \W\ was in Germany until April 10 or 11, 1945, contrary to what is stated in \cite{wideroe}, where  \W 's return to Oslo is  dated in  March.\footnote{Actually, in \cite{brustad98}  it is told that Wider\o e went back to Norway in April 1945.}  In the following we shall present the March--April events, following the description given  in the letters by \T  \ himself. In case of contradiction with  Amaldi \cite{amaldi81,amaldi82} or Wider\o e \cite{wideroe}, we rely on the letters}.

\begin{wrapfigure}{r}{0.5\textwidth}
\vspace{-0.8 cm}
\resizebox{0.5\textwidth}{!}{
\includegraphics{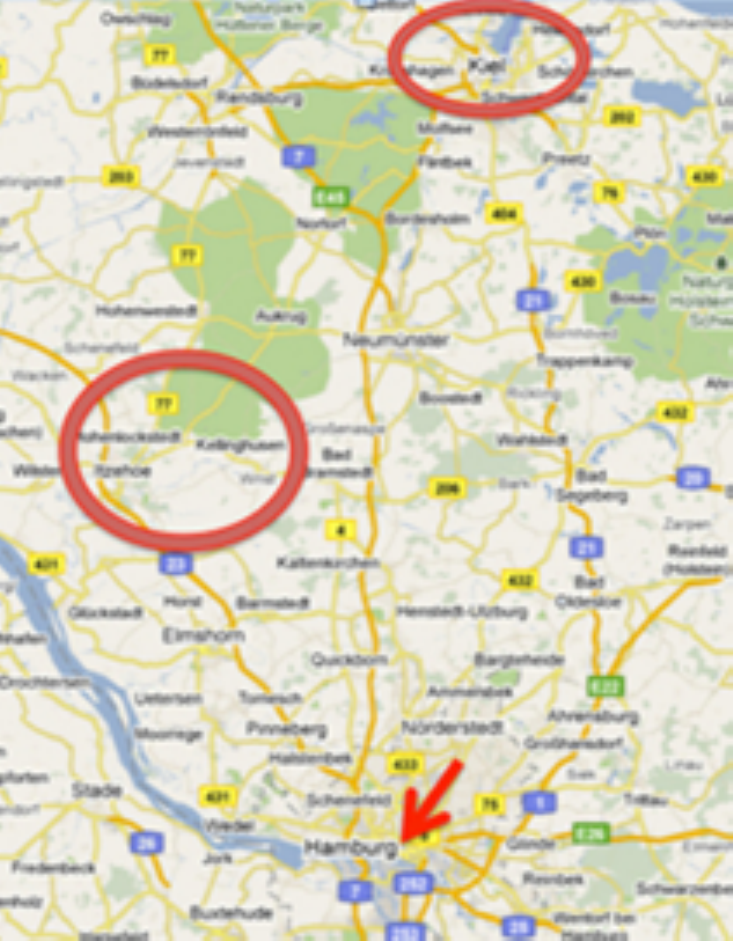}}
\caption{A map of the region around Hamburg (lower center):  on the upper right is Kiel, where Touschek and the other prisoners were being brought to by the SS guards on 11 April 1945,  in the left center Kellinghusen, where the betatron was brought from Hamburg in March 1945, and nearby the town of Itzehoe, mentioned in Touschek's letters to his parents.} 
\label{fig:german-map}\vspace{-0.5 cm}
\end{wrapfigure}

 According to  the above mentioned letters, on Thursday, 15th of March, \T \ was still in Kellinghusen, and on Friday, 16th of March,   left for  Hamburg and arrived there  by night, during an  air raid. 
On the following morning he was arrested and taken to prison to Fuhlsb\"uttel, near Hamburg,
 under suspicion of espionage.  
To render clear \T 's movements during these last days, we show in Fig.~\ref{fig:german-map} we  a map of the region.

While in prison, as  described in Touschek's letter to his father, \W \ would go to visit him, bring him books and continue discussing about the physics and their ideas. \W \ also, in his autobiography, remembers bringing him food and cigarettes and a copy of the classic book by Heitler {\it The Quantum theory of Radiation} \cite{heitler}.  On this book, it is said     that Touschek wrote, with invisible ink, a short note on ``Radiation damping in betatrons'' \cite{amaldi81,amaldi82}.\footnote{ The question of writing with   invisible ink is mysterious. No  mention of such episode can be found in \T  's letters. In Amaldi's original typescript, after  February 28th, 1978, there is mention of a note  written in "modo invisibile",  "invisible way", with   a  question mark. A related note, authored by Touschek,  dated 1945, and entitled ``Zur Frage der Strahlend\"ampfung im Betatron'' is listed in the catalogue of R. Wider\o e's papers preserved at the Library of the Wissenschaftshistorische Sammlungen of Eidgen\"ossische Technische Hochschule (ETH), in Zurich.}

From this narrative,  there  emerge   the intellectual development and studies which would lead \BT \ many years later  to the first  electron-positron storage ring  and to a program of administering radiative corrections to high energy electron positron scattering. But all this was yet to come. In early 1945, while the Allied were
progressing through Europe, Touschek met the most dramatic point of his life. We have his direct
account of a famous incident in a letter he wrote to his father shortly after the war ended. We have
extracted some details, as follows.
\par\noindent
``Kellinghusen, 1945, June 22nd
\par\noindent
Dear parents,  [\dots] I have not received any news from you for a very long time [\dots] [I shall now give you ] a brief update about what happened to me [\dots]. After 3 weeks in prison in Hamburg, where I was because of suspected espionage, the prison was evacuated [and] all the (200) prisoners were put in a long line towards Kiel [concentration camp]. In front of us, behind, and on the sides, marched the SS guards.
Near Hamburg [\dots]. I fell to the ground [\dots]  and the guards pushed me in the gutter, near the
road, and shot at me. One shot went through my left ear, the other through the padding  of my coat. [After they left me for dead] I went to the hospital, and was again made a prisoner and sent to Hamburg from prison to prison. This lasted about four weeks.''
We present  a translation of the complete  letter in Sec.~\ref{sec:letters}.  

 After being set free on April 30, and during several months --- at least until the end of 1945, according to what he wrote to his parents --- Touschek was practically a prisoner in Kellinghusen before the  whole situation of the Wider\o e betatron was clarified with the British authorities, who decided to take the machine to England, near London, as part of the booty of war \cite{kollath}.\footnote{Manuscripts signed by Touschek regarding the theory of betatrons, and marked with the name of Kellinghusen are kept at the Library of the Wissenschaftshistorische Sammlungen of Eidgen\"ossische Technische Hochschule, among Rolf Wider\o e's documents: B. Touschek,  Zur Theorie des Strahlentransformators; On the Starting of Electrons in the Betatron; Die magnetische Linsenstrasse und ihre Anwendung auf den Strahlen-Transformator (1945); Zur Frage der Strahlungsd\"ampfung im Betatron (1945).}
At the beginning of 1946  \BT\ went  to University of G\"ottingen to obtain his diploma, and then, for his final education, to Glasgow, where he obtained his doctorate in 1949. As discussed in the next subsection, this experience established Touschek's subsequent and seminal work of all the subsequent years. 

When, in the   1950s, in Frascati, near Rome, an electron synchrotron was  commissioned and finally designed and built a few years later, Touschek, who had moved to Rome in 1953 with an INFN position, was then ready and prepared to join the work taking place in the newly built INFN National Laboratories in Frascati.

\subsection{Diploma at G\"ottingen}

By the end of September 1945, Touschek must have received a letter from Sommerfeld, to whom he answered telling him he was working on  ``betatron calculations'', on ``neutrino theory'' and ``on radiation damping.''\footnote{B. Touschek to A. Sommerfeld from Kellinghusen, September 28, 1945. Deutsches Museum Archive, NL089,013.  Sommerfeld was very worried about his son Ernst, from whom he had not received any news since some time and actually nobody knew where he was.}

 In 1946, he went to G\"ottingen to continue  his studies and get his diploma. The university town of G\"ottingen, respected for its longstanding and high tradition in the natural sciences, had been chosen by the British authorities as the center for the reconstruction of West German science. Max Planck had arrived there as a refugee on June 4, 1945,  while Werner Heisenberg  settled there at the beginning of 1946, after returning to Germany from internment in Great Britain.
 He organized a new Institute for Physics and made it flourish with the help of Carl Friedrich von Weizs\"acker, Karl Wirtz and Erich R. Bagge. 

 The Siemens 6-MeV betatron built by Gund  had been brought to G\"ottingen, where Wolfgang Paul later extracted the electron beam in order to study the reaction $e+D\rightarrow e+p+n$. On June 26 1946, Touschek  obtained his ``Diplom-Physiker'' (Diploma in Physics) with a thesis on the theory of the betatron, made under the supervision of Richard  Becker, director of the Institute of Theoretical Physics and Hans Kopfermann,\footnote{See Touschek's Studienbuch and Diplom in  Bruno Touschek's personal papers preserved by Elspeth Yonge Touschek.}
and was appointed research worker at the Max Planck Institute of G\"ottingen, where he began to work under the direction of Heisenberg, with whom he continued to keep in contact during the following years.
Touschek had first seen Heisenberg giving a public lecture in Vienna in 1939, and later had met him in Berlin.  In an unpublished and undated manuscript, he recalls his impressions of him and talks about some problems they discussed together during \T 's G\"ottingen days.\footnote{ ``First impressions: Common sense. The word I remember to have heard him use was ``vernunftig''= reasonable. Then a phrase: ``Das wollen wir einmal versuchsweise nicht glauben.'' [$\sim$ 
 To try it out, let's not believe it for now]. In G\"ottingen I attended his lectures on the quantum theory of fields. It was not a good lecture course, but there was one lecture, among them, which for me was a complete eye-opener: the harmonic oscillator and  its quantization. I had learned Q. T. from Sommerfeld's ``Wellenmechanisch Erg\"anzungsband'' and  I had tried DiracÕs famous book, both of which lean heavily on wave mechanics. His lecture opened my understanding to the mechanical  approach [\dots]. The problem which bothered H. and  which he asked me to unravel was ``double  $\beta$-decay''. Haxel felt he could just do it (experimentally) and  I ran into the difficulty of distinguishing what was arbitrary and what was sound in Fermi's theory of ``weak interactions.'' I saw that clearly in 1947, but what I wrote then was riddled by stupid mistakes, which H. did not -- or did not want to -- see.'' B. Touschek, ``Remarks on the influence of Heisenberg on physicists,'' undated manuscript, Bruno Touschek's personal papers preserved by Elspeth Yonge Touschek.} 
  From letters of the period we learn that during the summer he
spent some time in Wimbledon, where he reported the British authorities
on the betatron project. On November 24, 1946, he announced to his
parents that an article on the double  $\beta$-decay, was ready for publication \cite{touschek1948a}.\footnote{Touschek to his parents, November 24, 1946. Actually the article was published much later, on January 1948. It appears
that this research argument was suggested by Urban when Touschek left from
Vienna, according to what the former wrote later to Amaldi on June 3, 1980, Amaldi Archive, Physics Department, Sapienza University of Rome, Box 524,
Folder 4, Subfolder 4.} 

\subsection{Glasgow}
  At the beginning of April 1947, \BT\ settled in Glasgow  with a scholarship of the Department of Scientific and Industrial Research. A the time,  there was active work  to design and construct a 300 MeV synchrotron. \BT \ joined the work, which was taking place under   Philip I. Dee, director of the Department of Natural Philosophy.

\begin{figure}[htb]
\begin{center}
\resizebox{1.0\textwidth}{!}{
\includegraphics{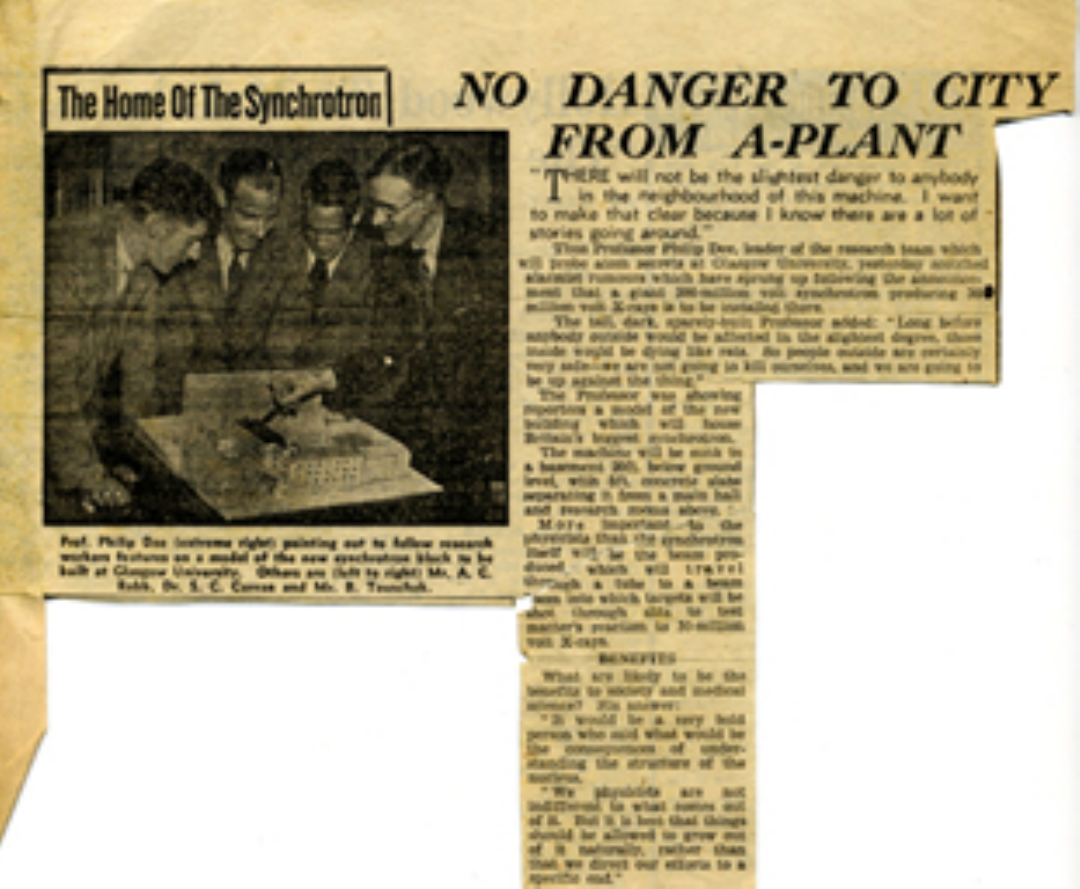}}
\caption{ B. Touschek in Glasgow and a comment from a local newspaper.}
\label{fig:cutting}
\end{center}
\end{figure}

This period in Touschek's life, albeit relatively short, from the spring of 1948 to the end of 1952, 
is very important  for his future work both in accelerators and in theoretical physics.
His work with Dee and J. C. Gunn clearly built upon his earlier interest with Wider\o e; during the Glasgow period Touschek studied in depth problems related to the working of the 30-MeV synchrotron, as well as to the construction of the 300 MeV machine.\footnote{ Correspondence and documents of the period clearly show his involvement both in the building  of the new synchrotron, and in his contribution to clarify problems with a betatron machine in Manchester. Bruno Touschek Archive, Physics Department, Sapienza University of Rome, Box 3, Folders 1, 4, 5, 6 and 7 and Box 1, Folder 1.}
  Later,  the work with Walter Thirring, who was also a young post-doc in Glasgow \cite{thirring2008}, was connected  with his  wider interest for electrodynamic  processes and with relativistic quantum electrodynamics, whose study constituted, at the time,  the new physics frontier.   In Fig.~\ref{fig:cutting}  we show a cutting from a local newspaper with a photograph of \T \  with Philip Dee and other colleagues in front of the model of the 300 MeV Synchrotron.

  Touschek submitted his first paper for publication on October 11 of 1947; it dealt with excitation of nuclei by electrons \cite{touschek1947}, a research which he continued to carry on during the following year and which became the basis for his Ph.D. dissertation. During the months December 1947 and January 1948, Touschek carried out researches on different issues. While absorption-measurements with the already existing 30 MeV electron synchrotron were being prepared, a ionization chamber had been ordered and a magnetic collimator was being designed, Touschek made accurate calculations for using carbon as an absorber. He also worked out a solution for the problem regarding statistics of `effective' track-lengths in an ionization chamber with radius comparable but larger than range of particles emitted from sources randomly distributed over the gas of the chamber.\footnote{At that time, Dee was particularly interested in studying the disintegration of atomic nuclei with the Wilson cloud chamber technique, for which in 1952 he won the Hughes Medal.}
  In the meantime, Touschek held ``triangular discussion'' between  Heisenberg and N. Hu in Copenhagen on questions regarding the analytic behavior of the S-matrix.\footnote{``Research carried out during the months December 1947 \& January 1948'', Bruno Touschek Archive, Physics Department, Sapienza University of Rome, Box 3, Folder 1; W. Heisenberg to B. Touschek from Cambridge: January 28, February 23, 1948; from G\"ottingen: April 20, 1948.  Bruno Touschek Archive, Box 1, Folder 1.} 
 
\begin{figure}[htb]
\resizebox{1\textwidth}{!}{
\includegraphics{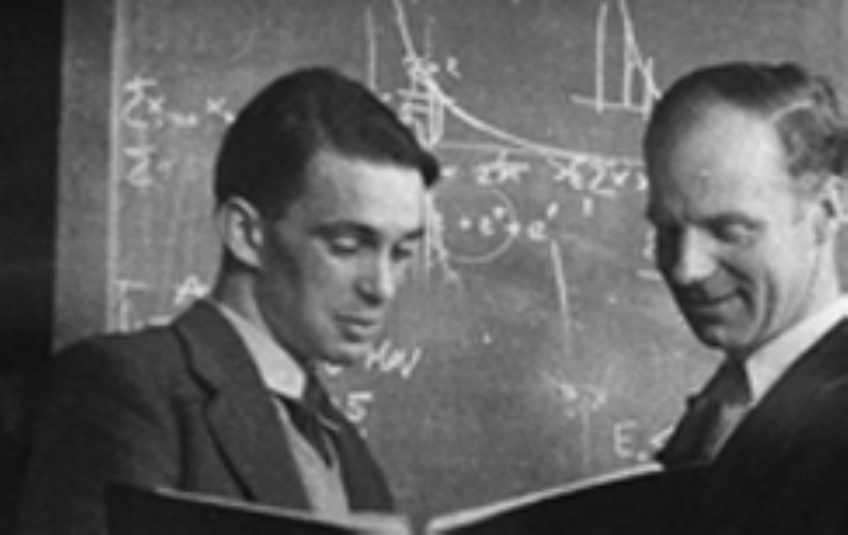}}
\caption{
 \T  \  in Glasgow with  Samuel C. Curran  (identification courtesy of  Glasgow University).}
\label{fig:glasgow}
\end{figure}

 Touschek's research report for the period February 1 to April 30, 1948, mentions submission of the review paper on the synchrotron \cite{touscheksynchrotron} and work connected with ``Heisenberg's theory of the $\eta$-matrix'' : ``In March -- after a conversation with Prof. Heisenberg in Manchester I started closer investigation on a model system (meson-field in interaction with an oscillator) which seemed rather promising. Heisenberg wrote on April 20th: `the example you described in your last letter seems to me to be an extraordinarily reasonable choice [\dots]. At present it appears, as if the model in question did not contain particle-production at all.'' As a sideline research Touschek worked on production of mesons in fission processes and determination of matrix-elemens for neutron-nuclear interaction.\footnote{``Research report for February 1s to April 30th 1948'', Bruno Touschek Archive, Physics Department, Sapienza University of Rome, Box 3, Folder 1.} 
 During the period 1947--1948 \BT\  published several papers on this issue, alone or in collaboration with other people \cite{touschek1948b,touscheksneddon1948,touschek1948c,touscheksneddon1948b,touschek1948d}. 
 On November 5, 1949, he was awarded his Ph.D. with a thesis entitled ``Collisions between electrons and nuclei'' which represented a review of the work on electron excitation and production of mesons by electrons carried out by Touschek in collaboration with I. N. Sneddon during the years 1947 to 1949.\footnote{One of us (L.B.) is indebted to  prof. D. H. Saxon, of Glasgow University, for sending a copy of Touschek's thesis. } During the year 1949 Touschek published a series of works on arguments reviewed in his thesis \cite{touscheksneddon1949,touschek1949}. In Fig.~\ref{fig:glasgow},
   we show Touschek  in Glasgow together
  with Samuel C. Curran\footnote{Sir Samuel Curran (1912-1998) was  Lecturer in the Physics Department  of University of Glasgow from 1945 until 1955. In 1959 he was appointed Principal of the Royal College of Science and Technology and was instrumental in leading the institution to Universiy status at the beginning of the 1950s (Courtesy of University of Glasgow).}.

 After his Ph.D. he became Nuffield lecturer and from the autumn 1950 he worked on the covariant formulation of the Bloch-Nordsieck method with his friend Walter Thirring who had arrived in Glasgow as Nuffield Fellow \cite{thirring1951}. During his Glasgow years  Touschek was in contact with Max Born, who had become Tait Professor in Edinburgh after leaving Germany in 1933. Touschek often visited him in Edinburgh  and  helped him to edit the new edition of  his {\it Atomic Physics},  a series of lectures that Born had given in Germany in 1933, later published in Glasgow in 1935. Born thanked him in the preface ``for scrutinizing and criticizing the whole script and for many valuable suggestions, in particular about $\beta$-decay and meson theory \cite{born}.\footnote{Touschek recalled that during this work he discovered the universality of beta interaction and prepared an appendix where he discussed analogies between nuclear $\beta$-decay and the decay of $\mu$ and charged $\pi$-mesons, but he later realized that in the meantime Giampiero Puppi had already done something similar. \BT ``Curriculum Vitae", Bruno Touschek's personal papers preserved by Elspeth Yonge Touschek.} 
In 1950 and 1951 he published works on the production of $\pi$-mesons, and a paper on ``A perturbation treatment of closed states in quantized field theories'' \cite{touschek1950,touschekgunn1951,touschekperturbation}. 

From letters to his parents, and from a letter written to Arnold Sommerfeld on October 5, 1950, we learn that he was unhappy in Glasgow, and that he hoped to find a different position.\footnote{B. Touschek to Arnold Sommerfeld, October 5, 1950, Deutsches Museum Archive, NL 089,013.} 
  On June 20, 1951, he writes to his parents that he is planning to buy a motorbyke to travel through Europe and visit Rome during the summer. On November 9 he is announcing that he has tried to obtain a position in Rome, and that Bruno Ferretti  (professor of theoretical physics at Sapienza University of Rome) has written him, but the matter was still to be perfected. Only in December of the following year Touschek will succeed in moving to Italy. On December 30, 1952, he is writing to his parents from Rome with enthusiasm: ``In the Institute I will occupy Blackett's room after his leaving [\dots]. Roman food and wine make me feel well and also the climate -- even now in winter -- has nothing to share with Glasgow. The Institute is really excellent. At the moment there are two Nobel Prize (Pauli and Blackett) and a possible candidate and other people are very interesting, too.'' 


\section{Touschek's impact on  theoretical developments in Rome and Frascati}
  \label{sec:resummation}
  We have described, in Sect.~\ref{sec:ada}, the events which followed \T  's arrival in Rome and the construction of AdA. In this section, we shall present his later work and illustrate the influence he exerted  \ on theoretical physics at the University of Rome and Frascati.

Soon after AdA had started working, Touschek already thought of a bigger 
machine, one 
which could do real physics, with a higher energy, ADONE, namely a 
bigger, better, more 
beautiful AdA. The proposal is signed by Touschek,  Carlo Bernardini and Giorgio Ghigo (AdA proposers),  together with   Fernando Amman and  
 Raoul Gatto. Amman became the director of the  ADONE project and had it ready for commissioning in 1968.

 Gatto\footnote{R. Gatto is presently Emeritus Professor at University of Geneva, Switzerland.  After a few years at U. of Rome, he moved to Florence, as  Professor of Theoretical Physics  and trained  many  students, also known as  ``gattini'' [kittens]. Some of the ``gattini'', such as  Guido Altarelli, Franco Buccella, Luciano Maiani, Giuliano Preparata, Gabriele Veneziano, to name just a few, were later to give important contributions to theoretical particle physics.  Gatto was also Chief Editor of Physics Letters B for many years.}   was at the time a young Assistant Professor in Rome, interested in all which was happening in Frascati.
After Panofsky's seminar and Touschek's excitement on the possibility  of  electron-positron collisions,  Gatto, together with Nicola Cabibbo,  had  started exploring the physics which could be probed by such collisions 
 \cite{Cabibbo1960a,Cabibbo1960b}.  In the appearing of Gatto and \T 's names together in the ADONE proposal, we see the close collaboration of theory, technology and experimentation which was present in Rome  and Frascati at the time. It also highlights another contribution of 
Bruno Touschek, namely his influence on the development and successes of the  Rome University  
theoretical  physics school.  As already mentioned, Touschek had many students, his first two having been  Nicola Cabibbo and Francesco Calogero, to be followed by  Paolo Di Vecchia, Sergio Ferrara, Giovanni Gallavotti, Giancarlo Rossi,  and many others.\footnote{A  list, of \T '\ students can be found in \cite{amaldi81}.}
 Through the 10 years spanning the proposal to build AdA and the construction of ADONE
from the late '50s until the late '60s, \BT\  was a referral for most of the physics  students enrolling in unprecedented number at the University of Rome, following the wave of interest generated by the launch of Sputnik.\footnote{ As mentioned, the  1959 broadcast in national television of physics lectures by the Director of the Frascati Laboratories,  Giorgio Salvini, also contributed to such interest.}
He was, in those years, a  charismatic personality, a great teacher, of brilliance and clarity,  a scientist who had known the famous physicists of the German school and  a driving figure. 
\begin{wrapfigure}{r}{0.5\textwidth}
\vspace{-0.5 cm}
\resizebox{0.5\textwidth}{!}{\includegraphics{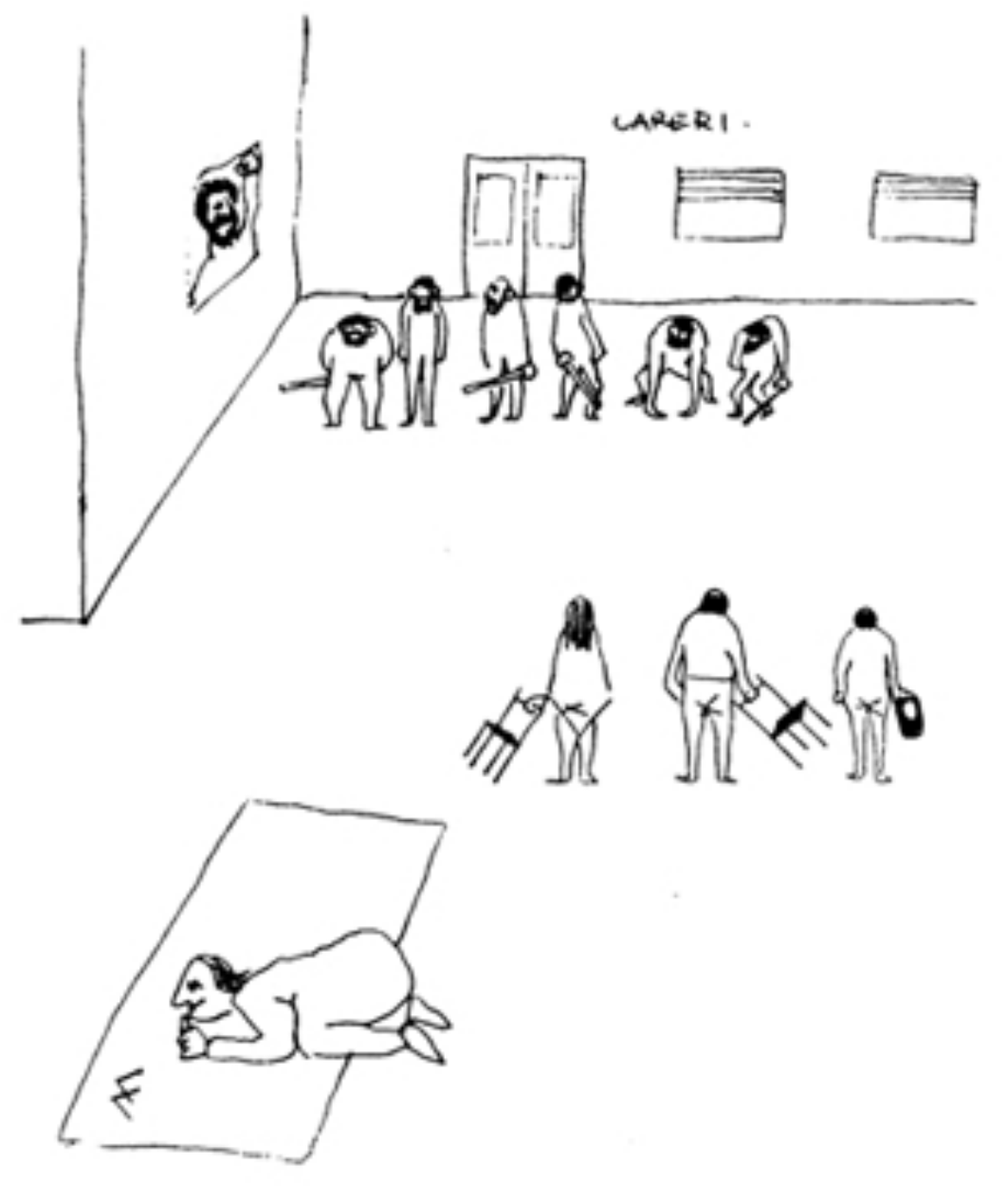}}
\caption{Undated  drawings by \BT \ describing the  1968 student unrest in University of Rome. }
\label{fig:careri}
\vspace{-0.45cm}
\end{wrapfigure}
Because of AdA and 
ADONE, many 
teoretical physics theses came to be, some with \BT, others with  Cabibbo or with  Gatto. At first, single 
and double bremsstrahlung processes were studied, but soon Touschek became concerned with the high energy limit of soft photon emission, a problem he had been interested when in Glasgow.
 It became clear to \T \ that, at ADONE's center of mass energy of $\sqrt{s}\le 3.0 \ GeV$, summation 
of an infinite number of soft  photons would be  a necessity for any precision measurement. Using  experience in QED processes accumulated  both during the war and in Glasgow in his work with Walter Thirring, \T \ prepared  a work on summation of soft photons in high energy reactions, which had a strong impact  on subsequent theoretical work in Frascati and in University of  Rome \cite{etim1968}.  Touschek's concern was indeed well founded: as the center of mass energies reached by electron-positron colliders increased,  resummation techniques\footnote{The term {\it resummation} was not used during \T's times and came in use later, in connection with Quantum Chromodynamics.}
 would soon become   an indispensable tool to extract theoretical quantities from the data, as it proved to be the case when
the $J/\Psi$ was discovered \cite{greco1975}.  Unfortunately, the 1968 student unrest  in Rome and other political problems in Frascati delayed ADONE's operation   for a whole year \cite{amman,valente}. In Fig.~\ref{fig:careri} we show one of the many drawings which \T \ dedicated to the 1968 unrest in Rome University.

The need for a special formalism to apply radiative corrections   to very narrow resonance production in  electron-positron 
annihilation had seemed  a 
rather virtual one in 1968, when he had suggested the problem to the young theorists of his group in Frascati, but it became a reality a few years later  with  the discovery of the $J/\Psi$. Another problem that Touschek was very keen in 
solving, but could not really complete,  was 
the angular distribution of resummed photon momenta. He did not live to see 
how important 
the problem became in Quantum Chromodynamics. In the following we shall discuss in  more details some of these issues.   At the end of this section, we shall also mention the influence of  \T 's teaching and experience in  the development of  statistical mechanics studies at University of Rome.

\subsection{The "Infra-red catastrophe" }

In 1948 Bruno Touschek was in Glasgow and became interested  in the
infrared
catastrophe phenomenon, namely 
 the fact 
that in any process in which charged particles scatter from an initial into a final state, i.e.   are created or destroyed,
the probability of emission of light quanta diverges as the photon
frequency goes to zero. 
The problem had been pointed out by Bloch and Nordsieck in a paper in 1937. 
In  \cite{bloch1937} Bloch and Nordsieck  observed that previous
methods of treating radiative corrections to scattering processes were defective in
that they predicted infinitely large low frequency corrections to the transition
probabilities. This was evident from the $d\omega/\omega$ spectrum for single photon emission in electron scattering in a Coulomb field as described by Mott and Sommerfeld \cite{mott1931,sommerfeld1931} and by Bethe and Heitler \cite{BH}. Bloch and Nordsieck had noticed that, while the ultraviolet difficulties are
already inherent in the classical theory, the infrared divergence has no
classical counterpart, and anticipated that only the probability for the
simultaneous emission of infinitely many quanta can be finite and that the
probability for emission of any finite number of them must vanish.
 They noticed that  for
emitted photons of  frequency larger than
a given $\omega_0$ the  probability of emitting each additional photon is
proportional to
$(e^2/\hbar c) \log [E/(
{\hbar}\omega_0)]$, 
which becomes large
as $\omega_0 \rightarrow 0$. Thus, the actual expansion is not
$(e^2/\hbar c)$, which would be small, but a larger
number, driven by the logarithm. This led them to analyze the scattering process in terms of what
came to be called Bloch-Nordsieck states, namely states with one electron plus the electromagnetic field, and to substitute the expansion in
$(e^2/{\hbar}c)$ with a  more adequate one. 

The important result of \cite{bloch1937}
 was that, albeit   the probability of emission of any finite number of quanta is zero in the $\omega \rightarrow 0$ limit, when summing on
all possible numbers of emitted quanta,  the total transition probability
and the total radiatated energy were finite. This led Bloch and Nordsieck  to anticipated that  the mean total number of quanta
had to be infinite. Thus the idea that any scattering process is always
accompanied by an infinite number of soft photons was introduced and proved
to be true in a non-covariant formalism.

In \cite{bloch1937}  one  sees the emergence of the concept of finite total energy,
with exponentiation of the single photon spectrum which is logarithmically divergent
such that the  probability for a finite number of emitted photons is always zero. On the
other hand, when summation is done over all possible photon numbers and
configurations, the result is finite.  Clearly there was still something
missing because 
there is no hint of how to really cure the infrared divergence. In addition the language
used is still non-covariant.

Before going to the covariant formulation, we notice that the crucial
argument   relies on the transition probability being proportional to
\begin{equation}
\Pi_{s\lambda}
e^{-
{\bar n}_{s\lambda}}
{{
{\bar n}_{s\lambda}^{n_{s\lambda}}
}
\over{
n_{s\lambda}!
}}
\end{equation}
namely to a product of Poisson distributions, each of them describing the
independent emission of $n_{s\lambda}$ soft photons.

The Bloch and Nordsieck  formulation and the subsequently
proposed solution were  framed in a non covariant language. 
When Touschek and Thirring met in Glasgow, after the war and with  second quantization and relativistic quantum field theory, they saw the need to reformulate the problem  in  a covariant formalism.
As Touschek and Thirring say in the introduction of their paper \cite{thirring1951},
the results they obtain were not new and had been discussed
by several authors, but the importance of the Bloch Nordsieck problem, as the
only one which admits an accurate solution,  justified  a general
reformulation. It should be stressed that  the simplification
which  enabled one to find an accurate solution rested  on the neglect
of the recoil of the source particles \cite {bloch1937}. Using this approximation
 Touschek and Thirring 
  calculated the average number of photons emitted in a  momentum  interval $\Delta$.
In their paper  Touschek and Thirring  first derive their results for a source scalar field, then
they generalize it to a vector source function $j_\mu(x)$ of a point-like
electron, i.e.
\begin{equation}
j_\mu(x)=e\int p_\mu(\tau) \delta(x-\tau p(\tau)) d\tau
\end{equation}
 where $p_\mu(\tau)=p_\mu$ for  the (proper) time  $\tau$ less than 0 and $p_\mu(\tau)=p'_\mu$ for
 $\tau$ larger than 0. Notice that the sudden change in momentum imposes the
 restriction that in order to apply the results to a real scattering process,
 the photon frequencies should always be much smaller than $ 1/\tau$,
where $\tau$ is the
 effective time of collision. Otherwise the approximation (of a sudden
 change in momentum) will break down. One then obtains
\begin{equation}
j_\mu(k)=e\big( {{p_\mu}\over{(pk)}}-{{p'_\mu}\over{(p'k)}}\big)
\end{equation}
and the average number of quanta ${\bar n}$ now becomes
\begin{equation}
{\bar n}={{e^2}\over{(2\pi)^3}}\int_\Delta d^4k \delta(k^2-\mu^2)\big[
{{(p\epsilon)}\over{(pk)}}-{{(p'\epsilon)}\over{(p'k)}}\big]^2
\end{equation}
where $\epsilon$ is a polarization vector.

\subsection{Schwinger's Ansatz on the exponentiation of the infrared factor and status of the field in early '60s}

The solution found by   Bloch and Nordsiek
and later brought into covariant form by  Touschek and Thirring  did not really solve the
problem
of
electron scattering in an external field and of how to deal with finite energy
losses.
 This problem was  discussed and solved in
the context of Quantum Electrodynamics, where
 the logarithmic divergence attributable to the
infrared-catastrophe from emission of real light quanta of zero energy
was compensated through the emission and absorption of
virtual quanta. This cancellation was taking place in the cross-section,
and not between amplitudes.    In a short paper in 1949 and, shortly
after, in the third of his  classic QED papers, Julian Schwinger \cite{Schwinger1949a,Schwinger1949b}
examined the  radiative
corrections to (essentially elastic) scattering of an electron by a Coulomb
field, computing second order corrections to the first order amplitude and
then cancelling the divergence in the cross-section between these terms and
the cross-section for real photon emission.
The result, expressed as a fractional
decrease $\delta$ in the differential cross-section for scattering
through an angle
$\theta$  in presence of an energy resolution $\Delta E$ of the scattered electron,
is of order
$\alpha$  and given by
\begin{equation}
\delta={{2\alpha}\over{\pi}}log({{E}\over{\Delta E}})\times F(E,m,\theta)
\label{js}
\end{equation}
where $F(E,m,\theta)$ in the extreme relativistic limit is just $\log (2E/m)$.
Schwinger notices that
$\delta$ diverges logarithmically in the limit
$\Delta E\rightarrow 0$ and
points out that this difficulty stems from the neglect of processes with
more than one low frequency quantum. Well aware of the Bloch and Nordsieck
result, he notices that it never happens that  a scattering event
is unaccompanied by the emission of quanta and proposes to replace the
radiative correction factor $1-\delta$ with $e^{-\delta}$, with
further terms in the series expansion of $e^{-\delta}$ expressing the effects
          of higher order processes involving  multiple emission of
          soft photons.

In 1949 however, given the
 energies available for scattering experiments, 
the exponentiation of the
radiative correction factor, was still far from being needed.   As  Schwinger  points out,
  the actual correction to  then available
experiments, could be estimated to be at most about  $10\%$. Almost twenty years had to
pass before the exponentiation became an urgent matter, and $\alpha log( E/m)$, the factor which 
 Touschek christianed  the {\it Bond factor}, 
 started to become
so large that  the first order correction, the double logarithm
$\alpha log(E/\Delta E) log (2E/m)$ would climb to $20 \div 30 \%$ and beyond.

In the 1950s, with Feynman diagrams technique available to the
theoretical physics community, many  higher order QED
calculations came to be part of standard theoretical physics handbooks.

Many important contributions to the radiative correction problem appeared
in the '50s and early '60s \cite{BF,jauch,sudakov1956,lomon1956,lomon1959,erikson}, with
a major step in the calculation of infrared radiative corrections
 done in 1961  by Yennie, Frautschi and Suura (YFS)  \cite{Yennie1961}.
In their classic paper,
they went though  the cancellation of the infrared divergence
 at each  order in perturbation theory in the cross-section and
obtained the final compact expression
for the probability of energy-momentum  loss in a high energy reaction between charged particles. 
In their paper they compute higher and
higher order photon emission in leading order in the light photon
momentum, showing that the leading terms always come from emission from
external legs in a scattering diagram. In parallel, order by order, they
extract the infrared divergent term from the virtual diagrams, making the
terms finite through the use of a minimum photon energy. They show that the
result is just as valid using a minimum photon mass, and finally eliminate
the minimum energy,  showing the result to be finite in the infrared.

\subsection{The {\it Bond factor } and the radiative corrections to $e^+e^-$ experiments}

\begin{figure}
\begin{center}
\resizebox{1.0\textwidth}{!}{
\includegraphics{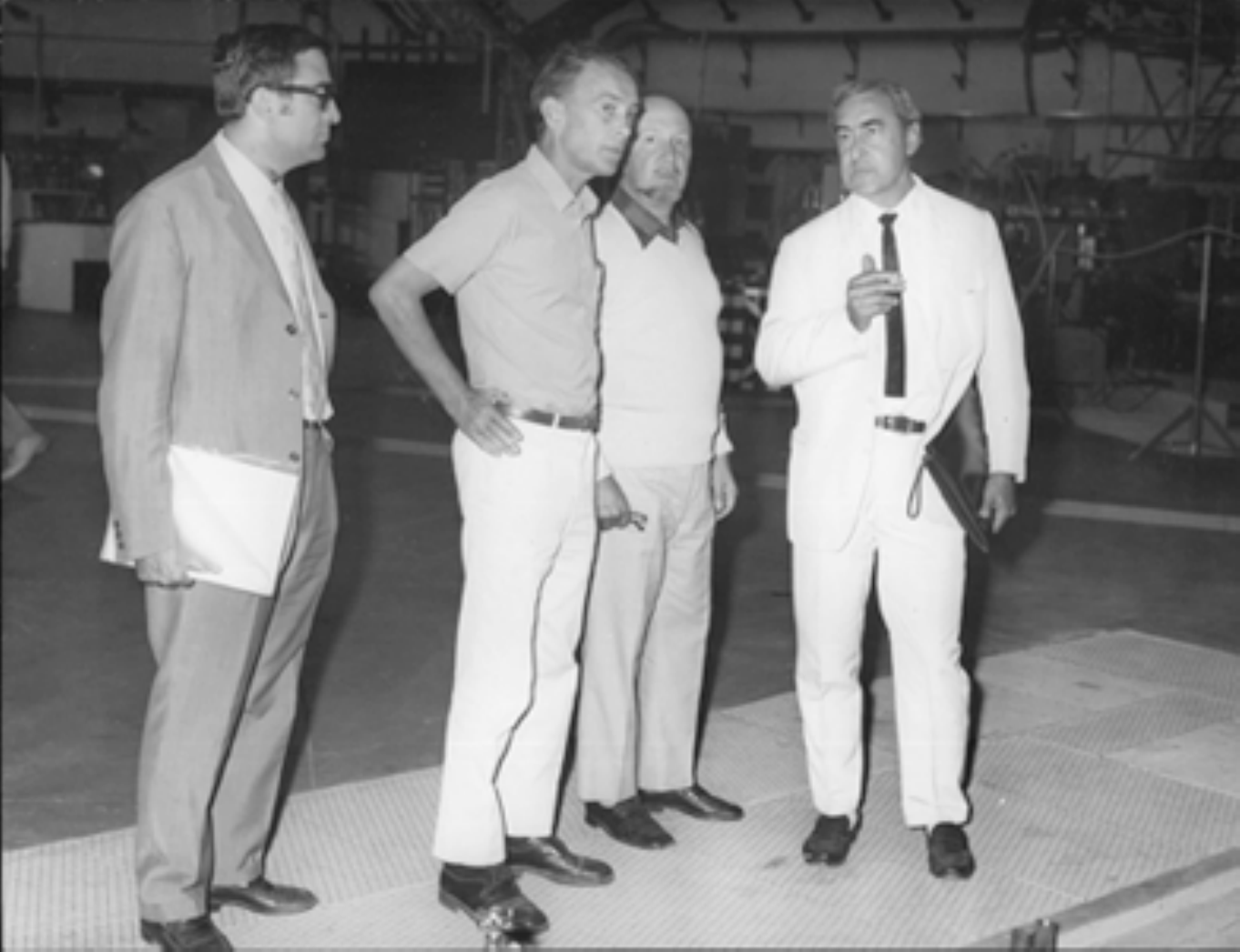}}
\caption{Bruno Touschek, during the construction of ADONE, in Frascati around 1965, with colleagues. At \T 's left is Italo Federico Quercia, Director of  Frascati Laboratories. }
\label{fig:touschekadone025}
\end{center}
\end{figure}

After the construction  of AdA, and the beginning of the construction of ADONE,  as Ugo Amaldi remembers in \cite{greco2005},
 Bruno   became
seriously
concerned with the success of ADONE experiments.  The major problem to solve concerned  the infrared radiative corrections to the proposed experiments.
This was at the time an ongoing preoccupation. In the United States,  Tsai  \cite{tsai1965} had
been  performing realistic radiative correction calculations to
colliding beam
experiments, which  however were restricted to first order
in $\alpha$. Touschek realized that, at a machine like ADONE,
the radiation factor, $\beta \propto \alpha \log (2E/m_e)$
would not be small and that it would be necessary to use   the exponentiation
advantage
 while also doing a calculation for a
realistic apparatus.  The problem was then  to 
combine the realistic approach by Tsai and the
theoretical formulation of the exponentiated infrared factors of \cite{Yennie1961}.
In Fig.~\ref{fig:touschekadone025} we show \BT \ in Frascati during the period of the construction of ADONE.

In the Spring of 1966,  the Frascati theory group  included  Giovanni De Franceschi, Paolo Di Vecchia,  Francesco Drago, Etim Etim, Giancarlo Rossi, Mario
Greco and G.P., one of the authors of this paper.  At the time, there was great interest in strong interaction physics,    with Finite Energy Sum
Rules and Current Algebra results. But  Touschek  knew that the success of experimentation on ADONE relied on precise radiative correction calculations. His approach was:
``We must do the administration of the radiative corrections to electron
positron experiments'', in his words, ``we must earn our bread and butter.''
He already had part of the work  in his mind. The paper \cite{etim1968} starts with some
fundamental considerations, which reflect the Bloch and Nordsieck approach
to the problem, namely that the picture of an experimentalist as counting
single photons as they emerge from a high energy scattering among charged
particles is unrealistic. Then Touschek,  to use his own words,  charges perturbation theory with being
unable to deal with the flood of soft photons which accompany any such
reactions. Originally   Bloch and Nordsiek had shown that, by neglecting the recoil of the
emitting electron, the distribution of  any finite number of quanta would
follow a  Poisson-type distribution, namely
\begin{equation}
P(\{n,{\bar n}\})={{1}\over{n!}}{\bar n}^n e^{-{\bar n}}
\label{poisson}
\end{equation}
and  Touschek and Thirring  had recast ${\bar n}$ in the covariant form. In \cite{etim1968}, Touschek
uses this distribution and adds to it the constraint of energy momentum
conservation. This is the major improvement, which has sometimes been
neglected in subsequent applications of the method to strong interaction
processes.

 We shall now repeat the argument through which Touschek   obtained the final 4-momentum probability distribution to have an energy momentum loss $K_\mu$.
   The final expression is the same as the one proposed earlier by Yennie, Frautschi and Suura, but the derivation is very different and, because of its semi-classical derivation,  its physical content more transparent. The paper also has a discussion on the energy scales which will become very important later, when dealing with resonant states, and in particular with $J/\Psi$ production.

In \cite{etim1968} the probability of having a total energy-momentum loss $K_\mu$ in a
charged particle scattering process, is obtained by considering all the
possible ways in which $n_\bk$ photons of momentum $\bk$ can give rise to a
 given total energy loss $K_\mu$ and then summing on all the values of
$\bk$. In this formulation one obtains a  total  energy-momentum loss  $K_\mu$
 through emission of $n_{\bk_1}$ photons of
momentum $\bk_1$, $n_{\bk_2}$ photons of momentum $\bk_2$ and so on.
Since the photons
are all emitted independently (the effect of their
 emission on the source particle is
neglected), each one of these distributions is a Poisson distribution,
and the  probabilty of a 4-momentum loss in the interval $d^4 K$ is written as
\begin{equation}
d^4P(K)=\sum_{n_\bk}\Pi_{\bk}P(\{n_\bk,{\bar n}_\bk\})
\delta^4(K-\sum_{k} k n_\bk) d^4K
\label{eq:d4p}
\end{equation}
where the   Bloch and Nordsiek's result of independent emission is introduced through the Poisson
distribution and four momentum conservation
is ensured through the 4-dimensional $\delta$-function, which
selects the distributions $\left\{ n_\bk,{\bar n_\bk} \right\}$ with
the right energy momentum loss $K_\mu$.
The final expression, obtained
  with methods of statistical mechanics was
\begin{equation}
d^4 P(K)={{d^4K}\over{(2\pi)^4}}\int d^4x \ exp [-h(x)+iK\cdot x]
\label{d4p}
\end{equation}
with
\begin{equation}
h(x)=\int d^3 {\bar n}_\bk \big(1-exp[-ik\cdot x]\big)
\end{equation}
This result is the same as in \cite{jauch,Yennie1961}, but it was obtained in a simple, almost intuitive  manner, and   it appeared extraordinarily simple  to the Frascati 
experimentalists, who were waiting for a precise
calculation to apply to the ADONE data.  Eq. (\ref{eq:d4p})  allows  to obtain immediately 
the correction factor for the energy. By performing an integration over the 3-momentum ${\bf K}$,
the distribution describing  a total energy loss
$\omega$ becomes
 \begin{equation} dP(\omega)=\frac{d\omega}{2\pi} \int_{\infty}^{\infty} dt \ exp[i\omega t -h(t)]
 ={\cal N} \beta {{d\omega}\over{\omega}}
\big( {{\omega}\over{E}}\big)^\beta\end{equation}
  where the already known result at the r.h.s. was obtained with a simple and elegant argument based on the fact that 
  $dP(\omega)=0$ for $\omega \le 0$.
 ${\cal N} $ is a normalization factor \cite{lomon1956,lomon1959,erikson}, and,
in the high energy limit,
\begin{equation}
\beta={{4\alpha}\over{\pi}}\big( \log{{2E}\over{m_e}} -{{1}\over{2}}\big)
\end{equation}
 Touschek named $\beta$  the {\it  Bond factor}, because
its numerical value in the range of ADONE energies was
around 0.07,  the  number made famous by the  James Bond movies of the 1960s.  This  was  just a small joke, but it held everybody's imagination for quite some time and is a typical example of \T  's capacity for humour and  making  light  of serious issues.

Although the radiative correction paper of 1967 was the last that Touschek
wrote on this subject, his influence on the field continued
in many ways and for longer than he himself may have imagined
 and came to know. While  working on the above paper, he had
suggested to his  young
collaborators  two more relevant papers,
one on the coherent state method \cite{greco1967}, which
indicates how to deal  with soft photon emission in the amplitude
rather than the cross-section, and  one on infrared corrections to
resonant process \cite{pancheri1969},  which appeared (independently) shortly after a similar work by V.N. Baier, the theoretical physicist from Novosibirsk \cite {baier1968}. Although he did not sign these papers, \T \ 
 had the original idea,  gave suggestions during the preparation of the work and commented on the final versions. Both papers
became  relevant to study production of   of narrow resonances at colliders. In later years, in addition to the work for the $J/\Psi$ \cite{greco1975}, the method he has inspired  was  used at  LEP for  the extraction of the parameters of the $Z_0$ \cite{greco1982}.
The influence of Touschek  is also reflected  in one of the earlier and most famous works on resummation in QCD \cite{altarelli1977}, whose authors were among the generation of theoretical physicists graduating with Gatto and Cabibbo.

\begin{figure}
\resizebox{0.41\textwidth}{!}{\includegraphics{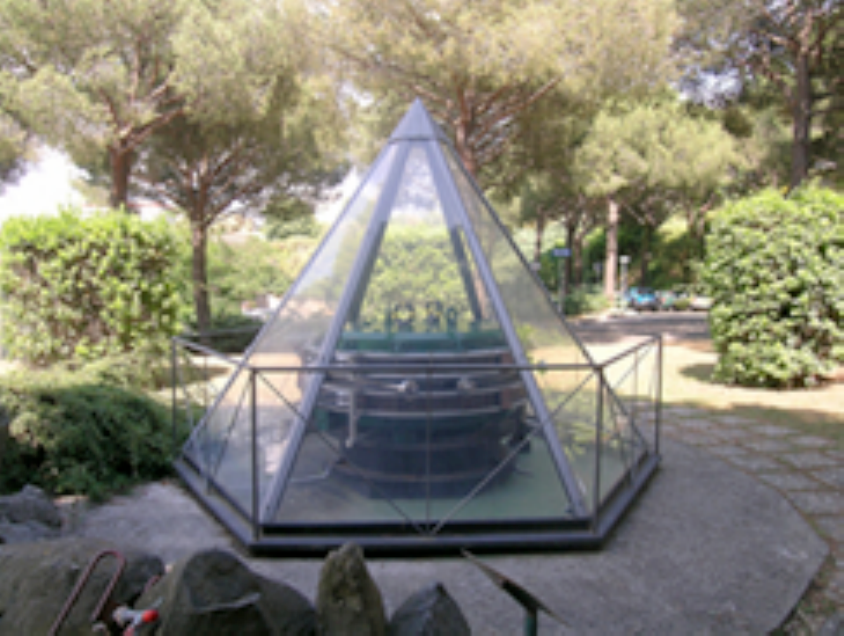}}
\resizebox{0.55\textwidth}{!}{\includegraphics{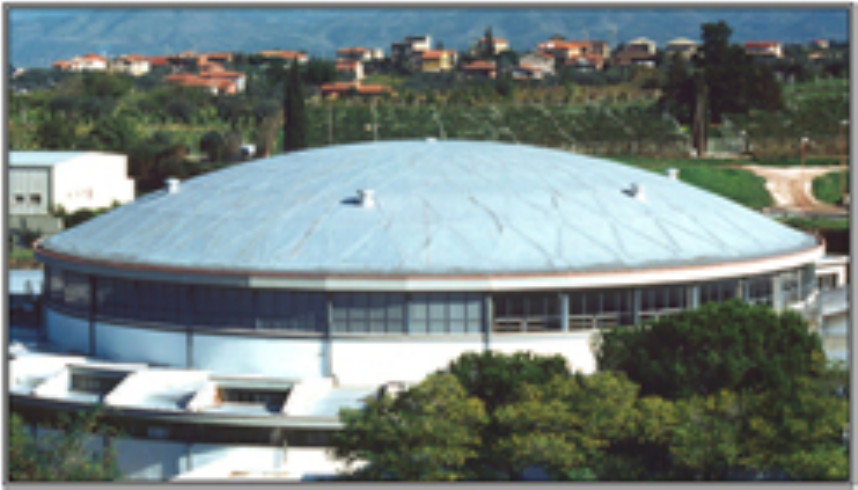}}
\caption{ At left: AdA now, in Frascati, as preserved under a plexigass container. On the right: view of the ADONE building, now housing  the accelerator DAFNE.}
\label{fig:cupola}
\end{figure}

Touschek's legacy in Frascati and in Italy is still very much alive today. In 1993 ADONE was definitely shut dow and construction for a new high luminosity $e^+e^-$ accelerator for precision physics at $1\ GeV$ c.m. energy was started. The project was approved by the Istituto Nazionale di Fisica Nucleare (INFN) under the presidency of Nicola Cabibbo and  the machine was named DAFNE, Double Accelerator For Nice Experiments.    DAFNE is housed in the ADONE building in Frascati, shown in Fig.~\ref{fig:cupola}. At left, in the same figure,   one can see   AdA, in the glass container which preserves it  on the INFN Frascati National Laboratory grounds.
DAFNE started producing physics in 1999 and plans  for an upgrade both in energy and luminosity have been presented \cite{kloe2}.

\subsection{The collective momentum distribution}
Touschek's theoretical legacy is present also in a large body of 
work, still very active today, namely the transverse momentum distribution
of soft  radiation.  Many days  were spent in 1967 by \T \ and his young collaborators
in trying to obtain a
closed form for the momentum distribution
\begin{equation}
{{d^3P(K)}\over {d^3{\vec K}}}=\int dK_0\ {{d^4 P(K)}\over {d^4K}}=
\int {{d^3{\vec x}}\over {(2\pi)^3}} exp[-h(x)-i{\vec K}\cdot {\vec x}]
\end{equation}
 Short of a closed expression, which cannot be obtained,   he   showed  \cite{etim1968} that the energy and the 3-momentum distributions factorize, i.e.
\begin{equation}
d^4P(K)=\beta {\cal N}^{-1}\frac{d\omega}{\omega}(\frac{\omega}{E})^\beta A({\bf u})d^3 {\bf u}
\end{equation}
with ${\bf u}={\bf K}/\omega $ and  the momentum function $A({\bf u})$ well approximated by a first order expansion in the fine structure constant $\alpha$.
Since   in the QED case the angular distribution of the collective radiation was estimated as being
well reproduced by the  first order contribution,\footnote{The usefulness of such expressions for experimentalists working on electron-positron collisions was recently remembered by Francois Richard, in a talk given at the {\it Bruno Touschek Memorial Lectures}, Frascati, November 30th, 2010.} the problem was 
left dormant. 

However,  in 1976,   focusing  on  the transverse momentum rather that the 3-momentum distribution,   the problem was revived  and the
reference to Touschek's work on radiative corrections and the memory of the
time spent by him on the problem came again to life. 
In \cite{ourkt},  the   transverse momentum distribution for soft radiation was obtained from 
the Bloch-Nordsieck method integrating  Eq.(\ref{eq:d4p})   in
both energy and longitudinal momentum as  
\begin{equation}
{{d^2\ P(K_\perp)}= {d^2{\bf K}_\perp}}  \int {{d^2{\bf x}_\perp}\over
{(2\pi)^2}} exp[-h({\bf x}_\perp)-i{\bf K}_\perp  \cdot {\bf x}_\perp]
\label{eq:d2p}
\end{equation}
Working within    an abelian theory, but with a large coupling
costant,    approximations were  proposed for  the function $h({\bf x}_\perp)$  such as to  allow  a closed form for Eq.(\ref{eq:d2p}). An  exponentially decreasing   $p_t$-distributions for  hadronic particles  in high energy hadronic collisions in agreement with experiments \cite{ourmean} was then obtained.
  
At the same time and all along,  resummation had also been   developed 
by the great Russian theoretical physics school, starting with Sudakov's paper \cite{sudakov1956} in QED. In 1978, from the Russian school there came  a seminal work on  the transverse
momentum distribution of QCD radiation \cite{ddt1,ddt2}, in which  
 soft gluons were resummed with a  running coupling constant. This was  followed very soon 
 by a similar  pivotal work from Frascati  and University of Rome \cite{Parisi79}.  In subsequent years, there appeared  stlll other  works on resummation traceable  to \T 's influence, 
 including 
the application of soft gluon resummation techniques to calculate the transverse momentum distribution of $\mu^+\mu^-$ pairs (Drell-Yan pairs)  \cite{chiappetta1981} and the  W-boson transverse momentum \cite{Altarelli:1984pt}. But then \T \ was no longer alive.  
 As a conclusion to this  description, we  stress  that   the problem of soft gluon resummation is still very much alive  today. The question of how to access  the infrared limit, where perturbative QCD fails and the large distance behaviour of the theory plays a major role, is still unsolved.

\subsection{The book  {\it Meccanica statistica}}

 As mentioned, the resummed expression for soft photon emission given in Eq.(\ref{eq:d4p}) was obtained using the methods of statistical mechanics, which \T \ had studied in G\"ottingen and in Glasgow. His deep understanding of this subject is reflected in the book {\it Meccanica statistica} \cite{rossibook}, which he wrote together with his student and collaborator  Giancarlo Rossi.\footnote{Parts of this section have been  contributed by G. Rossi, who graduated with Touschek  in 1967 and is now  Professor of Theoretical Physics at University of Tor Vergata, Rome.}  

The first version of the book 
dates back to the winter of 1967. It was the result of a long process which started two years before with the publication of the notes that 
 Giancarlo Rossi   took when, as a fourth year Physics student,  was following the course of Statistical Mechanics held by Touschek at La Sapienza in 1965. The notes, revised by Touschek himself, were published by a local Editor, ÒLa GoliardicaÓ, as lecture-notes for students. At the same time, having in mind the idea of finally publishing a book on Statistical Mechanics, Touschek started to write in English with the help of the beloved Olivetti Lettera 22 his version of the lectures he was delivering.

The final manuscript was published in Italian by Boringhieri in 1970 in the Series ``Programma di Matematica Fisica Elettronica'' with the title {\it Meccanica Statistica}. It finally resulted from the intersection of the English version written by Touschek with the one  Rossi had elaborated in Italian. 

In the long process of deciding the content of the book and the style of the text, Touschek's  guiding idea and  main concern were always clarity, as the book was supposed to be addressed to undergraduate students. For this reason the chosen language was plain and simple and it was decided to have at the end of each chapter a summary of the results and main ideas that were successively developed. Simplicity did not mean that all the subtleties inherent in the construction of Statistical Mechanics should be overlooked. Quite the contrary! Not only in the book standard subjects, like the construction of the various statistical ensembles, the proof of their equivalence or the derivation of the ÒMaster EquationÓ, have been inserted and discussed in a somewhat original though elementary way, but unsolved conceptual problems were also addressed. Among others it is worth mentioning two of them since they drew the attention of the physics community at large as witnessed by the significant interest they spurred in the specialized literature. The first topic is a simple proposal to understand the apparent antinomy between microscopic reversibility and macroscopic irreversibility. The second is the solution of the problem posed by definition of  temperature for a moving body in Special Relativity.

It is important to conclude these considerations on the birth and the content of the book {\it Meccanica Statistica} by observing that, despite the fact that Touschek had been lecturing on the subject for only 4 or 5 years at Sapienza University of Rome (at a certain point he moved to the course of ÒMetodi Matematici della FisicaÓ), his cultural legacy had a large  impact on the development of theoretical physics in Rome.
The roots of the many important contributions that Italian physicists have given to a number of research fields related to Statistical Mechanics (among which the theory of Spin Glasses and Complex Systems \cite{spinglass,Parisi88}, Lattice QCD \cite{rossichiral} and the emerging field of Biophysics) can  be traced back to the crucial influence  exerted by  Touschek
on a whole generation of physicists in the 60s and early 70s.

\section{Unpublished correspondence: Letters to his parents}
\label{sec:letters}
We present here an English translation of the two letters sent by \T \ to his parents from Kellinghusen.  These letters, as most of the others we have retrieved, are addressed to his father and to his stepmother, since \T 's mother had died when he was young, and the father had remarried. The letters were written after the war, on June 22 and November 17, 1945,  the first of which was received only on October 22, as remarked in a short note under the date. The two letters contain some repetitions, however, as they complement each other, we have opted for publishing here both of them.  Some parts of the letters are hard to read and we have inserted question marks when the sense was not clear. Also some references to personal matters have been omitted and indicated with [\dots].

\paragraph{June 22nd, 1945  (letter received by Touschek's parents on 
October 22nd, 1945)}

{\fontfamily
{cmtt}
\selectfont
Dear parents,

One of our drivers will leave today to go home to Br\"unn
 [the German name  of the Czech town  Brno]  and I hope he will  be able to send you this  letter. 
 
 I have not had news from you for such a long time and I am, obviously, worried about what may have happened during this time. I do not know what is the situation in Vienna, aerial attacks, battles in the city. It will take some time before I will be able to come.

Now a brief account  from  my side: I wrote  you my  last letter from Kellinghusen. The next day I left for   Hamburg and arrived there  by night, during the  air raid. The following day, I was taken to prison to Fuhlsb\"uttel under suspicion of espionage. The first week was terrible, and I was on the verge of committing suicide, in an isolated cell and guarded by the SS. Then Wider\o e gave me some relief: cigarettes, the possibility to complete a theoretical work, etc. After 3 weeks, the prison was evacuated and we (200 prisoners) had to set off on foot towards Kiel. There were SS soldiers behind us, in front, and on both sides. Near Hamburg (Langenhorn) I collapsed (I was no longer used to walking). They threw me into the ditch on the side of the road and then shot me. Without success. A bullet   went glancing brushing close to my left ear, the other went through the padding of my coat. Since Wider\o e had visited me and had told me that the courier carrying the papers for my release was on his way from Berlin, I went to Langehorn Hospital to get bandaged. Of course they imprisoned me again and sent me to Hamburg, from one prison to another. All this lasted about 4 weeks. In the meantime, \W \  had gone to Norway and nobody seemed to know where  I was. I wrote a series of secret messages [to let people know of my situation]. Prof. Lenz would smuggle some food into the prison for me, and finally  my working team at the Ministry learned of my existence.  A few days before the occupation of Hamburg, Holnack came to get me out of prison, although not quite legally. This was timely, since I would have been shot, in the best case.  I then went to Kellinghusen by car and I am still there. In the meantime, I have tried to recover. I can state that this attempt was successful. As an Austrian, I receive 4 times the normal ration and this is already something. I have done the interpreter and now speak fluent English. We want to save the {\it Strahlentransformator} [the betatron] and it may even be possible. Our name is MW-Research- Association.
 My plan is to do a PhD in a Northern country, probably Oslo, Wider\o e started discussing this with Prof. Hyleraas already during the war. But before I would like to go to Vienna for about 2 months. My life appears more secure now, as I compare it with other's. Anyway, there appear to be all the premises for a good start in the new world. [??].  I have received two letters [\dots] Another from Sommerfeld, in which he asks about his son, who during my imprisonment lived in my room in Kellinghusen, and who then very stupidly, so to speak, went to Berlin with the last [??] and now nobody knows  where he is. The English here are very correct and decent. How is it  going with your occupation troops? 
 
\noindent  P.S. Please, do manage to write to me.}
\newpage
\paragraph{\BT 's letter to his father dated November 17th, 1945}
{\fontfamily{cmtt}\selectfont

\noindent

Dear parents!
I was so relieved when I received your letters, one after the other and in inverted order. After all I heard on the Russian zone, especially in the last period I was terribly worried, without the possibility of unfortunately doing anything neither then nor now. In the next days though, a decision will be taken. To make the matter completely clear, I would like to tell you what happened starting from the beginning of March.
More or less in that period, I do not remember very well, I must have gone to Hamburg where I felt terribly cold. Relationships with Berlin were not the best: disapproval for my collaboration at the Betatron, many telephone conversations with Wider\o e, [...],
etc. Wider\o e told me that CHF M\"uller, where the Betatron was moved to, had become unbearable and that he had the intention of moving to Kellinghusen. On the 15th the situation had thus become tragic. We left together and I had a nice room  down there. On the morning of the 16th we were still sitting down in the veranda, it was a beautiful spring morning and we were reading the Physical Review in which we had found an article on the Strahlungsd\"ampfung (an article of two Russians) that later on had received quite a good attention. In the evening I came back home with the truck. The driver did not know the area very well, he drove over a child at Itzehoe who, thanks to God, got only scratched, and along the way two holes burst 10 kilometres from Hamburg. Around midnight I reached Hamburg during the alarm and after the alarm I had another telephone conversation, then I went back to sleep to be waken up at 7.30 in the morning by two gentlemen. I was so sleepy that when they said: ``Secret state police!''
I answered: ``Yes, but at midnight?''.
The two, though,  were  very kind. It took them an hour to search through the mess that my desk\footnote{\T s letter calls it  ``Augian Stables".} was, to search me for hidden firearms and to put everything in my bag. The bag, which was quite heavy, I had to bring it by myself to the Gestapo (15 minutes from Dammtor). They interrogated me for an hour, to tell me in the end that they did not know why they had to arrest me, the order had come from Berlin. I asked them to inform Seifert and the `assistant', whose name was Kneesch, called him. After  Kneesch told me I should not get angry if the tone downstairs was `rude but sincere', I was then brought in a basement.  We sat on a bench, the face towards the wall. The window in the courtyard was open and it was very cold. My neighbour was  not ``waterproof'' and was dripping   from above and below. After an hour we were brought by tram to Fuhlsb\"uttel. There the various rites of cleaning and lice searching. I was then brought to the cell where the jews were and where there was  a lot of good company. With a certain Waiblinger, who now studies, I am still friends. The only unreasonable thing was that there was no space  to sleep, the toilet stank and  obviously the people were not in good health. Apart from this, there was practically nothing to eat. The following day was Sunday.
The next Monday I was again brought to the Gestapo in a overfull tram wagon, better not discuss the treatment during these transportations. At the Gestapo Seifert, Dr. Wider\o e and Dr. Kollath were waiting for me. I  boldly made Wider\o e understand that I preferred to be imprisoned by myself and he explained to the Criminal Commissioner Windel that they would not have taken  the responsibility  if I had been put in a cell with the others. Apart from this, the future of the Reich, for better or worse, depended from a research on the influence of the radiation-damping just started by Mr. Touschek. Mr. Touschek should have the right to smoke, read and receive visits. The first week nothing came from these concessions. I was confined without a pencil. Wider\o e had put a couple of cigarettes in my pockets, but I  had no  matches. On Friday I  wanted to hang myself,  and on  Saturday, Wider\o e came. From then on the situation got better. I had a `decent' cell on  the first floor and  Wider\o e brought me Heitler's Quantum Theory of Radiation and I started research on radiation-damping. W. never forgot to bring me a pack of cigarettes with written on it the important sign `Propellent for you'. Apart from the terrible nutrition,  the worse thing was  to be forced to sit or stand up alternatively all day long. Furthermore, it was horribly cold. 
After a lot of coming and going which procured some free cigarettes to the SS on guard, I managed to obtain the permission to lay down in my cell, so I stayed horizontal for a whole week with Heitler and Joos under my arm and, in my mouth, a quote from Goethe's ``G\"otz.''\footnote{ The quote  is ``lick my ass''.}
I was treated relatively well, because  the frequent visits by important people gave me a certain respect. The VDE wrote me: we hope everything is going well and that you have found the ideal conditions to work in\dots  thinking I was in Kellinghusen. 
One Wednesday, about April 10th, W. visited me and told me the courier with my pardon papers was coming from  Berlin [??] The day after, despite my protesting, I was woken up at five o' clock.

At dawn we met in the corridors. At the beginning, the fact that we should march towards Kiel was only a murmur, but at 10 o'clock it was a fact. I tried to protest again with the SS, especially since I was waiting for my release. I had suffered the whole week of diarrhea, the worst thing that can happen to a prisoner. I was not able to stand. In all 200 of us were deported. We all received a big sack to carry. They were extremely heavy, loaded with books. The people were divided in groups of twos. The whole affair was very discomforting. I definitely broke down in Langenhorn. They made me roll down the trench near the street and then they shot  me,  one bullet pierced through the padding  of my coat, the other one went very near my ear. I waited for the guards to go away. In the meanwhile some people had assembled to see if I was dead or not. I wanted to find the way to phone Seifert and ask him for a car. In the meanwhile my head was  hurting me terribly  and  I  managed to go to the hospital of Langenhorn, I needed help. Thanks to Wider\o e's message I had no worries  about it. But they would  have been justified. They brought me to all kinds of prisons for other three weeks, during which I met the state actor Gmelin, now a good acquaintance of mine, and the brother of the mayor of Frankfurt, a certain Kurt Pfeiffer, who  had been put into office in Frankfurt by the liberating English armies. The last station was the prison of Altona which -- although I  have no musical talent - reminded one of die Fiedermaus operette. From there, Hollnack came to pick me up on the 30th of April. He then explained to me -- after  having done nothing for three weeks -- that without him I would have definitely been shot. In any case he then made himself very busy. I went with him to Kellinghusen. Also Kollath and Schumann were there.  [There was] Also a group of not very nice people that Hollnack had brought with him to Kellinghusen from various posts  of special service. Feeling obligated to Hollnack, I drew an agreement with him. We founded the MV-Research Society. I took the part of stipulating deals with the military regime, for a short period I was an interpreter and I also obtained for our enterprise  to be  occupied by the T-force, which, in this situation full of looting and vandalism, seemed necessary. After a short time H. started harboring  grandiose ambitions. When things were not  as he wanted, he started going into publishing news, into cultural relations, etc. At the end of June, I asked them to end our collaboration. A long discussion ensued on the subject. We settled for three months leave. H., who managed to save 200,000\footnote{ It is not specified in  the text whether the number refers  to marks.} from the war, always tends to enlarge his contribution. The guards call him Kellinghusen's Mussolini. At the end, I fought with everybody. In late August I went for a trip  to G\"ottingen [\dots].
I visit Professor Jensen in Hannover, I meet Dr. S\"uss at the University of G\"ottingen, I am brought in tour  as a person who had been considered  dead. Jensen offers me an assistant position and grants that he will talk to the head engineer in order to assure a position for me. Same offers in Hamburg. I  write a dissertation on the Betatron. A reunion with the T-force has decided that things should remain a State secret, so that its use for a thesis   is out of the question. I will be able to leave Kellinghusen only after an Allied commission has decided in regard to the Betatron. From then on, I am seated in Kellinghusen, practically  as a prisoner. The food is bad, I have a cold, I am, as I was earlier, very badly nourished and have nothing to dress with. Part of my belongings has been stolen at the Gestapo, and here there are only useless `buying stamps'. The German offices only work for the Nazis and obviously the English  do not care for such low happenings. Of course there are exceptions. The commission should come in the next days.  After that, I will try to obtain  some vacation in Vienna. Please write to me in detail from what you live on at the moment, what you need, etc. Maybe I can also organize something from here too. Write soon. Say hello  to everyone, Peter etc.

\noindent P.S. I wish I were in Vienna already, I cannot  wait.}

\section{Selected  drawings}
\label{sec:drawings}

\BT 's  drawings were a cherished possession of all his friends and colleagues. His ability to draw emerged very early and  is clear from the  drawings included in the letters to his parents and which we have reproduced in earlier sections.  Many of \T 's drawings have been published in \cite{amaldi81}, but a large number of additional drawings are among \T 's documents kept by \T 's family and are not known.    
In this section we shall reproduce six  drawings of \BT . 
These  drawings have never appeared in print  and have been retrieved from the documents in possession of the Touschek family. The first two  drawings has been extracted from \T 's letters to his parents, respectively dated April 20, 1942 and  January 18, 1944. The others are undated  and give us Touschek's keen and incisive eye on academic life of the time.
\begin{figure}[htb]{!}
\resizebox{1\textwidth}{!}{\includegraphics{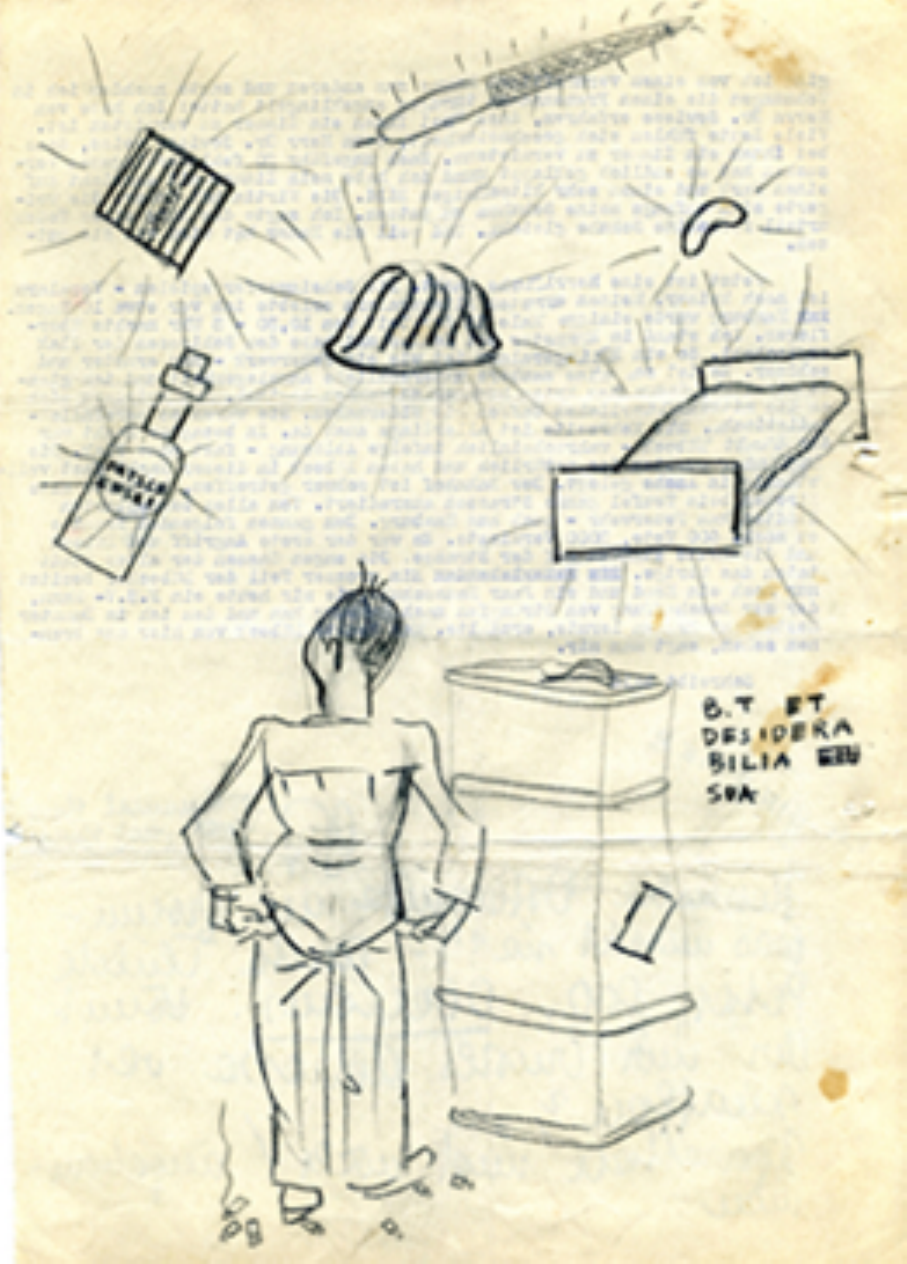}}
\caption{This drawing, included in the letter dated   April 20, 1942,  is entitled {\it Bruno's desiderabilia}.}
\end{figure}

\begin{figure}
\vspace{2cm}
\resizebox{1\textwidth}{!}{\includegraphics{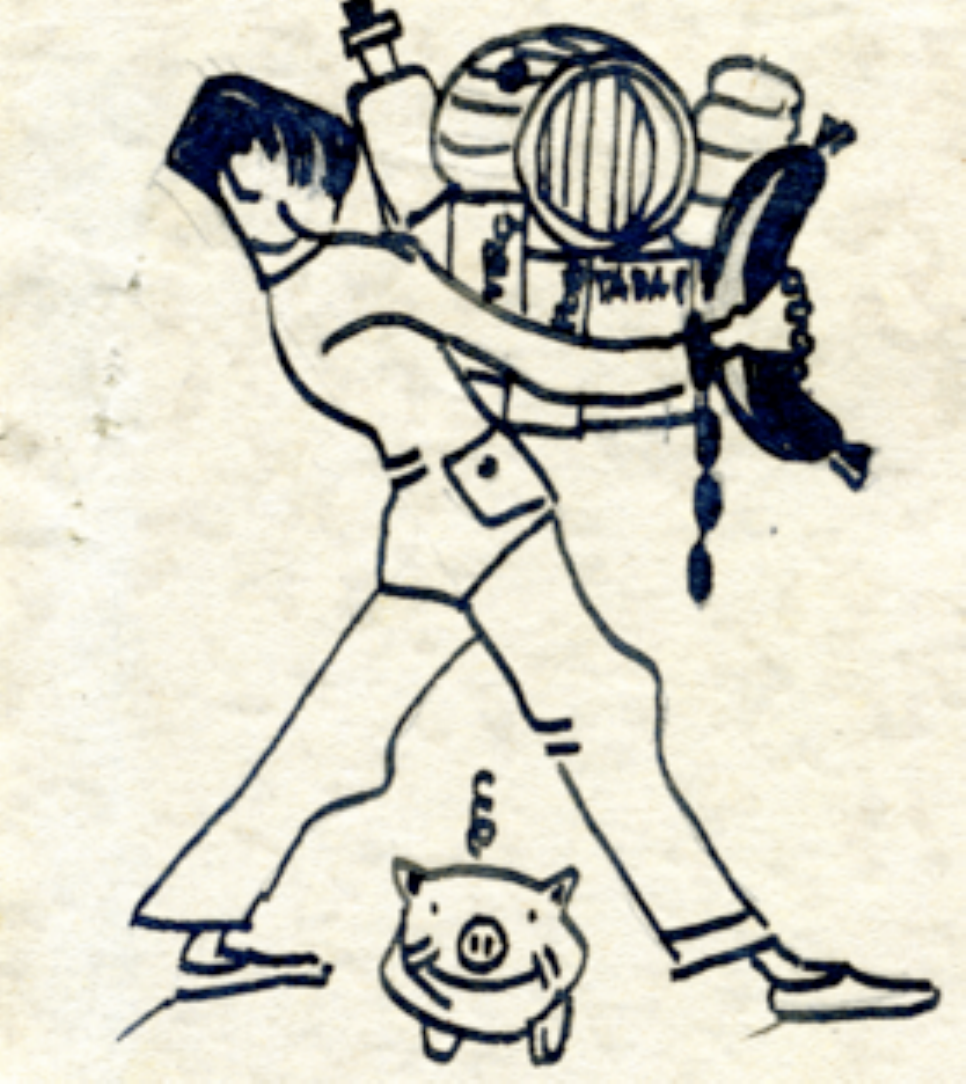}}
\caption{This drawing was  included in the letter dated   January 18, 1944 and represents a (virtual) offer of gifts for his father's birthday.}
\end{figure}

\begin{figure}
\vspace{5cm}
\resizebox{1.0\textwidth}{!}{\includegraphics{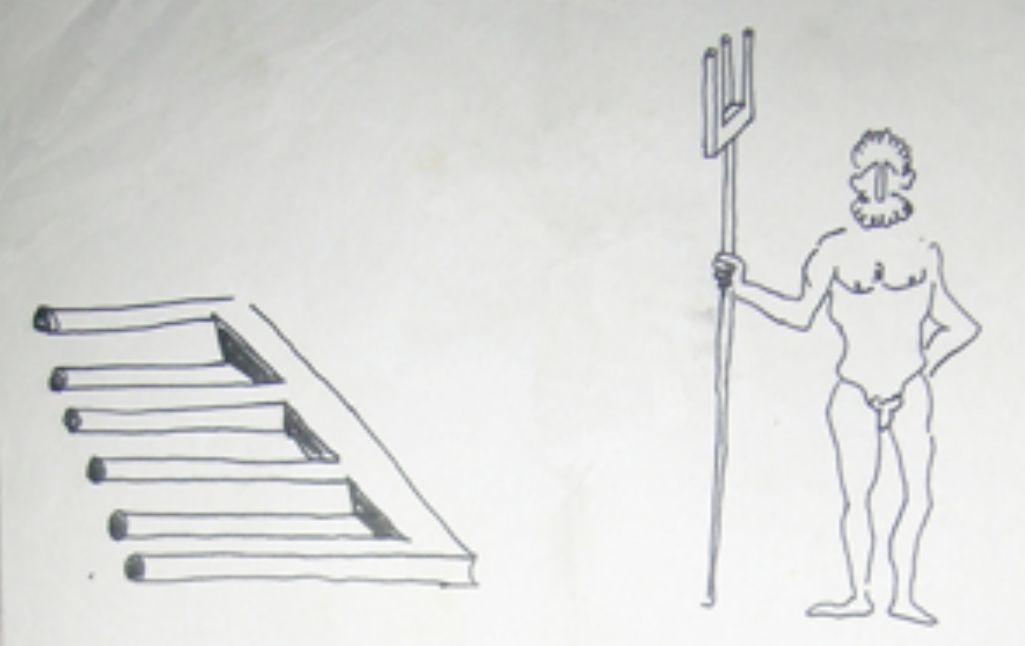}}
\caption{ One of Touschek's many variations on drawing optical illusions. Possible date is in the 1960s.}
\end{figure}

\begin{figure}
\resizebox{1.0\textwidth}{!}{\includegraphics{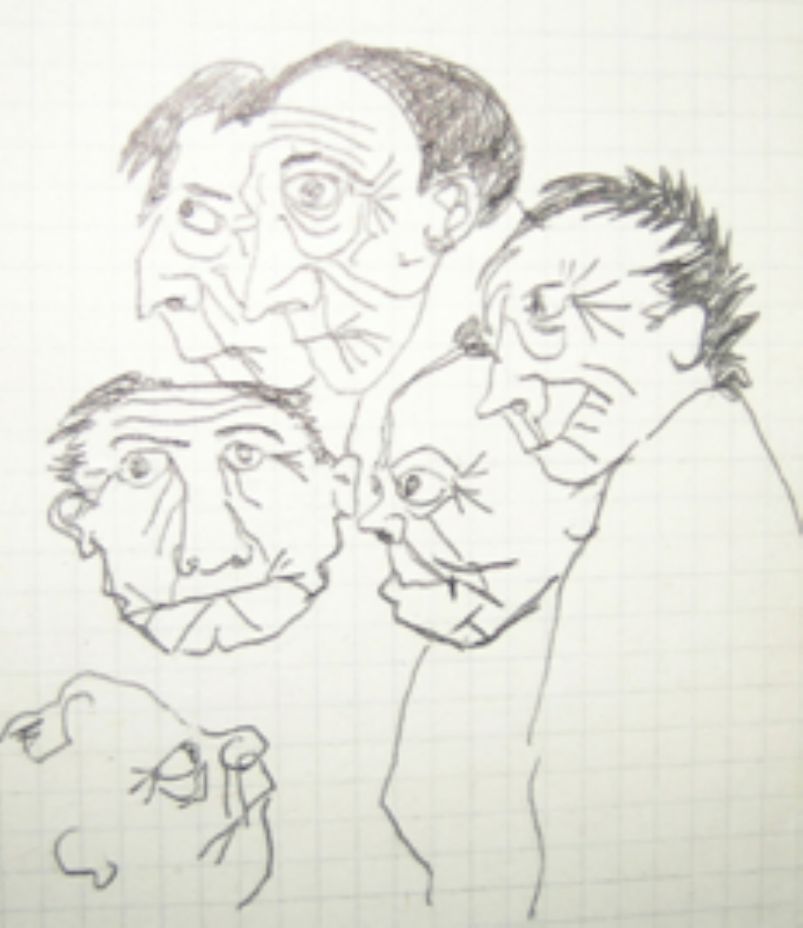}}
\caption{Caricatures of colleagues at meetings}
\end{figure}

\begin{figure}
\begin{center}
\resizebox{1\textwidth}{!}{\includegraphics{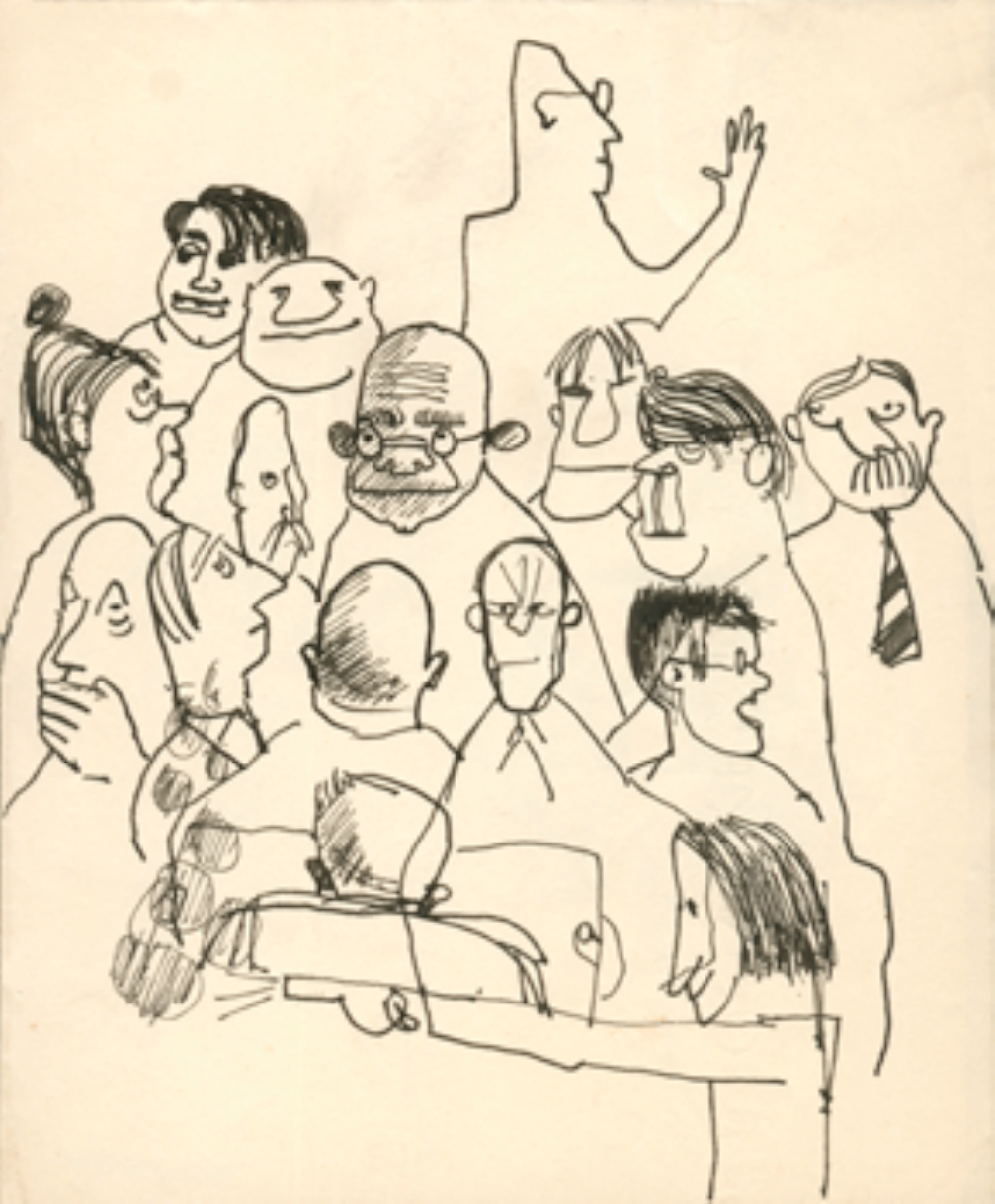}}
\caption{Sarcastic description of a conference meeting, date around 1960.}
\end{center}
\end{figure}

\begin{figure}
\hspace{2cm}
\resizebox{1.0\textwidth}{!}{\includegraphics{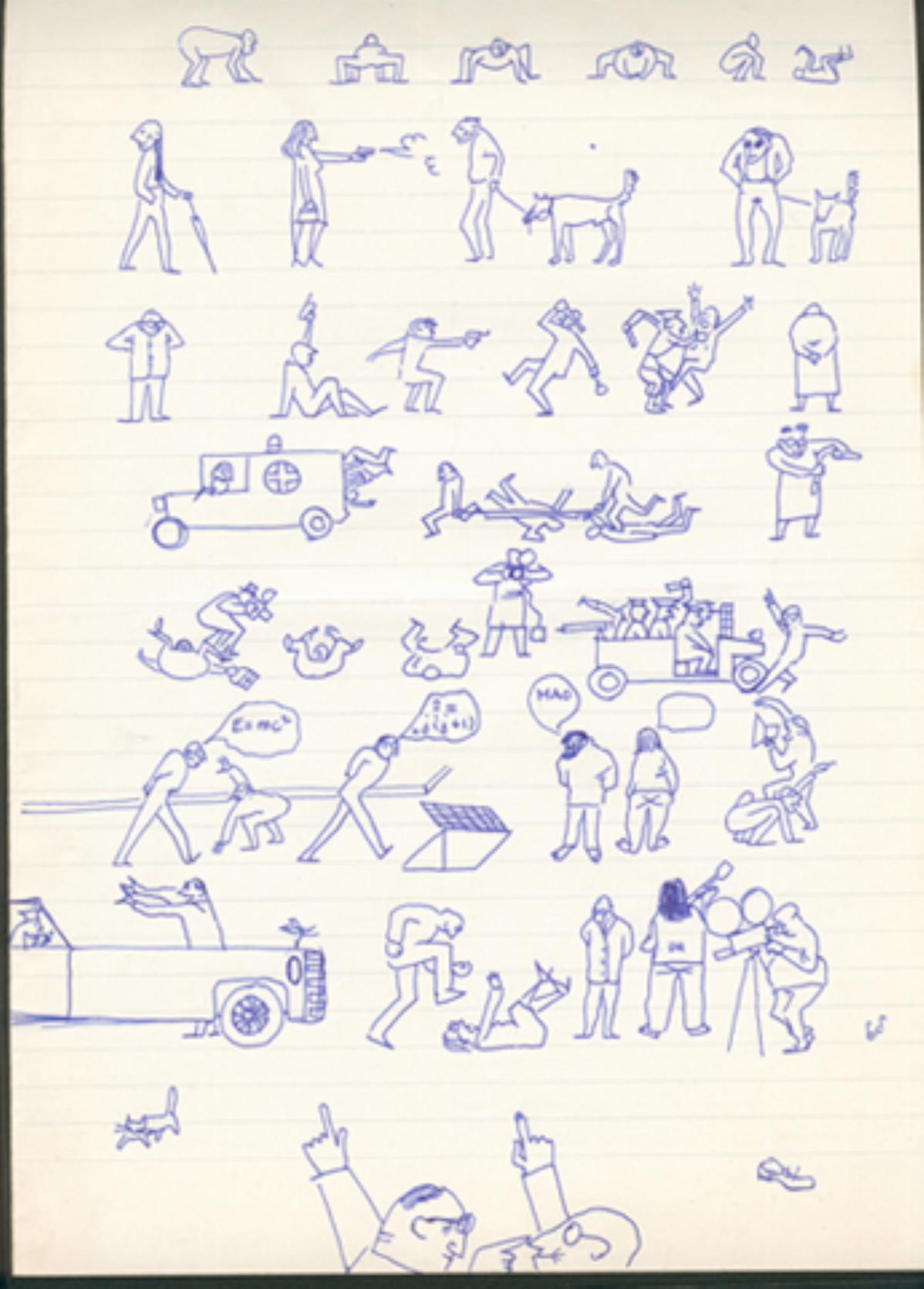}}
\caption{Cartoon dated around 1968.}
\end{figure}

\newpage

\section*{Conclusions}
In this brief note, we  have tried to recall   the scientific and human path which led \BT\ to propose and build the first electron-positron storage ring  AdA, in 1960, in Frascati, Italy. { Touschek's legacy is present today in   INFN  plans for an electron-positron machine at high luminosity operating at the center of mass  energy of the $\Upsilon$-resonance, a so called Super-B factory \cite{superB}.} We also described  the role he played  in developing the theoretical background for  administering high precision radiative corrections to electron-positron experiments and his influence on a generation of young theoretical physicist he trained in the 1960s and early 1970s.  Our  work has utilized existing sources, but also an unpublished and so far unknown series of letters written by \T \ to his parents during the war. Due to the complexity of the material and space restrictions, only a fraction of the content of these letter has been included here. It is our intention to publish separately the full  collection of  letters and drawings.

\section*{Acknowledgments}
Above all, the authors thank the Touschek family, Elspeth Yonge Touschek and Francis Touschek, for making available  many unpublished documents and the collection of letters written by Touschek to his family.  We thank the family also for allowing publication of photos and drawings of which they own the copyright. The authors are deeply   indebted to Carlo Bernardini  for discussions and 
historical and personal notes. We acknowledge conversations with  the late Nicola Cabibbo and exchanges with Raoul Gatto.  We thank Jacques {Ha\"issinski} for discussions and correspondence about AdA in Orsay and Robert Chebab and Simon Eydelman for  the Russian work on colliders physics.  One of us, G.P. thanks Earle Lomon for discussions on the status of QED radiative corrections in the 1950s. L.B. is very grateful to Stefan Wolff for important remarks on the problem of the application of the Nuremberg's laws. 
We thank Neelam and Yogendra Srivastava  for help in proof reading the manuscript,  Jennifer Di Egidio for translating \BT's  letters reproduced here from German into Italian and  Amrit Srivastava for translating them  from Italian  into English. We also thank  Olaf Holm and  Ruggero Ferrari for help in translating parts of other letters, and  Rainer Mauersberger for providing   understanding of some German terms in Touschek's letters and a critical reading of the translation of the two letters reproduced in this article.  For the digitalization, archival and library assistance work, many thanks are 
due to Debora Bifaretti, Orlando Ciaffoni, Antonio Cupellini, Claudio Federici,  Luigina Invidia,  and  Danilo Babusci  from SIS of INFN Frascati  National Laboratories,  and to Antonella Capogrossi of the Library of the Physics Department of Sapienza University of Rome.   We are  grateful to Giovanni Battimelli and to the Archives of the Physics Department of Sapienza University in Rome, as well as to Micheal Eckert   and to the Deutsches Museum Archive in Munich for granting access to the Archives and  permission to publish portions of \BT's letters and  documents. L.B.  acknowledges hospitality of  the Forschungsinstitut f\"ur Technik und Wissenschaftsgeschichte of Deutsches Museum during completion of this paper. G.P.  is grateful to  Brown University Physics Department for hospitality  during the preparation of parts of this article.

Finally, we wish to thank W. Beiglb\"ock for encouraging us in writing this article and providing invaluable criticism and suggestions during its preparation. 
\appendix
\section{Chronology of Touschek's life during the war years}
In this appendix we  present a brief chronology of \BT 's life during the war years.
\begin{description}
\item {\bf 1939-1940} : \BT \ attends classes in University of Vienna, passing the relative examinations
\item {\bf May 1940}: \BT \  is not allowed anymore to attend the university
\item {\bf 1940-1941}: from Urban's letter to Amaldi,  \BT \ appears to have been working at home or at Urban's home together with other students, studying books borrowed by Urban from the University library, among them Sommerfeld's treatise {\it Atombaum und Spektrallinien}
\item{\bf  Fall 1941}: Urban goes to Munich taking \BT \ with him and
introducing him to Arnold Sommerfeld, who helps him to find a job in Hamburg
and writes letters of introduction to Wilhelm Lenz and Paul Harteck.
 Urban's letter to Sommerfeld is  dated  December 1941 and Touschek's letters from Germany to his parents are already from February 1942 
\item {\bf March 1942} : \BT \ is in Hamburg, looking for lodgings 
\item {\bf Spring-Fall 1942}: \BT \ works in Hamburg for a firm developing ``drift tubes'' and follows courses  in Hamburg
\item {\bf September 1942}: \RW \ submits his paper for a 100 MeV  betatron to Archiv f\"ur Elektrotechnik" 
\item {\bf November 1942}: \BT \ moves to Berlin  and starts working at L\"owe Opta, following Max von Laue's courses at Berlin University
\item  {\bf Spring 1943}:  \BT \    working in Berlin  reads \RW's paper on the betatron submitted to {\it Archiv f\"ur Elektrotechnik}
\item {\bf Spring--Summer 1943} : \BT \  enters into correspondence with \RW \ who invites him to join the work on the betatron
\item {\bf August 1943}: \RW \ moves to Hamburg to start work on a 15 MeV betatron financed by the Ministry of Aviation
\item {\bf April 1944}: \BT \ moves back to Hamburg to accelerate work on the betatron, but continues to keep his flat in Berlin and keeps traveling between the two cities
\item {\bf Summer 1944}: \BT \ travels to Vienna to and from Berlin, and describes this with the ``train drawing'', shown in one of the figures
\item  {\bf March 13th} : \T \ writes to his parents from Kellinghusen, where the betatron has been moved.
\end{description}

In the following we present  the chronology of events between March 13th  and May 1945, extracted from the two letters reproduced in Sect.~\ref{sec:letters}.
\begin{description}
\item {\bf March 15th} : in Kellinghusen (postwar letters from here on)
\item  {\bf March 16th} : \T \ and \W \ are in Kellinghusen, on the veranda, it is a beautiful day; late in the night  \T\ reaches Hamburg during an air raid
\item{\bf March 17th} :  Early in the morning \T \ is arrested, brought to the Gestapo and then to prison 

\item{\bf March 19th} :  \W \ and Kollath arrive and obtain for  him to be in a  cell separate from other prisoners
\item {\bf March 23rd} : \T \ is desperate and contemplates suicide
\item {\bf March 24th} : \W \ comes visiting 
\item {\bf April 10 or 11}: \W \ comes and tells \T\ that papers for his release have been signed
\item {\bf April 12th} :  \T \ is in march toward the  Kiel concentration camp, together with other 200 prisoners; he is shot at and left for dead, after going to a hospital, he is again imprisoned
\item  {\bf April 30} : \T\ is set free
\item {\bf May 3rd} : The British troops occupy Hamburg
\end{description}

\end{document}